\documentclass[12pt,a4paper]{article}

\usepackage{amssymb}
\usepackage[figuresright]{rotating}
\usepackage{graphicx}
\usepackage{xspace}
\usepackage{units}
\usepackage{xcolor}
\usepackage{subfig}
\usepackage{multirow}
\usepackage{cleveref}
\usepackage{authblk}

\def\Xmax{\ifmmode {X_\mathrm{max}}\else{$X_\mathrm{max}$}\fi\xspace}%
\def\sigmaXmax{\ifmmode {\mathrm{RMS}(X_\mathrm{max})}\else{RMS$(X_\mathrm{max})$}\fi\xspace}%
\def\meanXmax{\ifmmode {\langle X_\mathrm{max}\rangle}\else{$\langle X_\mathrm{max}\rangle$}\fi\xspace}%
\def\eV{\ifmmode {\mathrm{\ e\kern -0.1em V}}\else
\textrm{e\kern -0.1em V}\fi\xspace}%
\def\gcm{\ifmmode {\mathrm{g/cm}^2}\else{g/cm$^2$}\fi\xspace}%
\newcommand{\energy}[1]{\unit[$10^{#1}$]{\eV}}
\newcommand{\Conex}{{\scshape Conex}\xspace}
\newcommand{\Corsika}{{\scshape Corsika}\xspace}
\newcommand{\Sibyll}{{\scshape Sibyll2.1}\xspace}

\newcommand{\Qgsjet}{{\scshape QGSJetII}\xspace}

\title{Comparison of the moments of the \Xmax distribution predicted
  by different cosmic ray shower simulation models}

\author[1]{Carlos Jose Todero Peixoto\thanks{toderocj@ursa.ifsc.usp.br}}
\author[1]{Vitor de Souza\thanks{vitor@ifsc.usp.br}}
\author[2]{Jose Bellido\thanks{jose.bellidocaceres@adelaide.edu.au}}
\affil[1]{Instituto de F\'{\i}sica de S\~ao Carlos - Universidade de S\~ao Paulo - Brazil}
\affil[2]{University of Adelaide, Adelaide, S.A. 5005, Australia}

\begin{document}

\maketitle

\begin{abstract}
In this paper we study the depth at which a cosmic ray shower reaches its
maximum (\Xmax) as predicted by Monte Carlo simulation. The use of
\Xmax in the determination of the primary particle mass can only be
done by comparing the measured values with simulation predictions. For
this reason it is important to study the differences between the
available simulation models. We have done a study of the first and
second moments of the \Xmax distribution using the \Corsika and \Conex
programs. The study was done with high statistics in the
energy range from $10^{17}$ to \energy{20.4}. We focus our analysis in
the different implementations of the hadronic interaction models
\Sibyll and \Qgsjet in \Corsika and \Conex. We show that the
predictions of the \meanXmax and \sigmaXmax depend slightly on the
combination of simulation program and hadronic interaction
model. Although these differences are small, they are not negligible
in some cases (up to 5 g/cm$^2$ for the worse case) and they should be
considered as a systematic uncertainty of the model predictions for
\meanXmax and \sigmaXmax. We have included a table with the suggested
systematic uncertainties for the model predictions.  Finally, we present a
parametrization of the \Xmax distribution as a function of mass and
energy according to the models \Sibyll and \Qgsjet, and showed an
example of its application to obtain the predicted \Xmax distributions
from cosmic ray propagation models.

\end{abstract}

\section{Introduction}
\label{sec:introduction}

The cosmic ray composition at the highest energies is probably the
most difficult and most meaningful question yet to be solved in the
present astroparticle physics scenario. Due to the unknown strength
and structures of the  magnetic fields in the Universe, anisotropy
studies are also intrinsically dependent on the
mass composition and a better identification of the sources
is probably only going to be possible if the cosmic ray composition is
known beforehand.

The most reliable technique to infer the mass composition of showers with
energy above $10^{17}$ eV is the
determination of the \Xmax and posterior comparison of the measured
values with predictions from Monte Carlo simulation.  This is because
above $10^{17}$ eV fluorescence detectors can measure \Xmax with a
resolution of 20 \gcm. The evolution of
the detectors, the techniques used to measure the atmosphere, the
advances in the understanding of the fluorescence emission and the
development of innovative analysis procedures have resulted in a high
precision measurement of \Xmax and its moments. The Pierre Auger
Observatory~\cite{collaboration_measurement_2010}, the HiRes
Experiment~\cite{bird_evidence_1993}, the Telescope Array~\cite{ta_er}
and the Yakutsk array~\cite{yaku_er} quote systematic uncertainties in the
determination of the \meanXmax to be 12, 3.3, 15 and 20 \gcm,
respectively.  Considering the quoted errors and taking into account
that the data have to be compared to simulation predictions, it is very
important to understand the details and reduce the differences between
simulation programs.
The proposed experiment JEM-EUSO~\cite{bib:jem:euso} is also going to
use the fluorescence technique to detect air shower from the
space. For this reason,
we extended the analysis done in this work up to \energy{20.4} within
the energy range aimed by JEM-EUSO.

The dependency of \Xmax with primary energy and mass (A) has been
analytically studied in a hadronic cascade
model~\cite{matthews_heitler_2005}.  Monte Carlo
programs can simulate the hadronic cascade in the atmosphere using
extrapolation from the measured hadronic cross sections at somewhat
lower energy. It has been shown before that different hadronic
interaction models do not agree in the prediction of the \meanXmax and
other parameters~\cite{knapp_extensive_2003}.

In this paper we study in detail the dependence of \meanXmax and
\sigmaXmax as a function of energy and primary mass. We compare the
result of two hadronic interaction models. We have done a high
statistics study and we show that the discrepancies between models and
programs are at the same level of quoted systematic uncertainties of the
experiments. The analysis done here points to the need of a better
understanding of the interaction properties at the highest energies
which can be achieved by ongoing analysis of the LHC data which
already resulted in updates of the hadronic interaction models. At the same
time the results presented in this paper point to discrepancies
between different implementations of the same hadronic interaction
model which need to be better understood.

We also present a parametrization of the \Xmax distribution as a
function of mass and energy. Several theoretical models have predicted
the mass abundance based on astrophysical
arguments~\cite{bib:allard,bib:biermann,bib:berezinsky,bib:hillas}. In
order to compare the predicted abundance with measurements, one has to
convert the calculated flux for each particle into \Xmax. Until now,
this could only be done using full Monte Carlos simulations. We
present here a parametrization of the \Xmax distribution  to
allow the conversion of astrophysical models into \Xmax
measurements. Parametrizations of \meanXmax as a function of
energy and mass have been already
studied~\cite{matthews_heitler_2005}. What we present here is a step
forward, we show the parametrization of the \Xmax distribution which
is good enough to calculate the first and second moments of the
distribution.

In section ~\ref{sec:simulation}, we study the dependence of the
results regarding simulation limitations like thinning and nucleon
types. In section~\ref{sec:results}, we compare the models. In
Section~\ref{sec:par:xmax} we show the parametrization of the \Xmax
distribution.  Section~\ref{sec:conclusion} concludes our analysis.

\section{Shower Simulation}
\label{sec:simulation}

In this work we have used
\Conex~\cite{bergmann_one-dimensional_2007,pierog_first_2006}  and
\Corsika~\cite{heck_monte-carlo_1998} shower simulators. \Conex uses a
one dimensional hybrid approach combining Monte Carlo simulation and
numerical solutions of cascade equations. \Corsika describes the
interactions using a full three dimensional Monte Carlo algorithm. By
using analytical solutions,
\Conex saves computational time. On the other hand, \Corsika makes use
of the thinning algorithm~\cite{hillas__1981,hillas__1985} to reduce
simulation time and output size.

Both approaches have negative and positive features. \Corsika offers a
full description of the physics mechanisms and a three dimensional
propagation of the particles in the atmosphere. However, it is very
time consuming, limiting studies which depend on large number of events
at the highest energies. The thinning algorithm introduces spurious
fluctuations that have to be taken into account in the final
analysis. \Conex is fast, but on the other hand it offers only a one
dimensional description of the shower. The use of intermediate
analytical solutions might also reduce the intrinsic fluctuation of
the shower.  In the following sections both programs are
compared in detail concerning the \Xmax calculations.

The hadronic interaction models for the highest energies were
developed independently of the programs that describe the showers. For
each shower simulator many hadronic interaction models are
available. We have used
\Qgsjet.v03~\cite{ostapchenko_re-summation_2006,ostapchenko_nonlinear_2006}
and \Sibyll~\cite{fletcher_sibyll:_1994} in this work. For the low
energy hadronic interaction we have used GHEISHA~\cite{bib:gheisha} in
all simulations.

Showers have been simulated with primary energy ranging from
$10^{17.0}$ to $10^{20.4}$ eV in steps of $\log_{10}{(E/eV)} = 0.1$. We
have simulated seven primary nuclei types with mass: 1, 5, 15, 25, 35,
45 and 55.  For each primary particle, primary energy, and hadronic
interaction model combination, a set of 1000 showers has been
simulated. The zenith angle of the shower was set to
60$\textordmasculine$ and the observation height was at sea level
corresponding to a maximum slant depth of 2000 \gcm allowing
the simulation of the entire longitudinal
profile of the showers. The longitudinal shower profile was sampled in
steps of 5 \gcm. The energy thresholds in \Corsika and \Conex were
set to 1, 1, 0.001 and 0.001 GeV for hadron, muons, electrons and
photons respectively.

\subsection{Fitting the longitudinal development of the shower}

For all studies in this paper, \Xmax was calculated by fitting a
Gaisser-Hillas~\cite{t._k._gaisser__1977}  function to the energy
deposited by the particle through the atmosphere. We chose a four
parameter Gaisser-Hillas (GH4) function given by:

\begin{equation}
  \frac{dE}{dX}(X) =  {\frac{dE}{dX}}^{max} \left ( \frac {X - X_0} {
    \Xmax  -X_0 } \right ) ^ { \frac {\Xmax - X_0} {\lambda} } \exp{
    \left ( \frac{\Xmax -X} {\lambda} \right ) }
\end{equation}

in which $\frac{dE}{dx}^{max}$, $X_0$, $\lambda$ and \Xmax are the
four fitted parameters and $X$ is the slant atmospheric depth. The first
guess of the \Xmax parameter in the fitting procedure was chosen to be
the maximum of a three degree polynomial interpolated within the three
points in the longitudinal profiles with largest $\frac{dE}{dX}$. The
full simulated profile was fitted.

We have studied the effect of fitting a different function to the
longitudinal profile. Instead of a Gaisser-Hillas function with four
parameters, we have also fitted a Gaisser-Hillas function with 6 parameters
(GH6). In this function, $\lambda$ is defined as $\lambda = a \times
X^2 + b \times X + c$ in which $a$, $b$ and $c$ are also fitted.  The
differences in \Xmax and \sigmaXmax calculations for both fitted
functions were smaller than 2.5 \gcm and 0.6 \gcm,
respectively. Anyway, there might be a systematic effect in the
determination of the \Xmax due to the fitting procedure and the fitted
function chosen to describe the longitudinal profile, which is not
investigated in this paper.

\subsection{Thinning analysis}
\label{thinning}

In order to save time and output size, \Corsika uses a thinning
algorithm~\cite{hillas__1981,hillas__1985}. The thinning factor
$f_{thin}$ defines the fraction of the primary particle energy $E_0$
below which not all particles in the shower are followed. Particles with energy below
$E_{cut}$, where $E_{cut} = f_{thin} * E_0$, are sampled, some are
discarded and others followed. Each active particle in \Corsika has a
weight attribute which compensates for the energy of the rejected ones
such as that energy is conserved.

\begin{figure}
  \centering
  \subfloat[THIN = $10^{-5}$]{\includegraphics[width=0.3\textwidth]{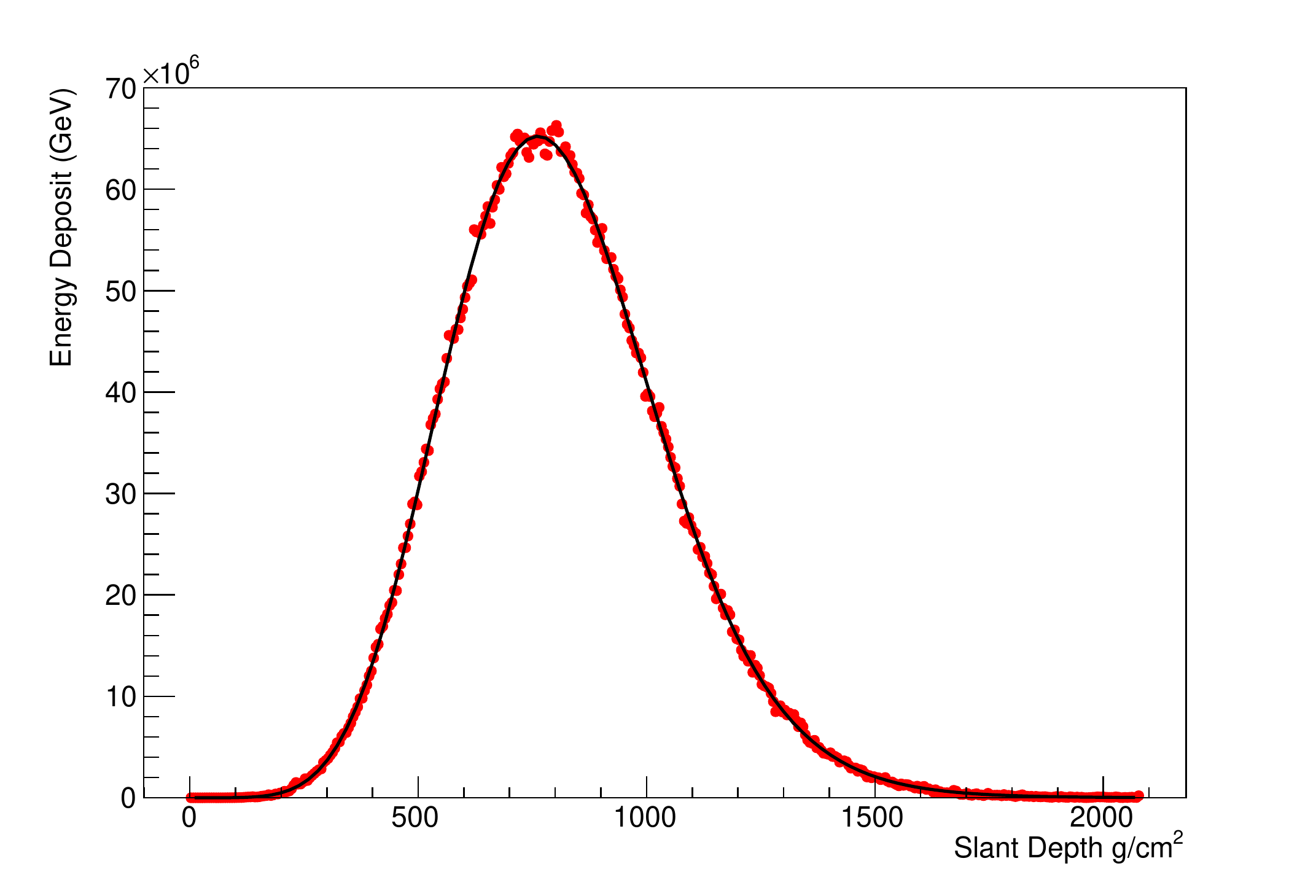}}
  \subfloat[THIN = $10^{-6}$]{\includegraphics[width=0.3\textwidth]{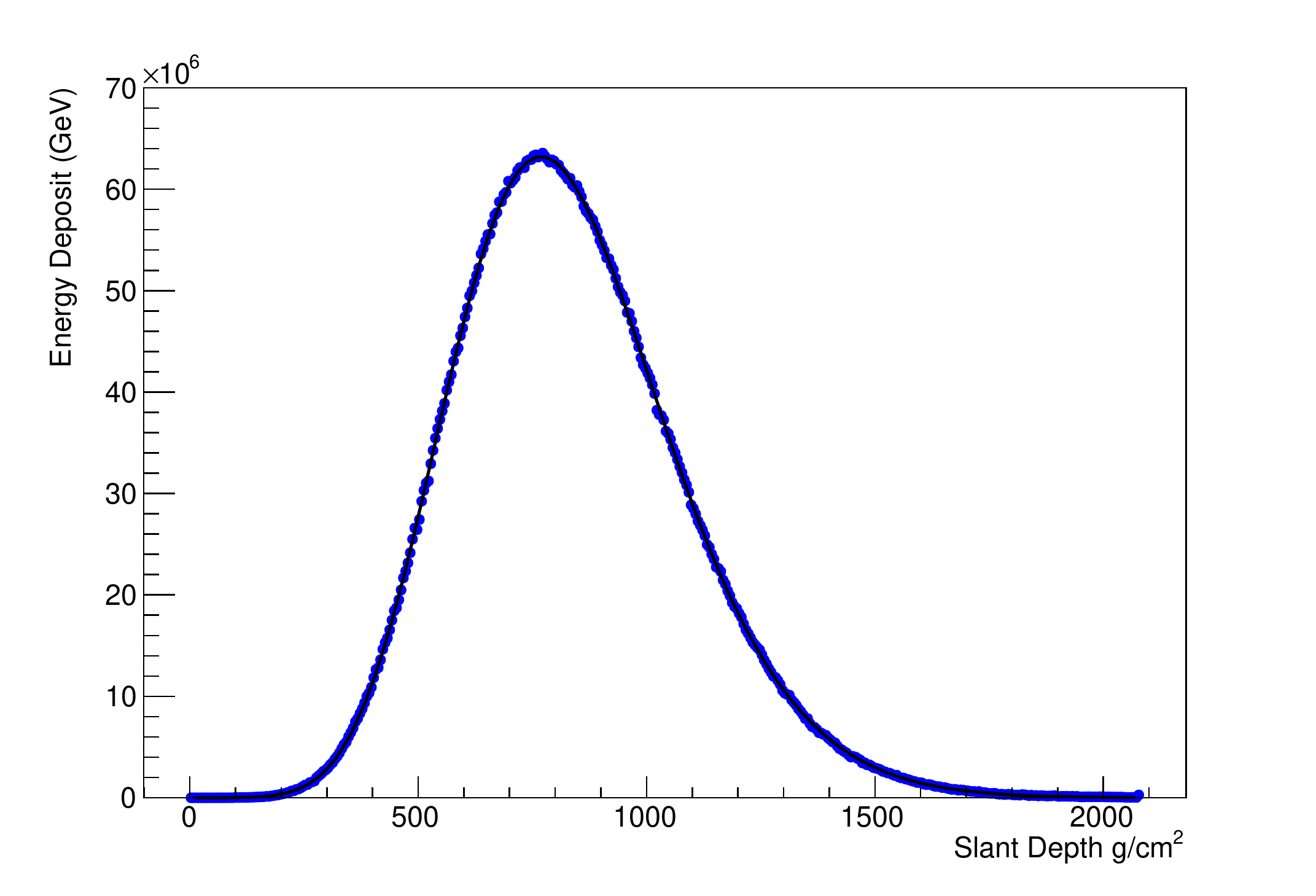}}
  \subfloat[THIN = $10^{-7}$]{\includegraphics[width=0.3\textwidth]{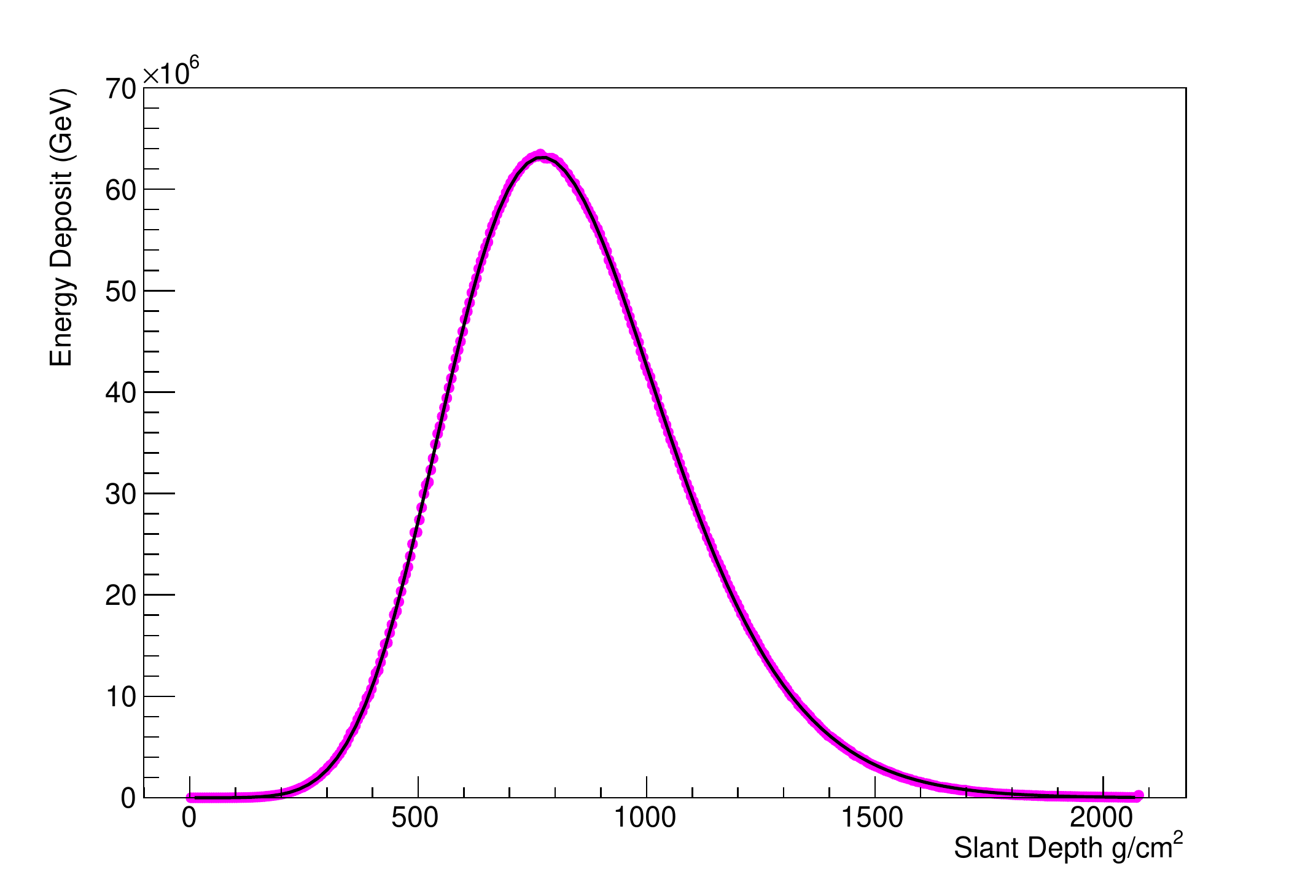}}\\
  \caption{Longitudinal development of proton showers with energy
    $10^{19}$ eV for different thinning factors. Each figure shows the
  development of one shower.\Sibyll was used for this study.}
  \label{fig:thin:long}
\end{figure}

The thinning algorithm causes artificial fluctuations in the
calculation of the shower development which needs to be taken into
account. For example, figure~\ref{fig:thin:long} shows the
longitudinal development of one shower simulated with three thinning
factors $10^{-5}$, $10^{-6}$ and $10^{-7}$. The first interaction
altitude was fixed at 60 km and the target is Nitrogen nuclei. \Sibyll
was used for this study. It
illustrates how the fluctuations of the simulated longitudinal profile
increase with increasing thinning factor.

\begin{figure}[!htb]
  \centerline{\includegraphics[width=0.7\textwidth]{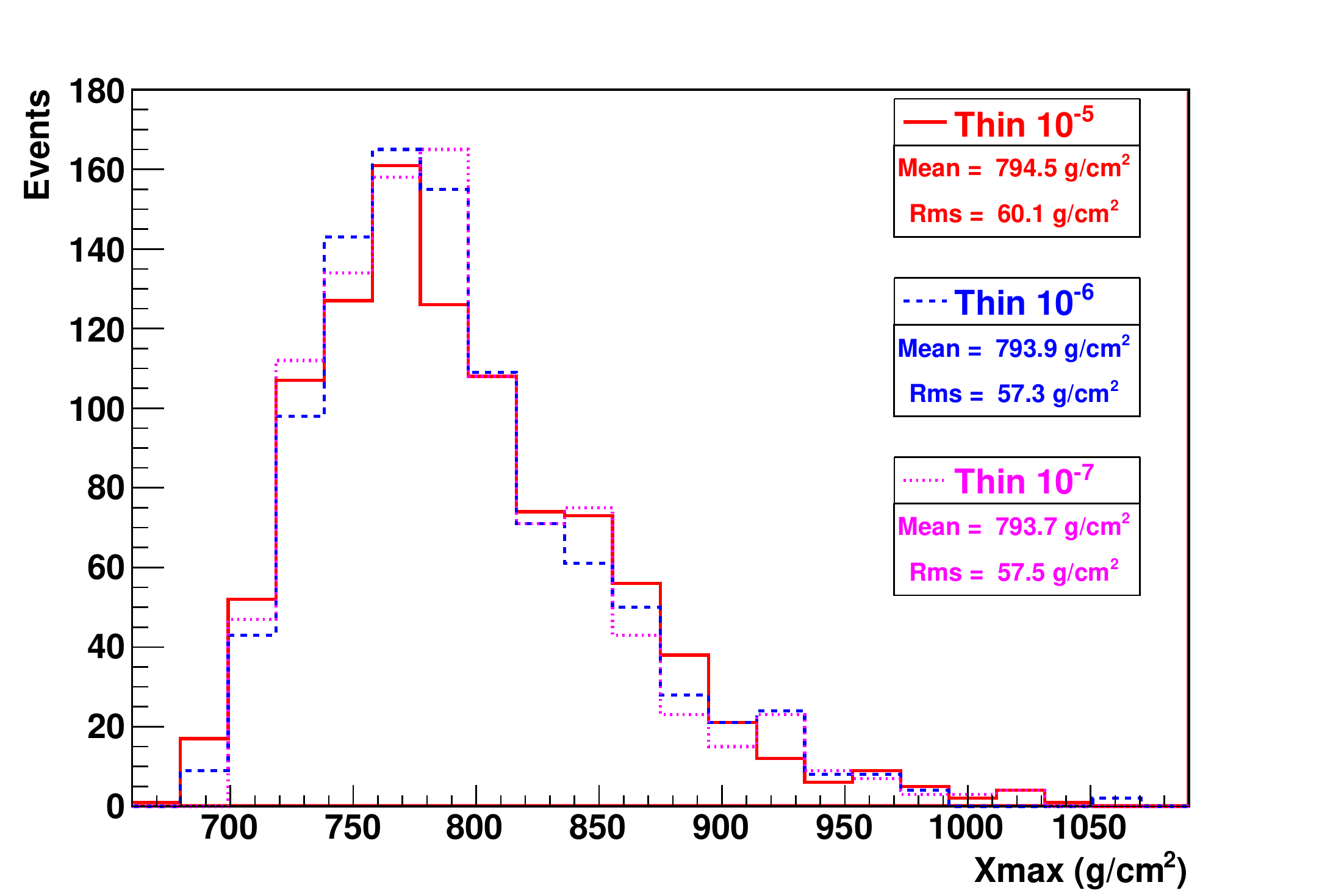}}
  \caption{\Xmax distribution for 1000 proton showers with $10^{19}$
    eV simulated with different thinning factors. \Sibyll was used for this study.}
  \label{fig:thin:xmax}
\end{figure}

Figure~\ref{fig:thin:xmax} shows the \Xmax distribution for 1000
simulated shower for three thinning factors. In this example the first
interaction point and target were not fixed. \Sibyll
was used again for this study.
Figure~\ref{fig:thin:xmax} shows that the \meanXmax and \sigmaXmax are
very similar for any thinning factor used. The maximum difference is
0.4 \gcm for \meanXmax and 2.8 \gcm for \sigmaXmax. Based on this
study we chose to simulate all showers with thinning factor $10^{-5}$.

\subsection{Isobaric Nuclei Analysis}
\label{sec:nucleon}

Some of the simulation used in this paper have been produced
for another study and have been re-used in this work to save
computational time. The previously simulated showers have exotic
primary particle with unstable number of protons and neutrons. In
this section we compare the longitudinal development of showers
started with exotic nuclei with
the shower started with stable nuclei with the same total number of
nucleons. Our intention is to show that the development of a shower
at high energies does not depend on the number of protons and
neutrons independently but depends only on the total number of
nucleons considered.

\Corsika and \Conex simulators differentiate isobaric nuclei by
allowing the determination of the number of protons and neutrons
of the primary nuclei. However the treatment of the first
interaction is done by the hadronic interaction models. \Qgsjet does
not differentiate isobaric nuclei interactions, \Sibyll does
differentiate.

At the highest energies the energy loss of protons and
neutrons is negligible when compared to the total energy and therefore
only the total number of nucleons should influence the development of
the shower. On the other hand, Coulomb dissociation should also be
taken into account~\cite{coloumb:dissociation}. Nevertheless, none of
the hadronic interaction models available include this effect and
therefore the development of the simulated shower should not depend on
the number of protons and neutrons in the nuclei.

\begin{figure}
  \centering

  \subfloat[Atomic Weight: 1 - SIBYLL.]{\includegraphics[width=0.3\textwidth]{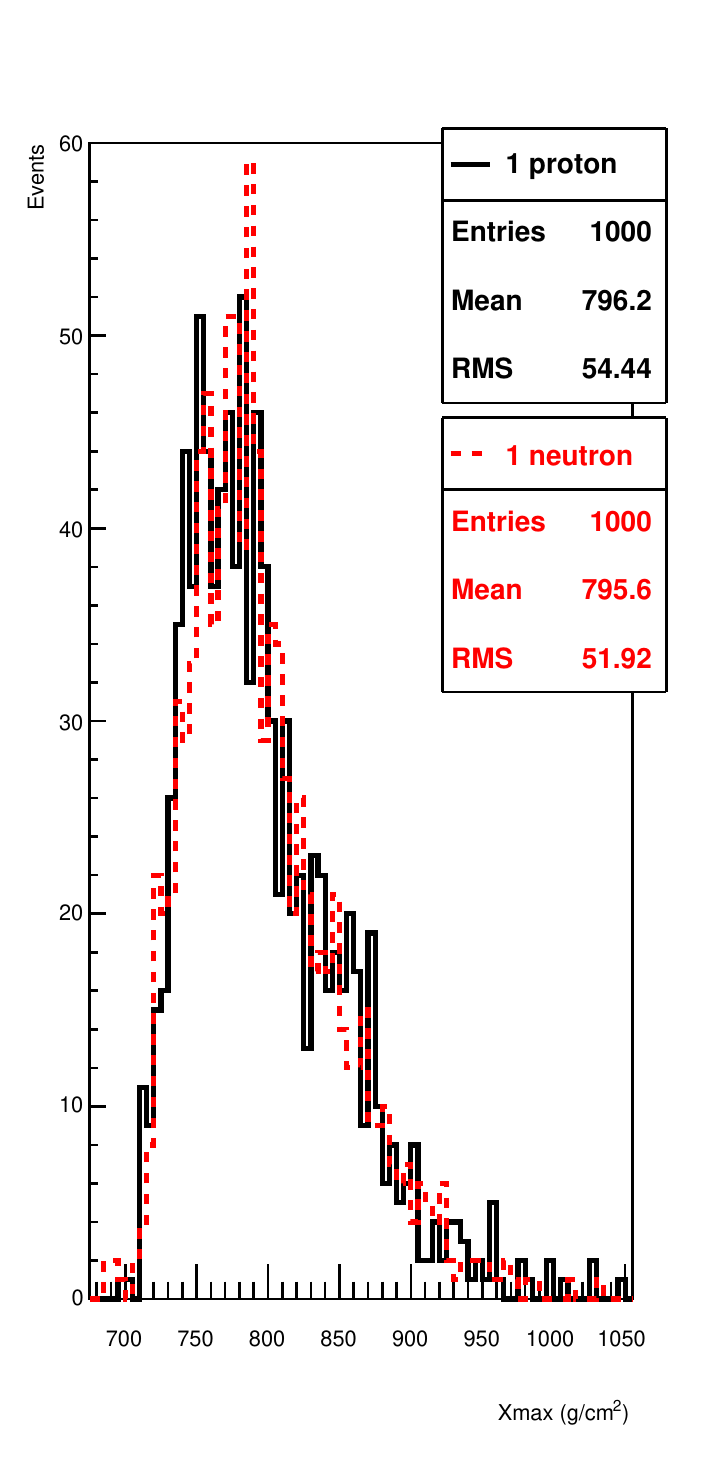}}
  \subfloat[Atomic Weight: 24 - SIBYLL.]{\includegraphics[width=0.3\textwidth]{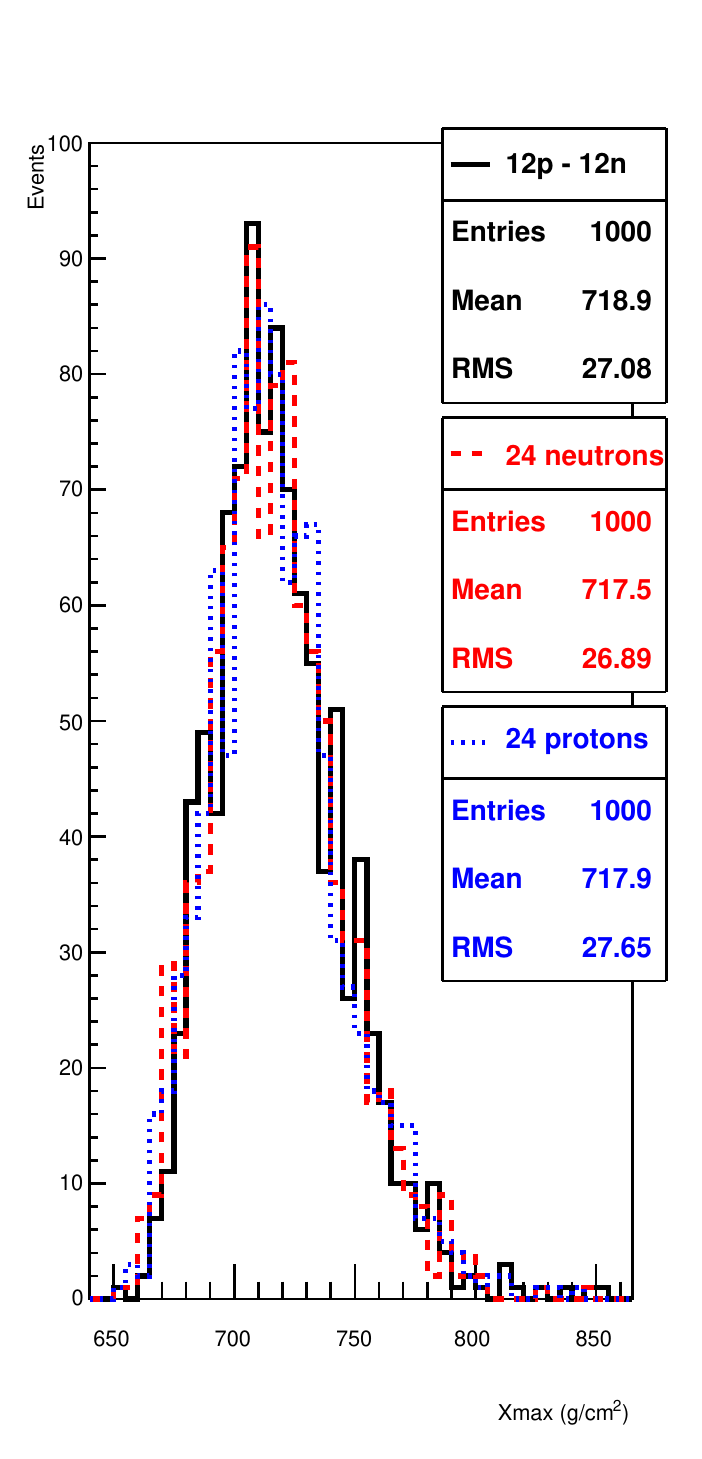}}
  \subfloat[Atomic Weight: 56 -
    SIBYLL.]{\includegraphics[width=0.3\textwidth]{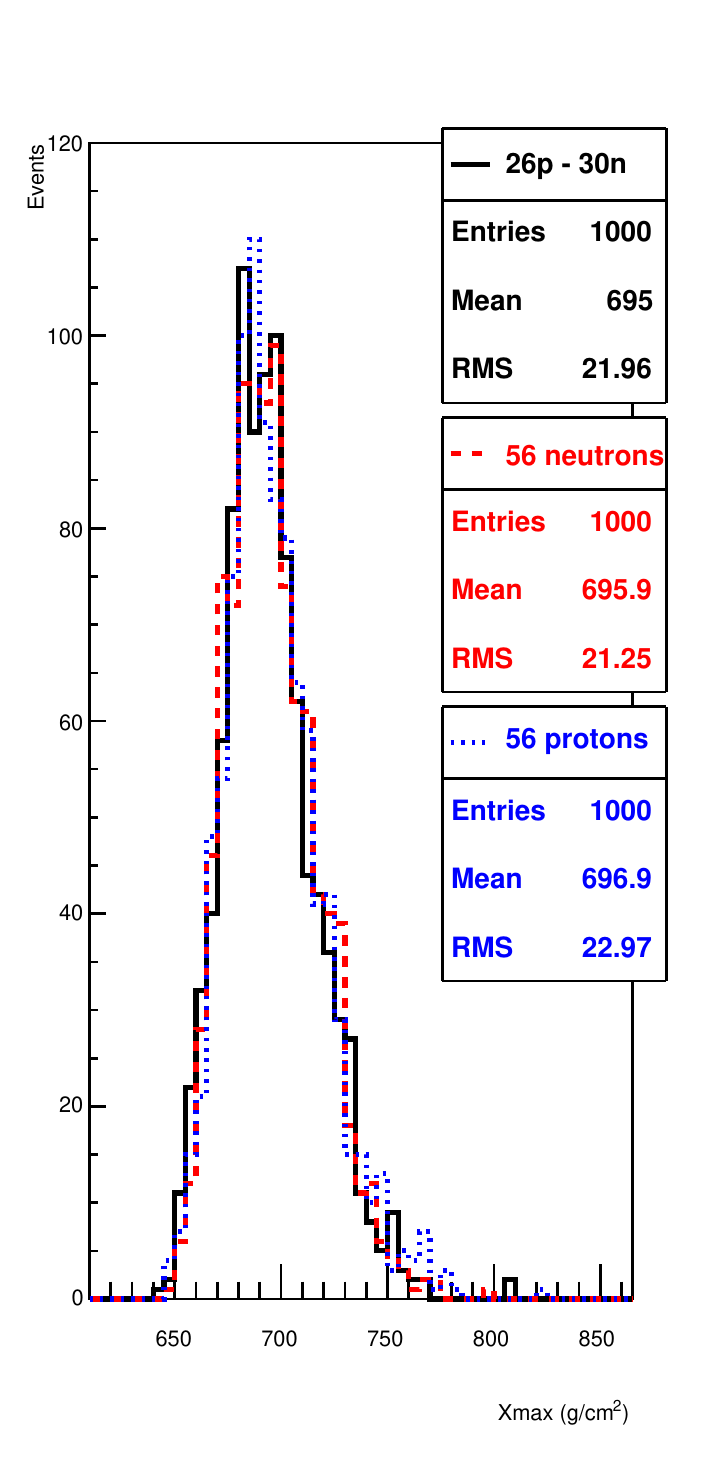}}\\

  \subfloat[Atomic Weight: 1 - QGSJETII.]{\includegraphics[width=0.3\textwidth]{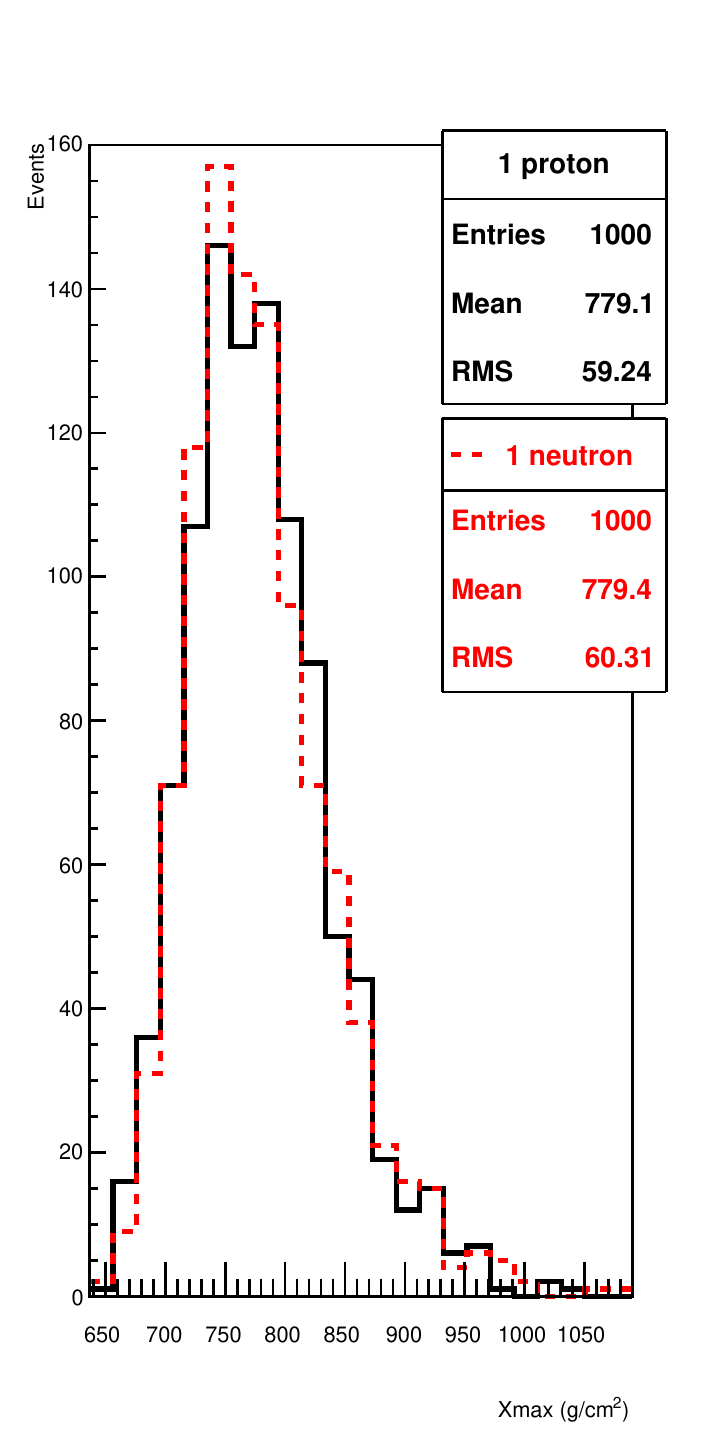}}
  \subfloat[Atomic Weight: 24 - QGSJETII]{\includegraphics[width=0.3\textwidth]{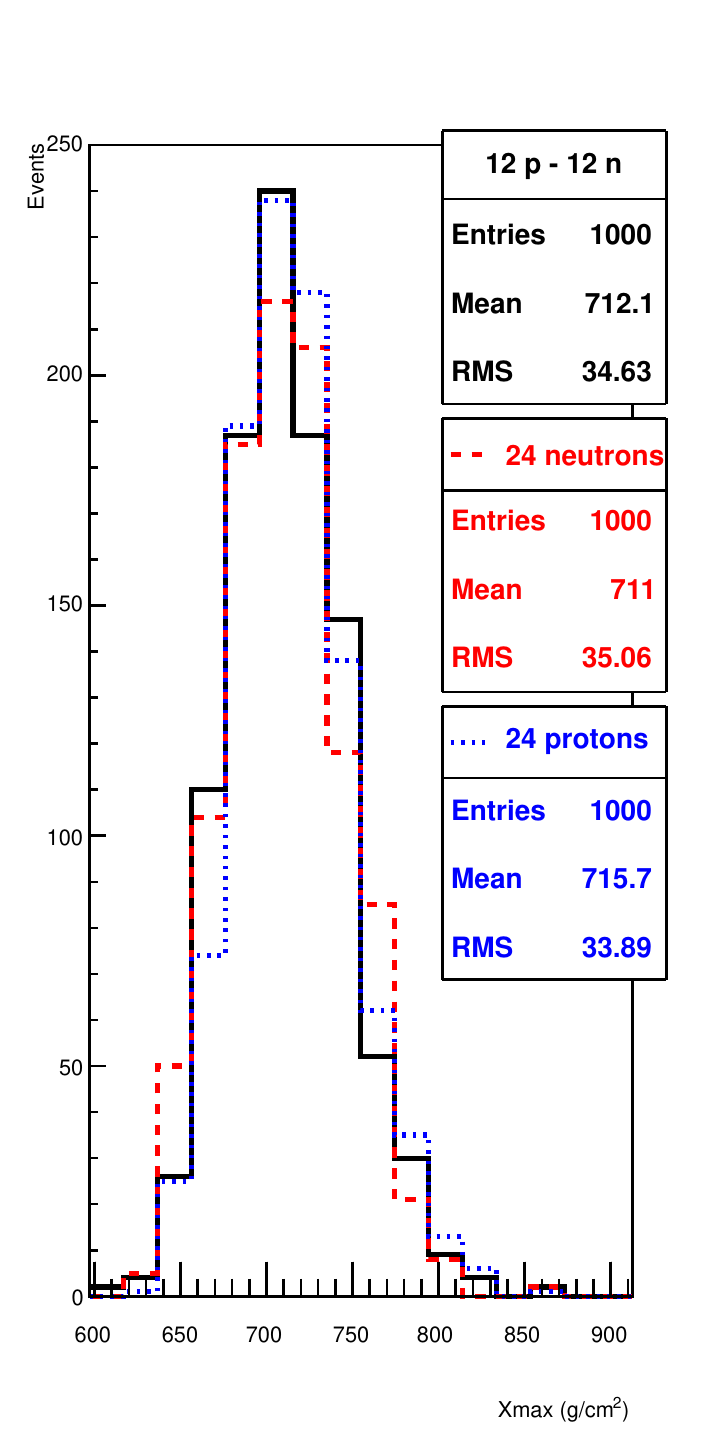}}
  \subfloat[Atomic Weight: 56 - QGSJETII]{\includegraphics[width=0.3\textwidth]{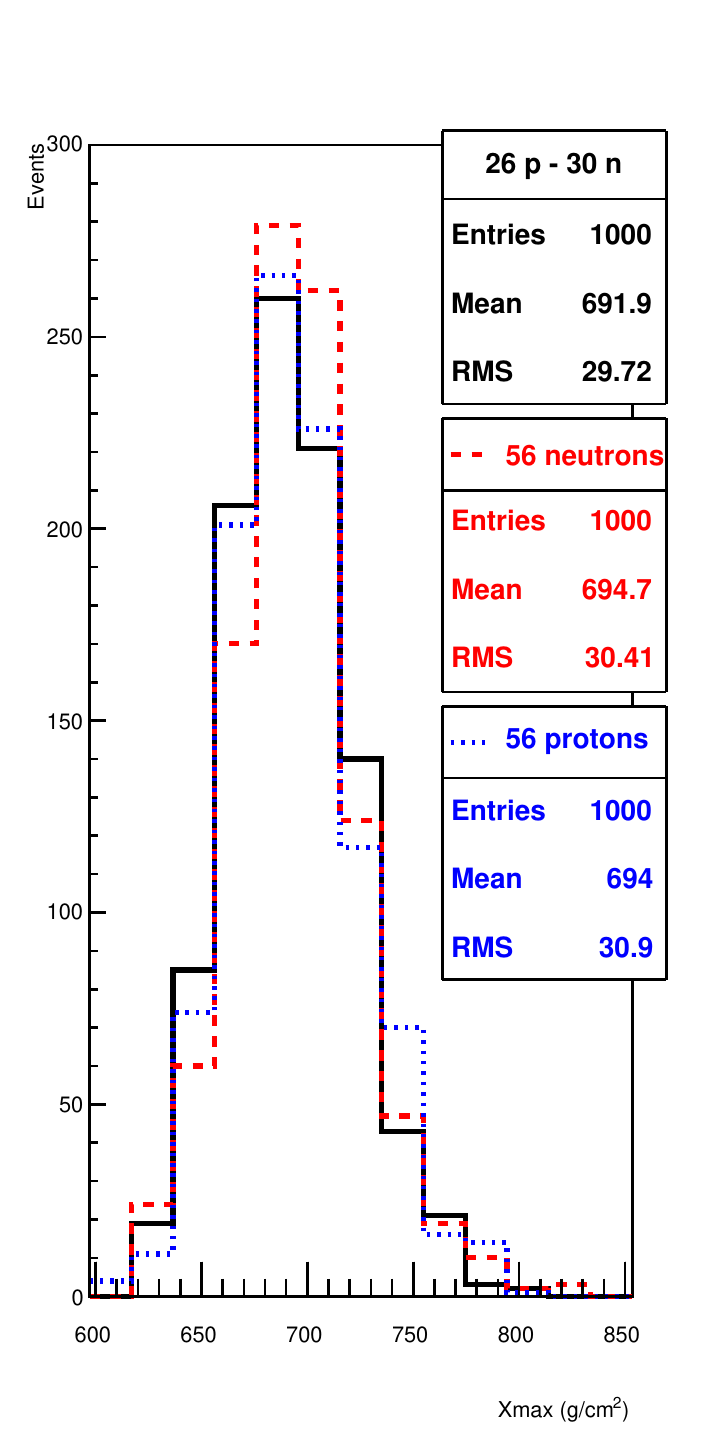}}\\

\caption{Study on the \Xmax distribution for nuclei with different
  nucleon constitutions.}
\label{fig:xmax:nucleons}

\end{figure}

Figure~\ref{fig:xmax:nucleons} shows the \Xmax distribution for
$10^{19}$ eV showers. We have simulated nuclei with different numbers
of proton and neutron constituents. We can conclude that, at the energy
range of interest, the predicted \meanXmax and the \sigmaXmax does not
depend on the number of protons and neutrons which form a nucleus with
given mass A. The maximum difference in the \meanXmax was 2.9 \gcm for
A = 24 and \Qgsjet and in the \sigmaXmax was 2.5 \gcm for A = 1 and
\Sibyll.

\section{Results}
\label{sec:results}

%MEAN ANALYSIS

\begin{figure}
  \centering

  \subfloat[\Conex - \meanXmax \emph{versus} energy]{\includegraphics[width=0.4\textwidth]{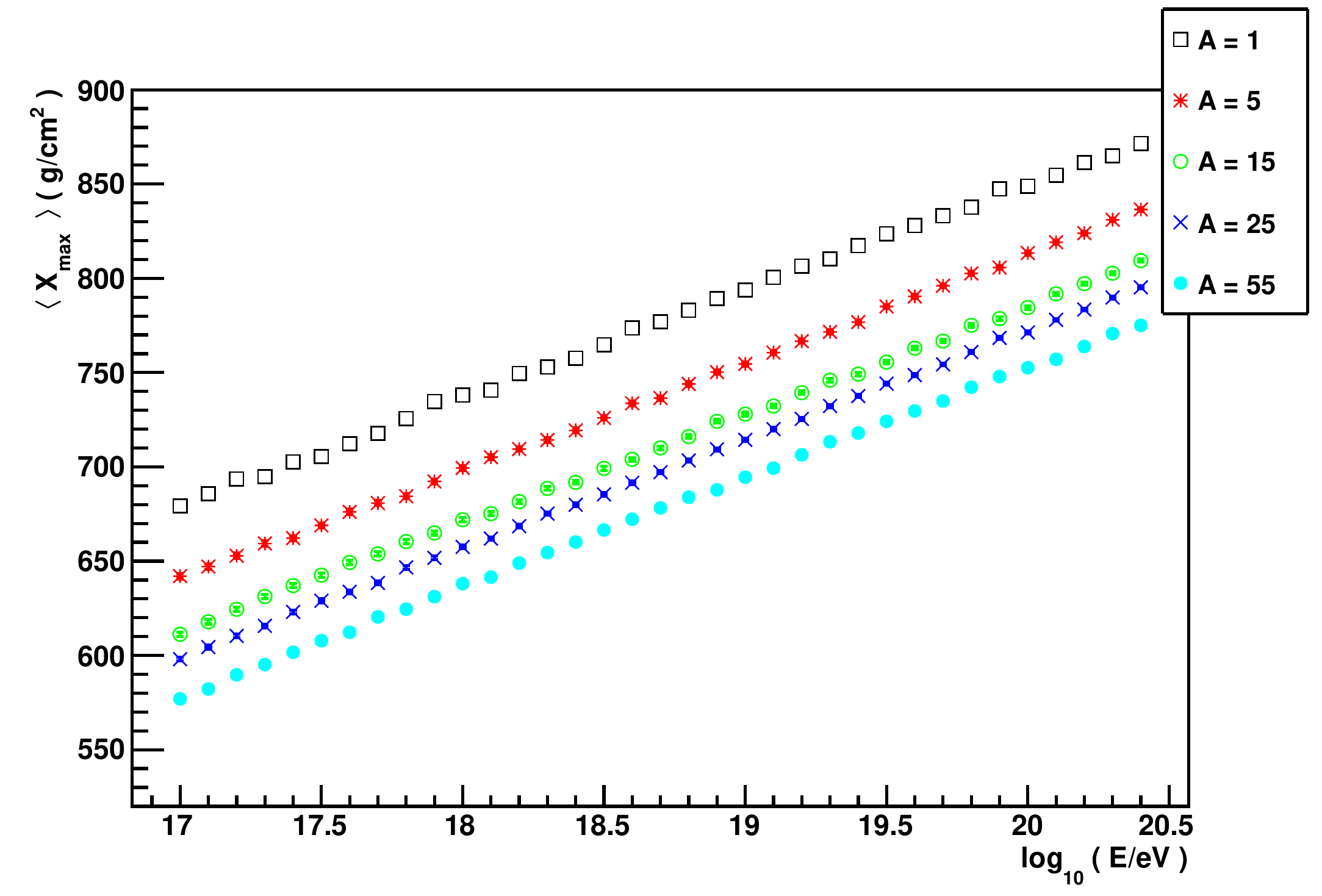}}
  \subfloat[ \Corsika - \meanXmax \emph{versus} energy
   ]{\includegraphics[width=0.4\textwidth]{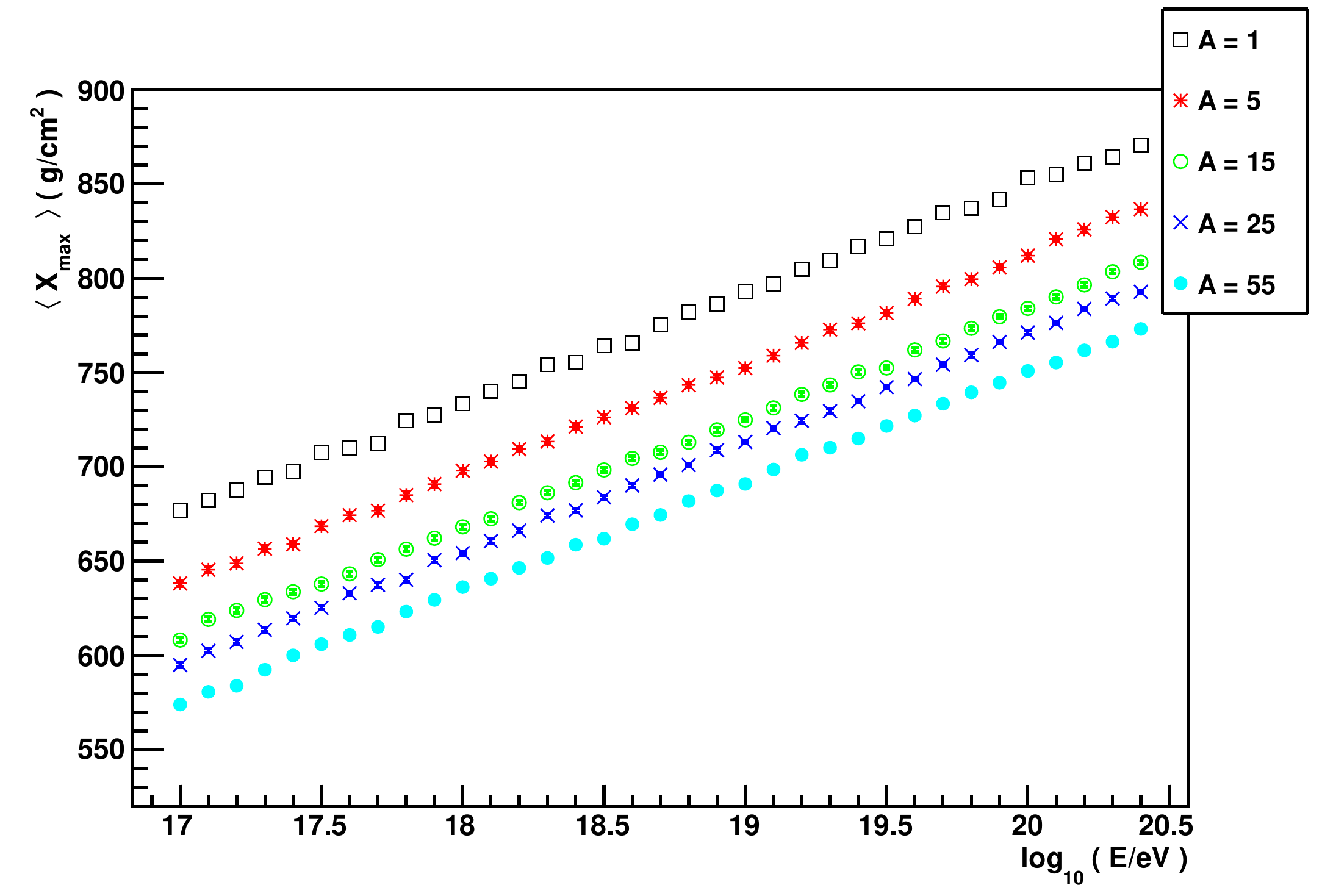}}\\

  \subfloat[\Conex - \meanXmax  \emph{versus} mass.]{\includegraphics[width=0.4\textwidth]{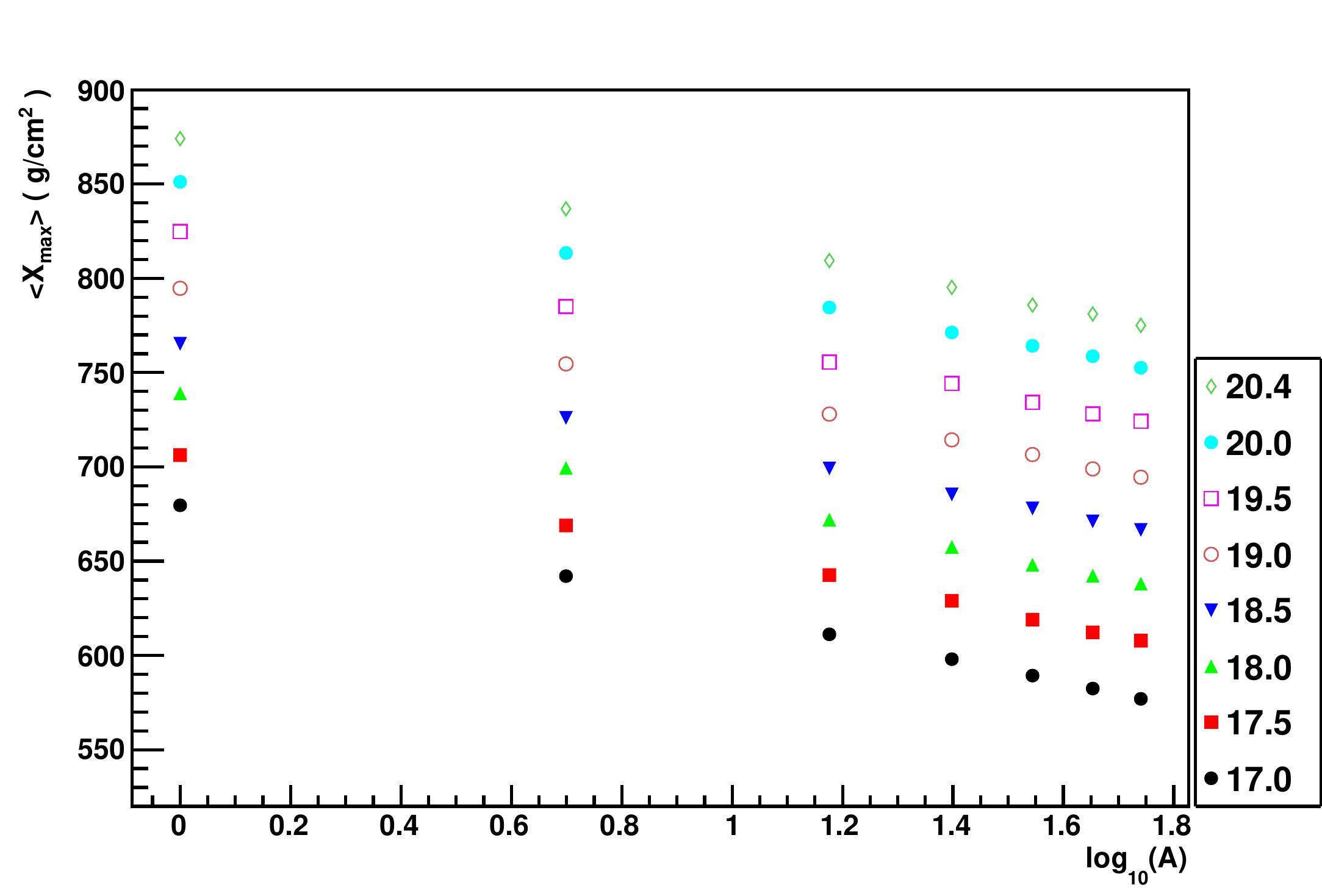}}
  \subfloat[\Corsika - \meanXmax  \emph{versus} mass.]{\includegraphics[width=0.4\textwidth]{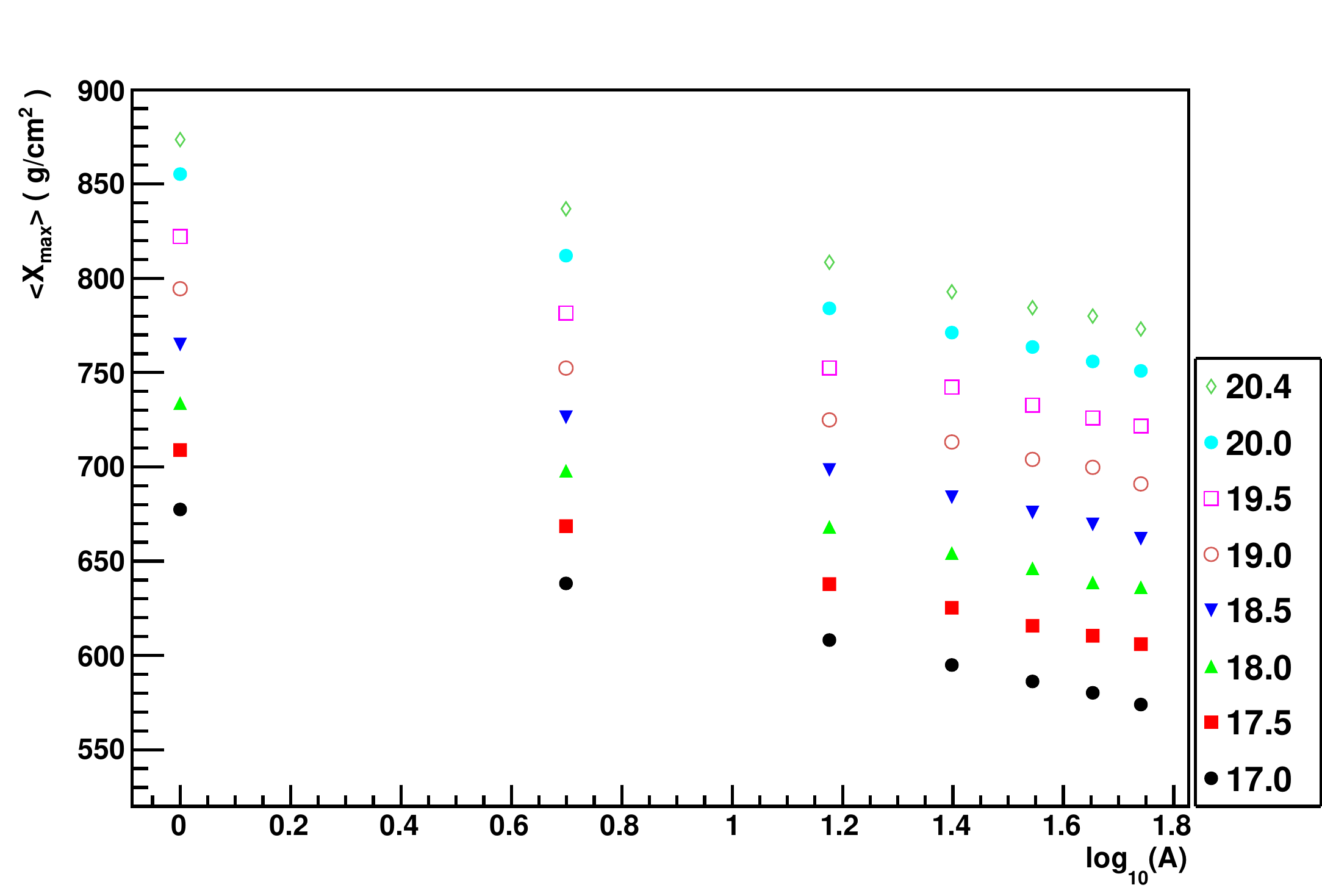}}\\

  \subfloat[Difference between \Corsika and \Conex \emph{versus}
    energy. Lines are linear fit to 1p and 55p data.]{\label{fig:mean:diff:sibyll:energy}\includegraphics[width=0.4\textwidth]{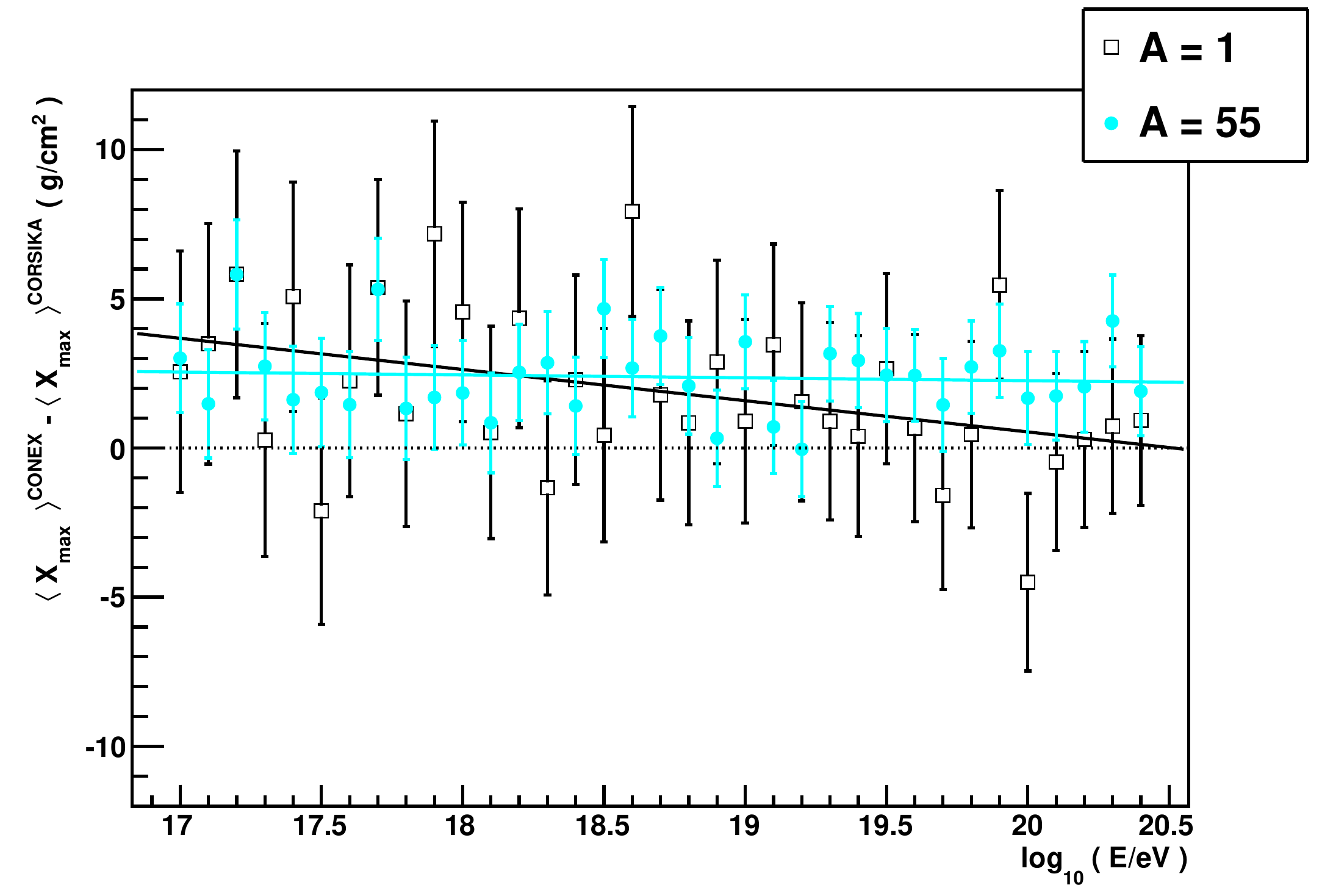}}
  \subfloat[Difference between \Corsika and \Conex \emph{versus} mass.]{\label{fig:mean:diff:sibyll:mass}\includegraphics[width=0.4\textwidth]{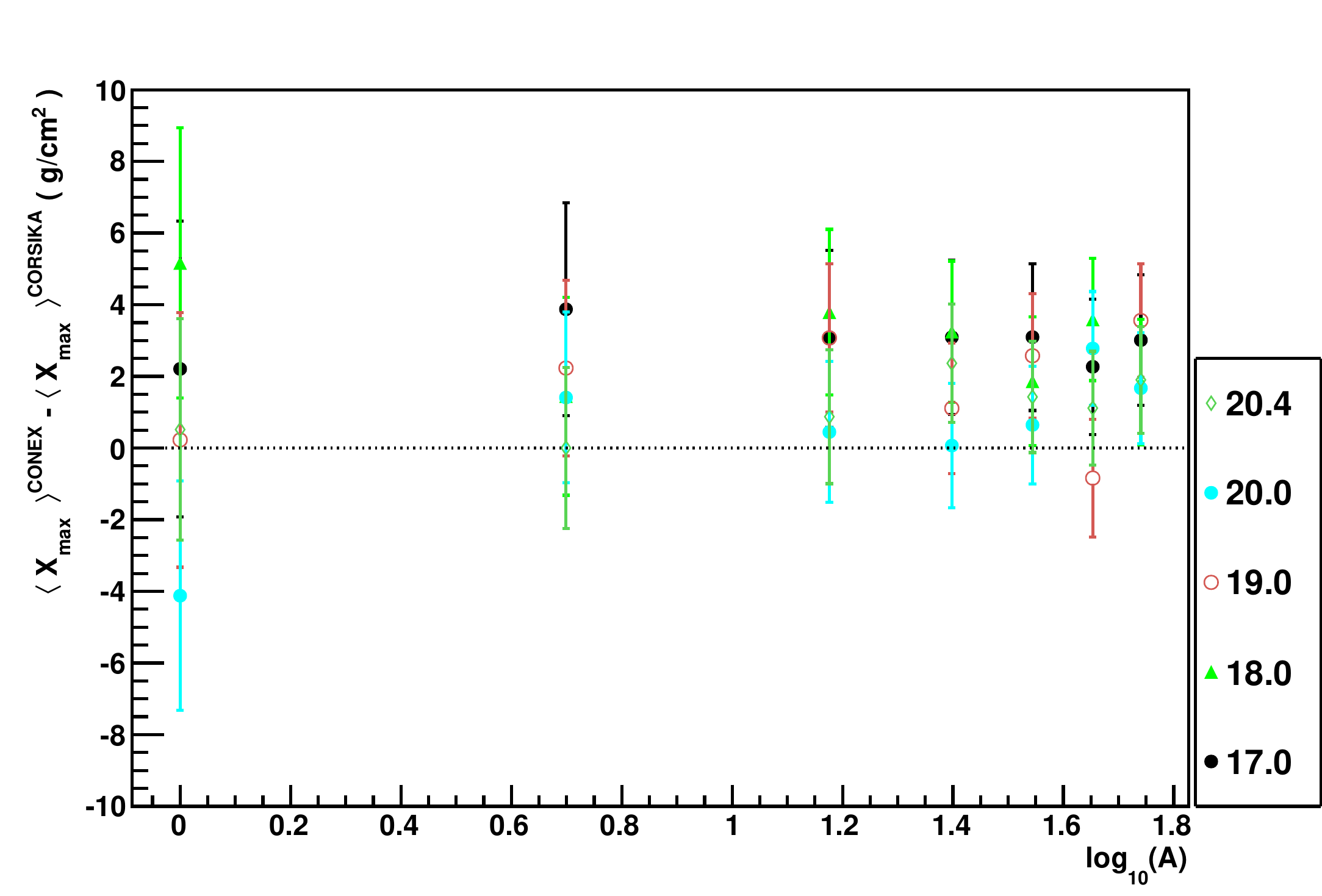}}\\

  \caption{\meanXmax as a function of energy and mass as calculated by
    \Corsika and \Conex using \Sibyll. Showers have been simulated
    with primary energy ranging from $10^{17.0}$ to $10^{20.4}$ eV in
    steps of $\log_{10}{(E/eV)} = 0.1$ and primary nuclei types with
    mass: 1, 5, 15, 25, 35, 45 and 55. A set of 1000 showers has been
    simulated for each combination. Not all energies and primaries are
    shown for clarity.}
  \label{fig:mean:xmax:corsika:conex:sibyll}

\end{figure}

\begin{figure}
  \centering

  \subfloat[\Conex - \meanXmax \emph{versus} energy]{\includegraphics[width=0.4\textwidth]{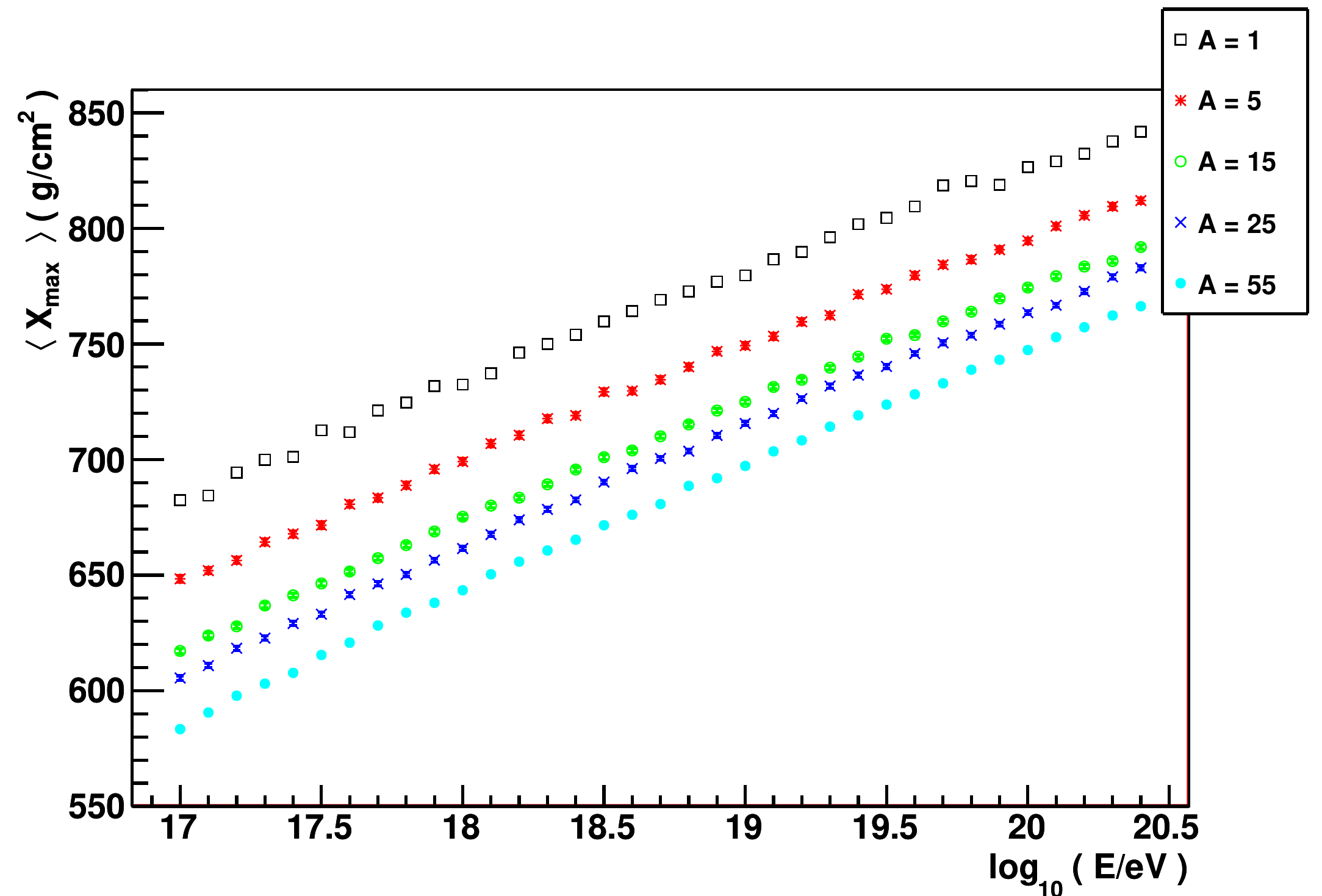}}
  \subfloat[ \Corsika - \meanXmax \emph{versus} energy
   ]{\includegraphics[width=0.4\textwidth]{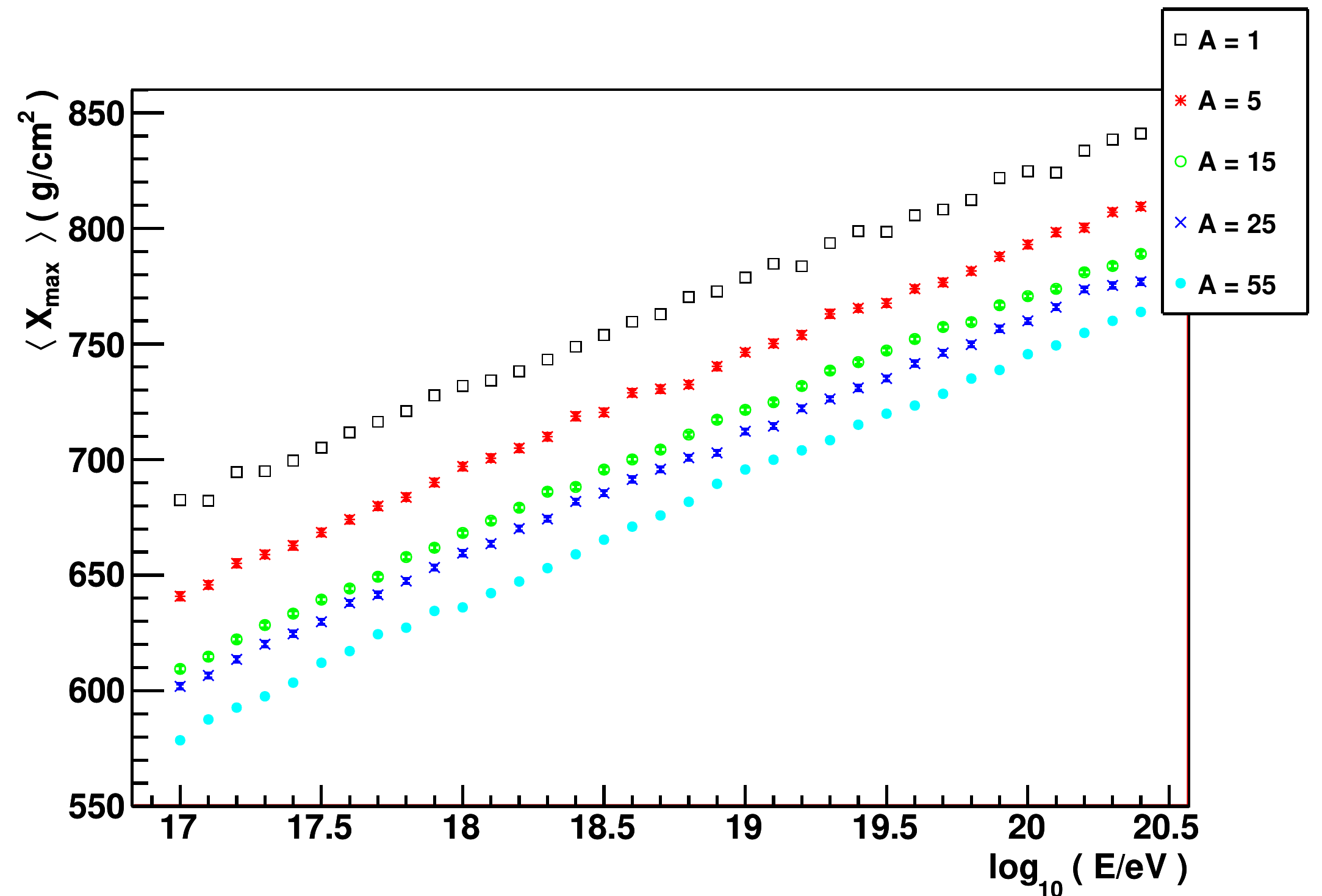}}\\

  \subfloat[\Conex - \meanXmax  \emph{versus} mass.]{\includegraphics[width=0.4\textwidth]{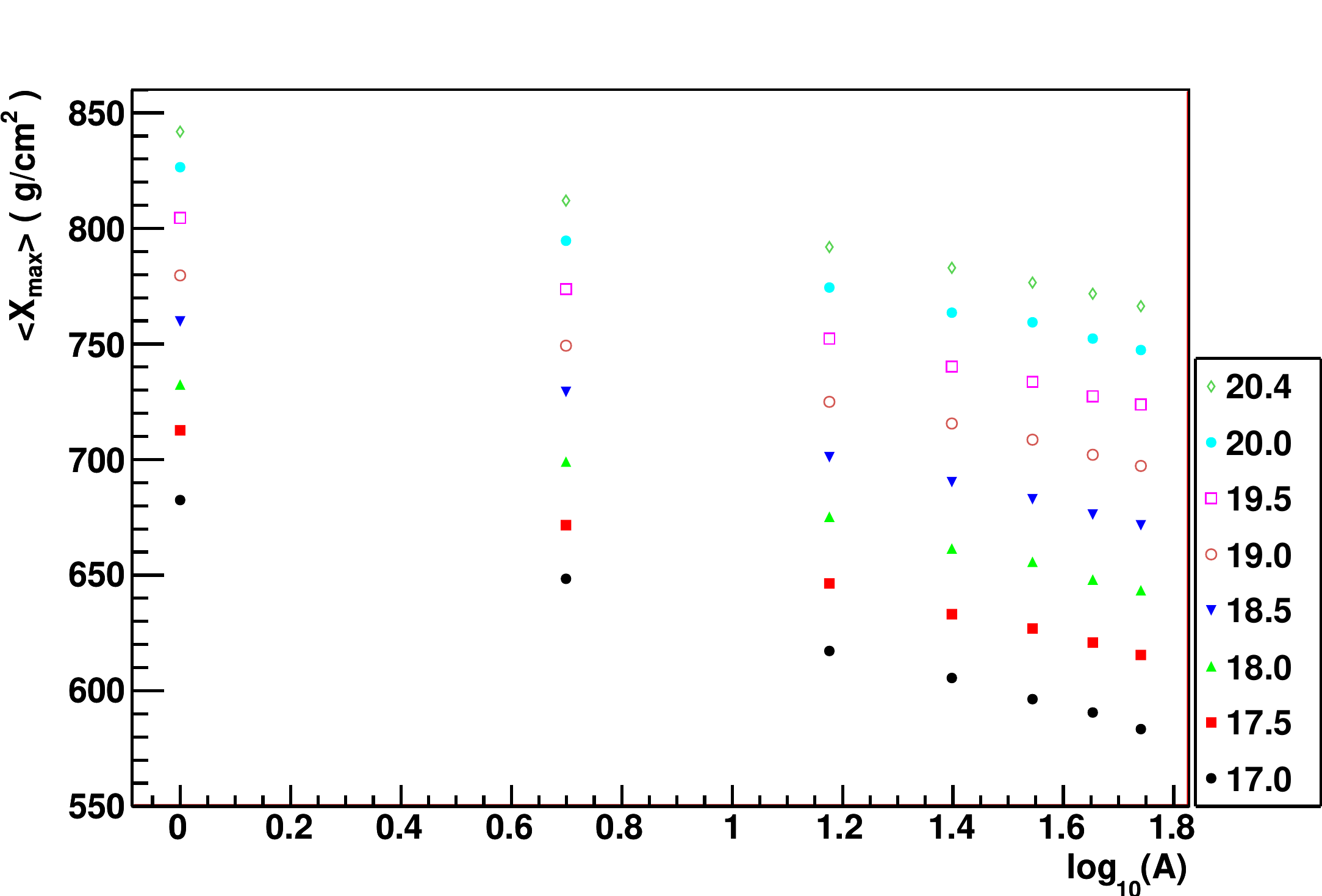}}
  \subfloat[\Corsika - \meanXmax  \emph{versus} mass.]{\includegraphics[width=0.4\textwidth]{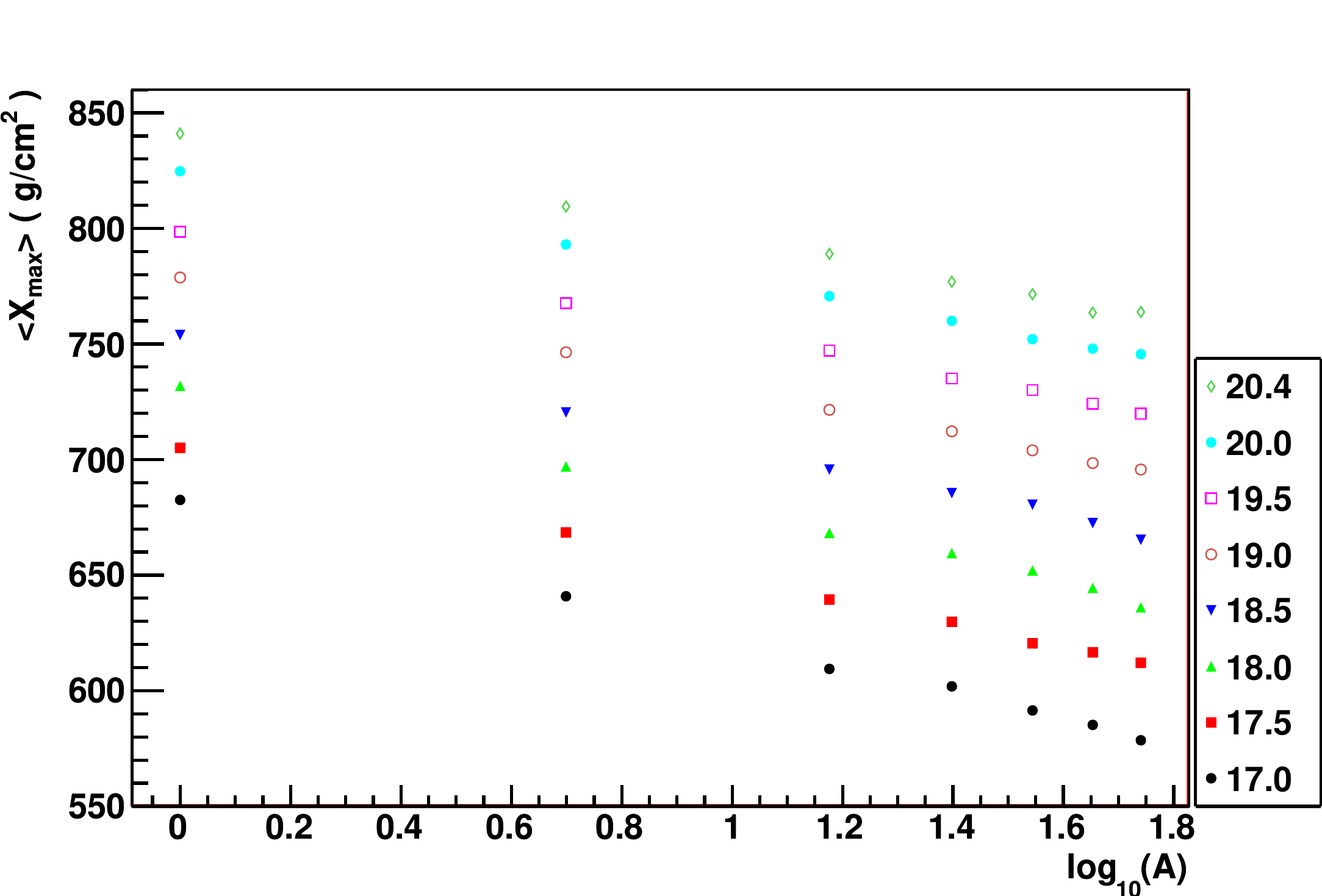}}\\

  \subfloat[Difference between \Corsika and \Conex \emph{versus}
    energy. Lines are linear fit to 1p and 55p data.]{\label{fig:mean:diff:qgsjet:energy}\includegraphics[width=0.4\textwidth]{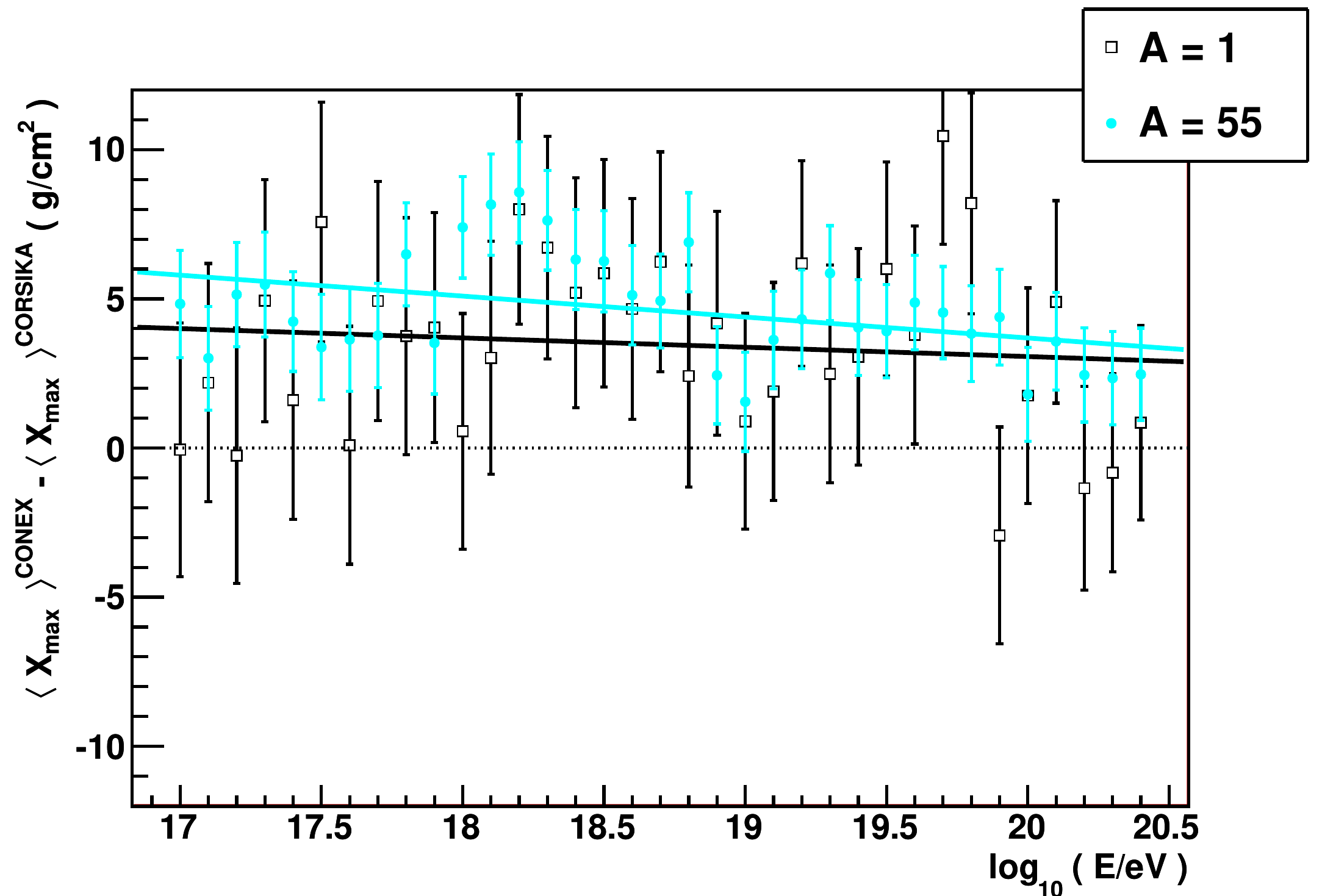}}
  \subfloat[Difference between \Corsika and \Conex \emph{versus} mass.]{\label{fig:mean:diff:qgsjet:mass}\includegraphics[width=0.4\textwidth]{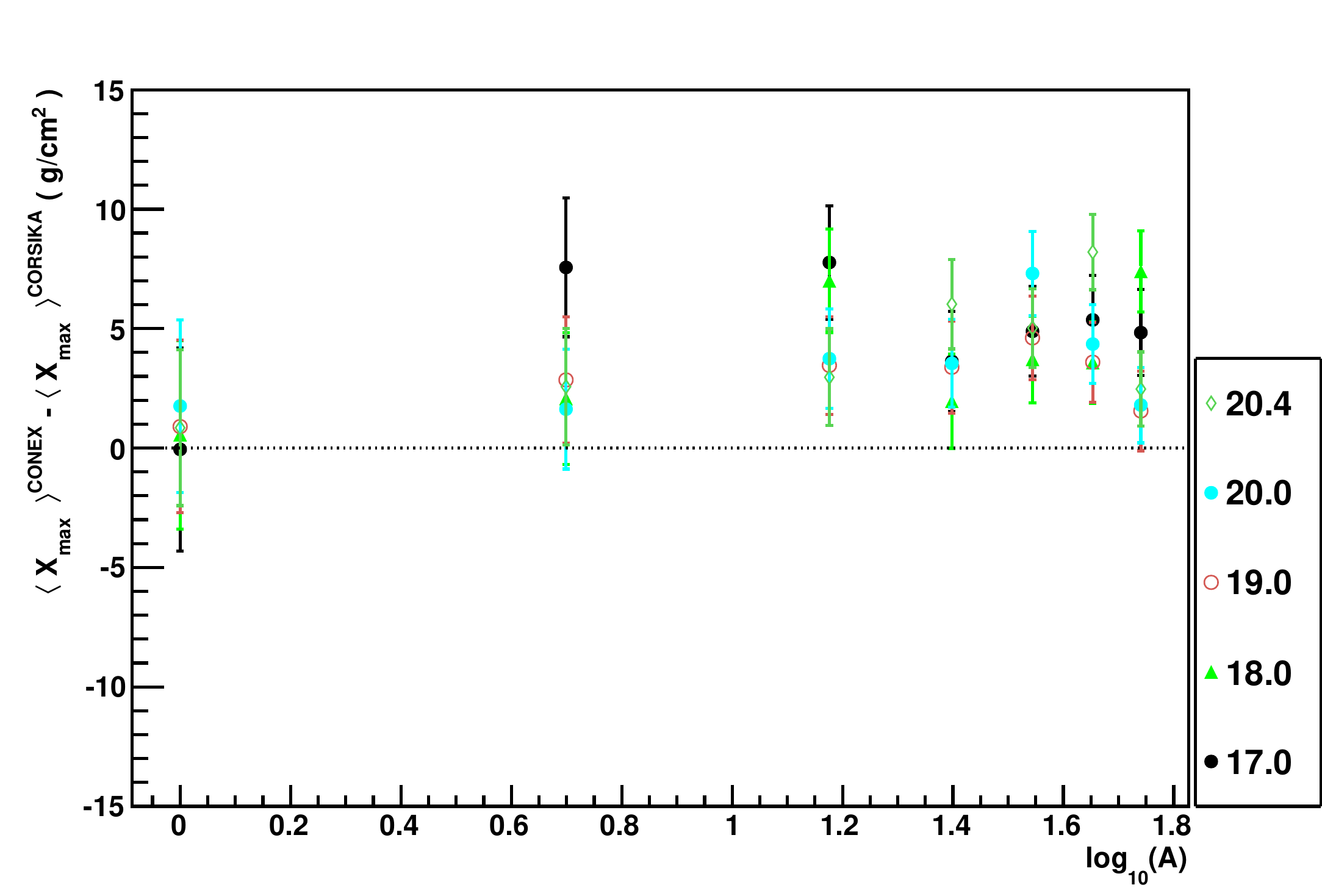}}\\

  \caption{\meanXmax as a function of energy and mass as calculated by
    \Corsika and \Conex using \Qgsjet. Showers have been simulated
    with primary energy ranging from $10^{17.0}$ to $10^{20.4}$ eV in
    steps of $\log_{10}{(E/eV)} = 0.1$ and primary nuclei types with
    mass: 1, 5, 15, 25, 35, 45 and 55. A set of 1000 showers has been
    simulated for each combination. Not all energies and primaries are
    shown for clarity.}
  \label{fig:mean:xmax:corsika:conex:qgsjet}

\end{figure}

Figures \ref{fig:mean:xmax:corsika:conex:sibyll} and
\ref{fig:mean:xmax:corsika:conex:qgsjet} show the comparisons of
\meanXmax using both simulators for \Sibyll and \Qgsjet
respectively. Figures ~\ref{fig:mean:diff:sibyll:energy} and
~\ref{fig:mean:diff:sibyll:mass} show the difference between \Corsika
and \Conex predictions as a function of energy and mass respectively
when both programs used \Sibyll. The same is shown in figures
~\ref{fig:mean:diff:qgsjet:energy} and
~\ref{fig:mean:diff:qgsjet:mass} for \Qgsjet. The differences between
the \meanXmax predicted by \Corsika and \Conex are smaller than 7 \gcm
in the parameter space studied by us. \Conex tends to simulate
showers slightly deeper than \Corsika.

% RMS ANALYSIS

\begin{figure}
  \centering

  \subfloat[\Conex - \sigmaXmax \emph{versus} energy.]{\label{fig:rms:sibyll:conex}\includegraphics[width=0.4\textwidth]{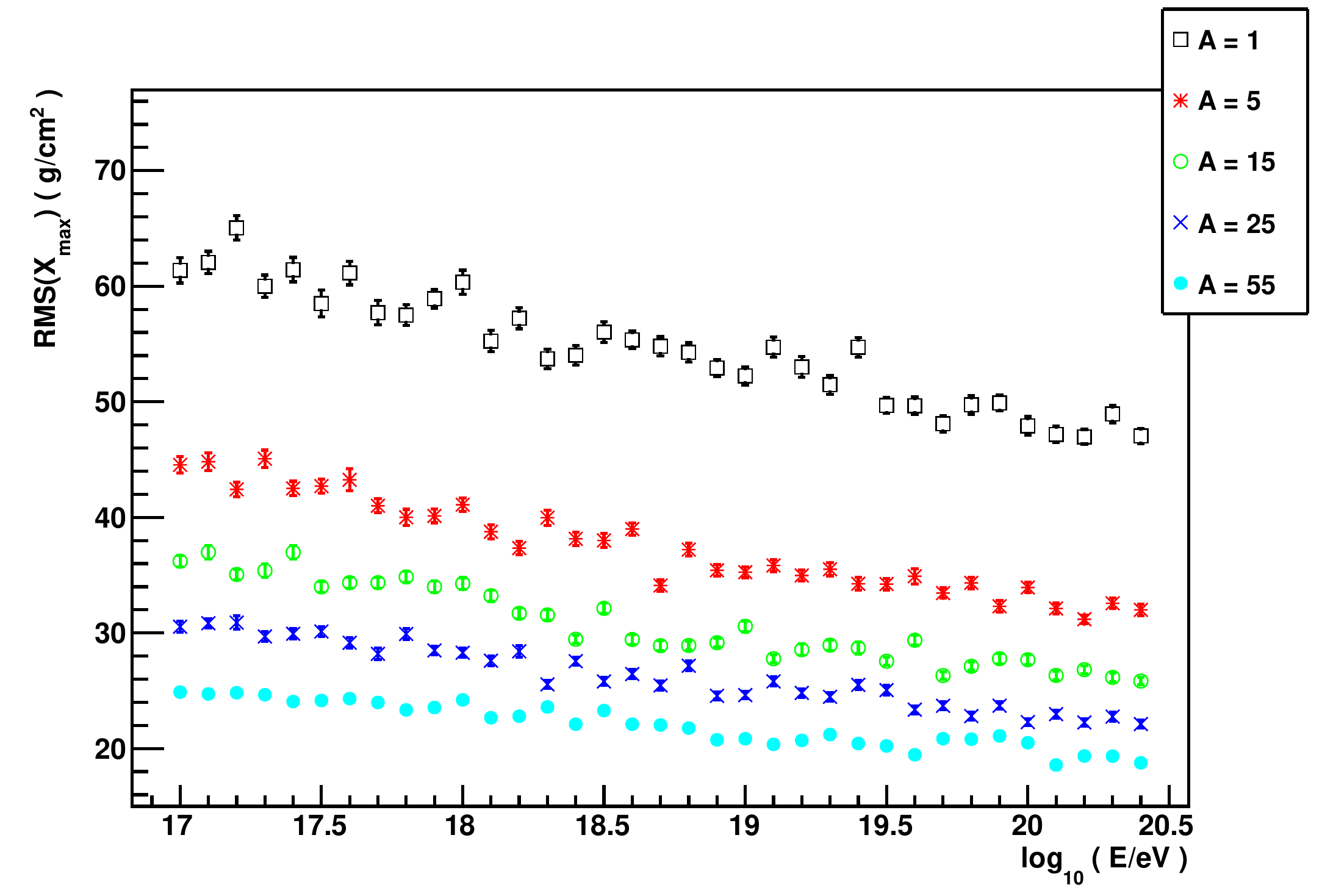}}
  \subfloat[\Corsika - \sigmaXmax \emph{versus} energy.]{\label{fig:rms:sibyll:corsika}\includegraphics[width=0.4\textwidth]{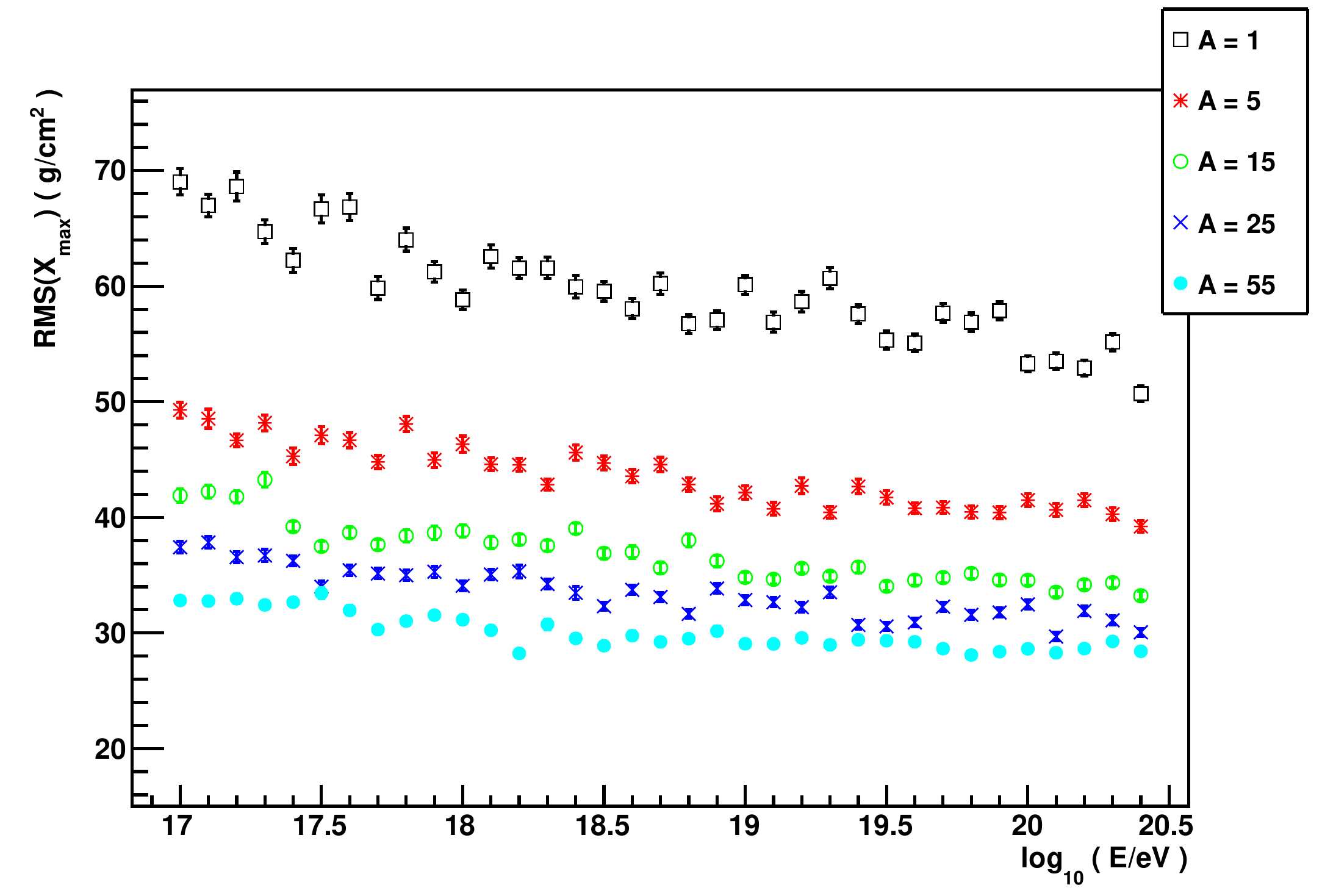}}\\

  \subfloat[\Conex - \sigmaXmax \emph{versus} mass.]{\includegraphics[width=0.4\textwidth]{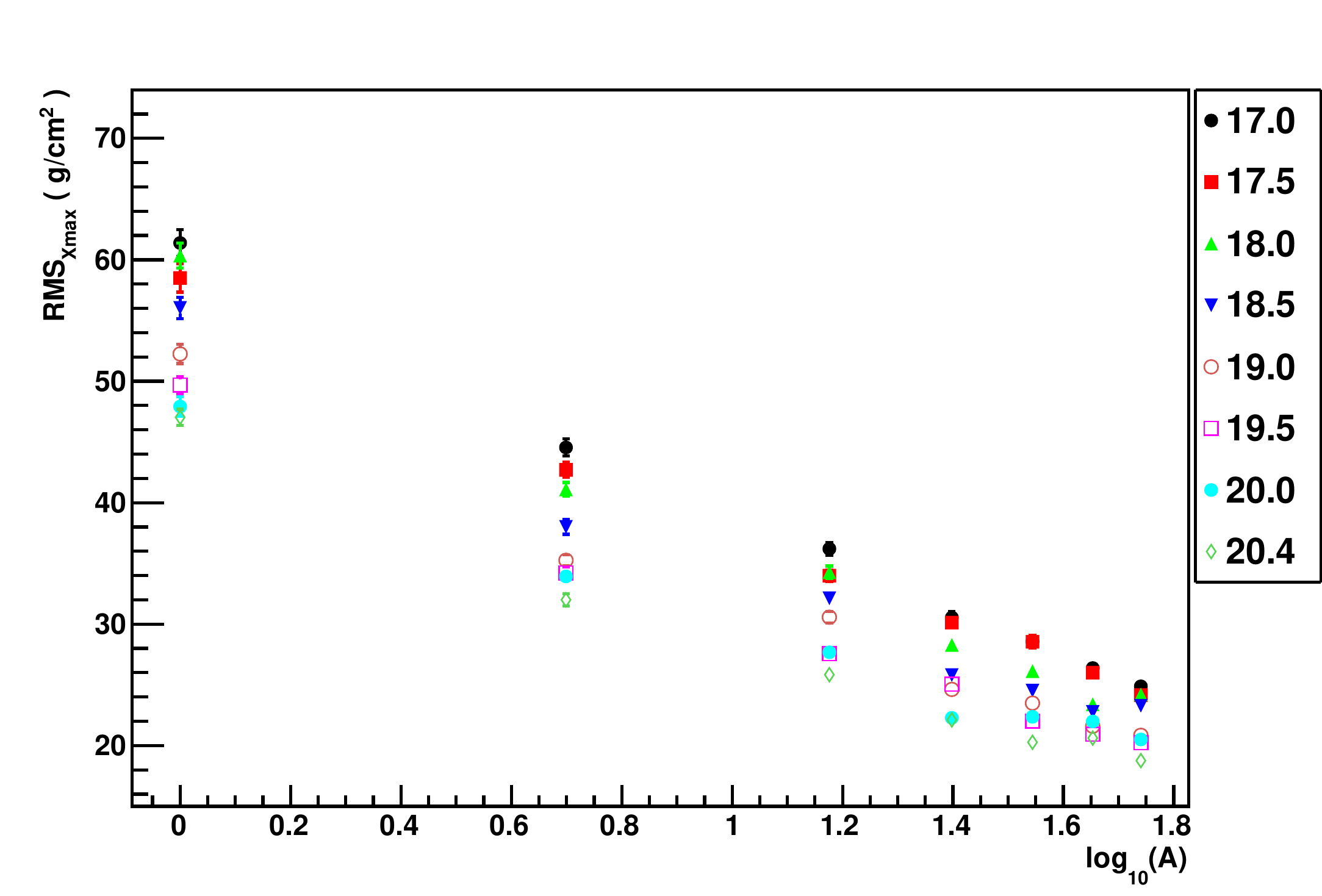}}
  \subfloat[\Corsika - \sigmaXmax \emph{versus} mass.]{\includegraphics[width=0.4\textwidth]{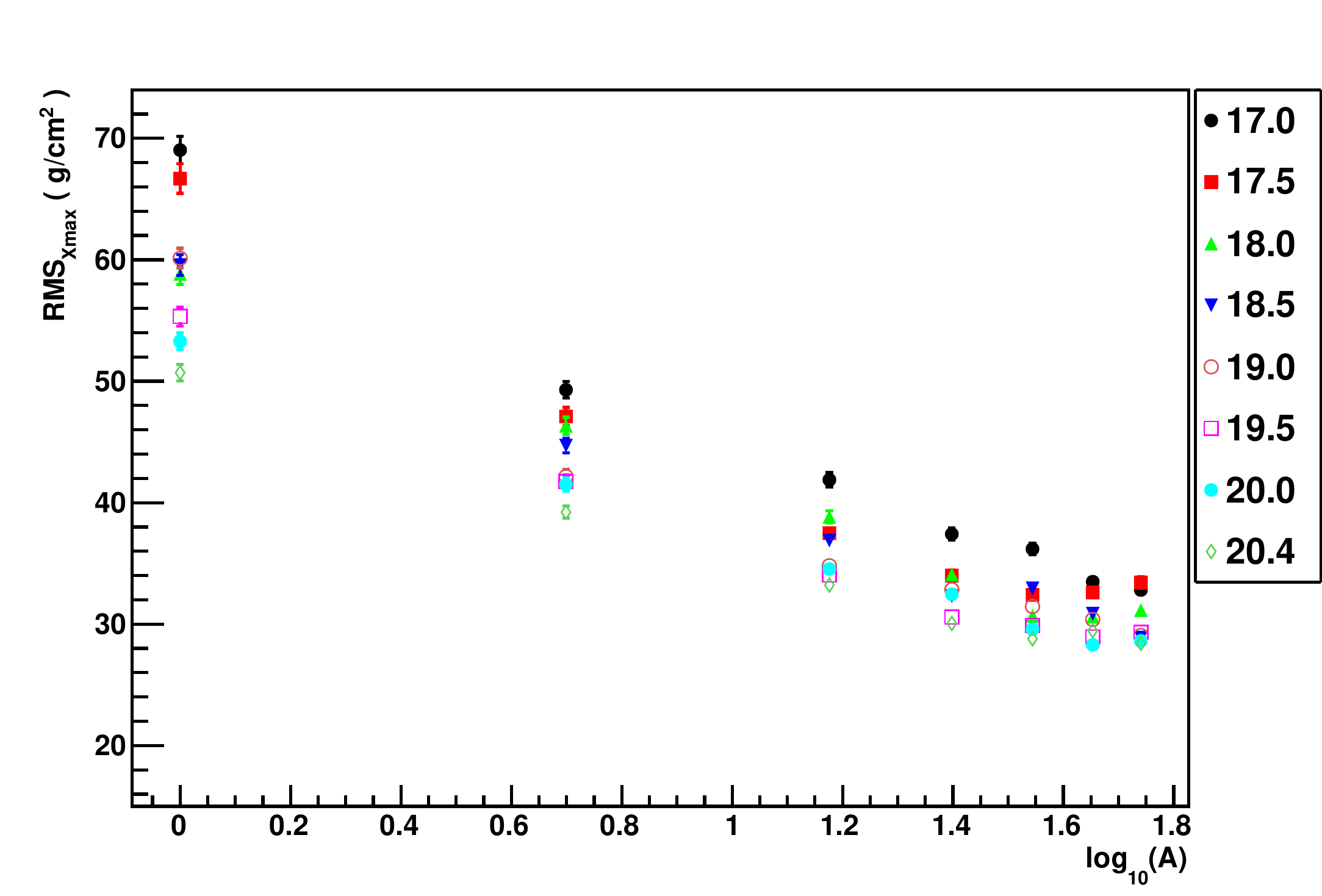}}\\

  \subfloat[Difference between \Corsika and \Conex \emph{versus}
    energy. Lines are linear fit to 1p and 55p data.]{\label{fig:rms:diff:sibyll:energy}\includegraphics[width=0.4\textwidth]{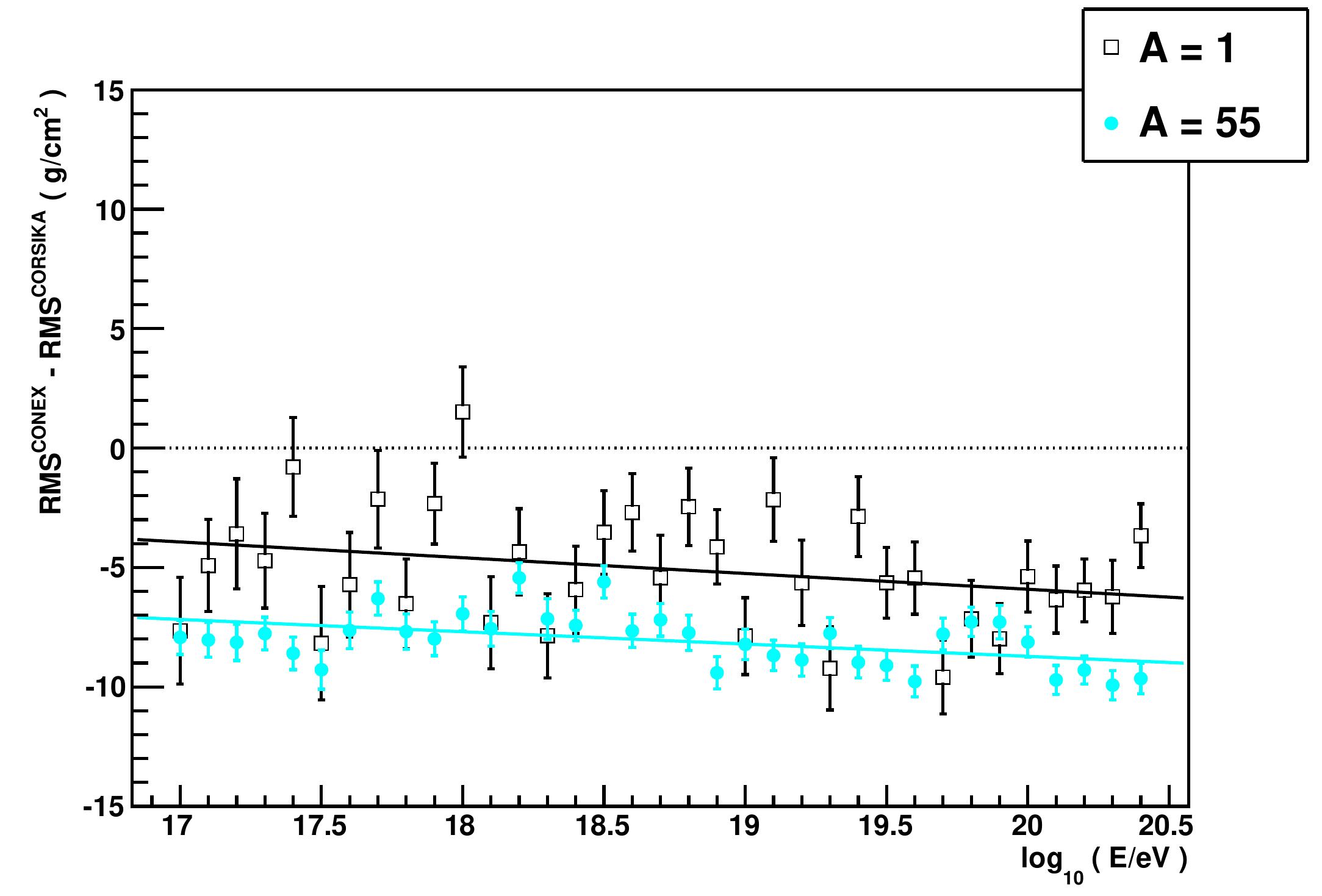}}
  \subfloat[Difference between \Corsika and \Conex \emph{versus} mass.]{\label{fig:rms:diff:sibyll:mass}\includegraphics[width=0.4\textwidth]{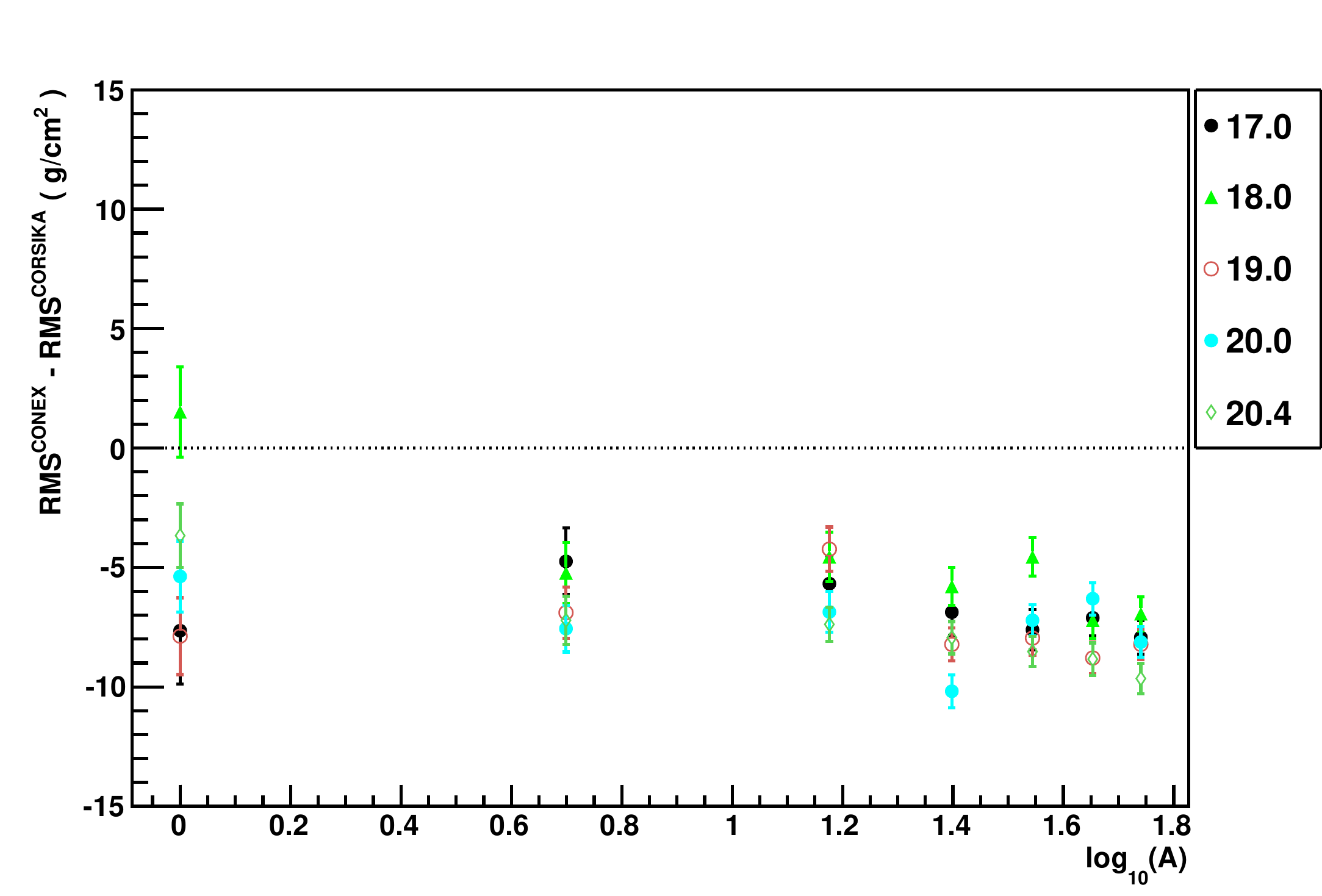}}\\

  \caption{\sigmaXmax as a function of energy and mass as calculated by
    \Corsika and \Conex using \Sibyll. Showers have been simulated
    with primary energy ranging from $10^{17.0}$ to $10^{20.4}$ eV in
    steps of $\log_{10}{(E/eV)} = 0.1$ and primary nuclei types with
    mass: 1, 5, 15, 25, 35, 45 and 55. A set of 1000 showers has been
    simulated for each combination. Not all energies and primaries are
    shown for clarity.}
  \label{fig:rms:xmax:corsika:conex:sibyll}

\end{figure}

\begin{figure}
  \centering

  \subfloat[\Conex - \sigmaXmax \emph{versus} energy.]{\label{fig:rms:qgsjet:conex}\includegraphics[width=0.4\textwidth]{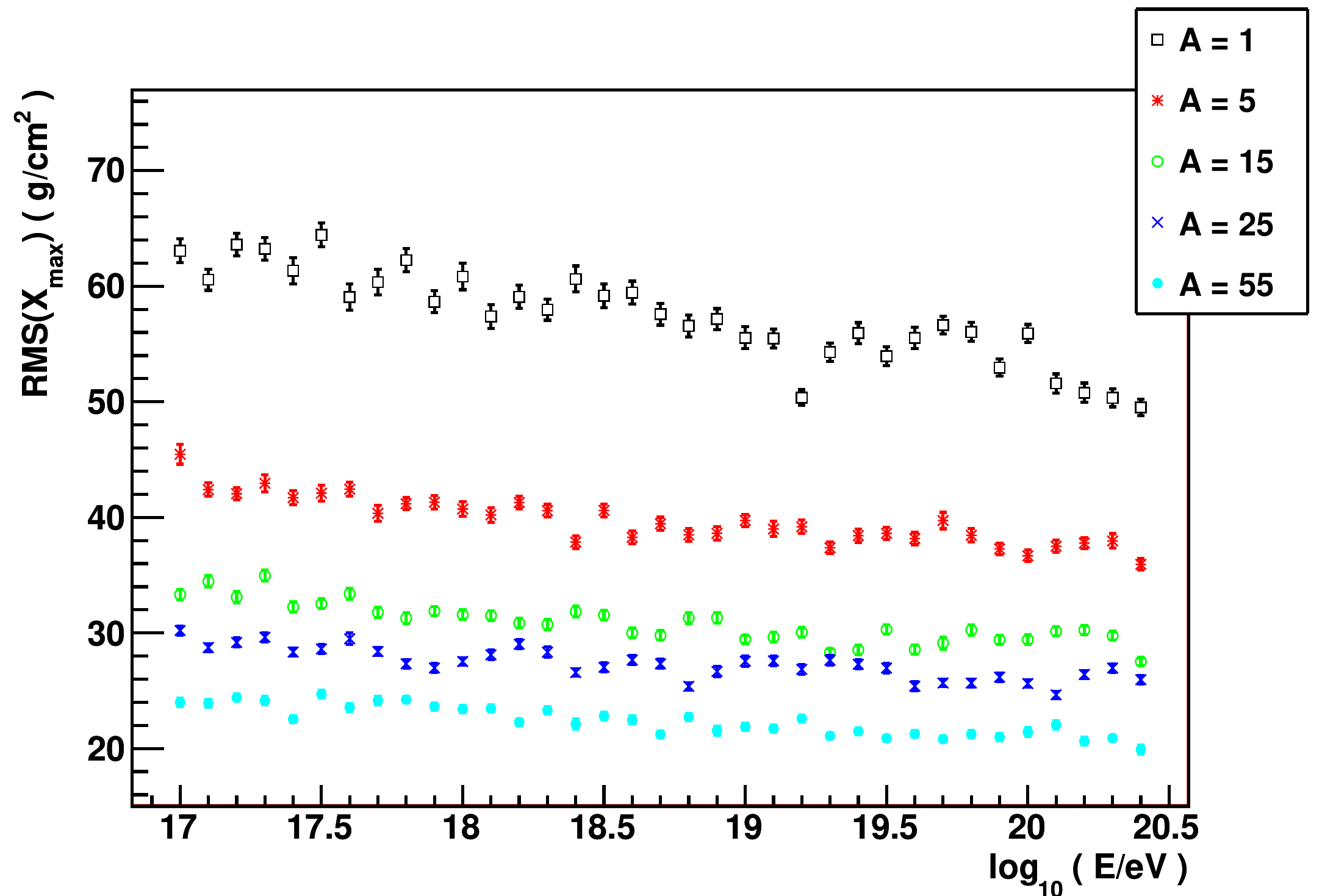}}
  \subfloat[\Corsika - \sigmaXmax \emph{versus} energy.]{\label{fig:rms:qgsjet:corsika}\includegraphics[width=0.4\textwidth]{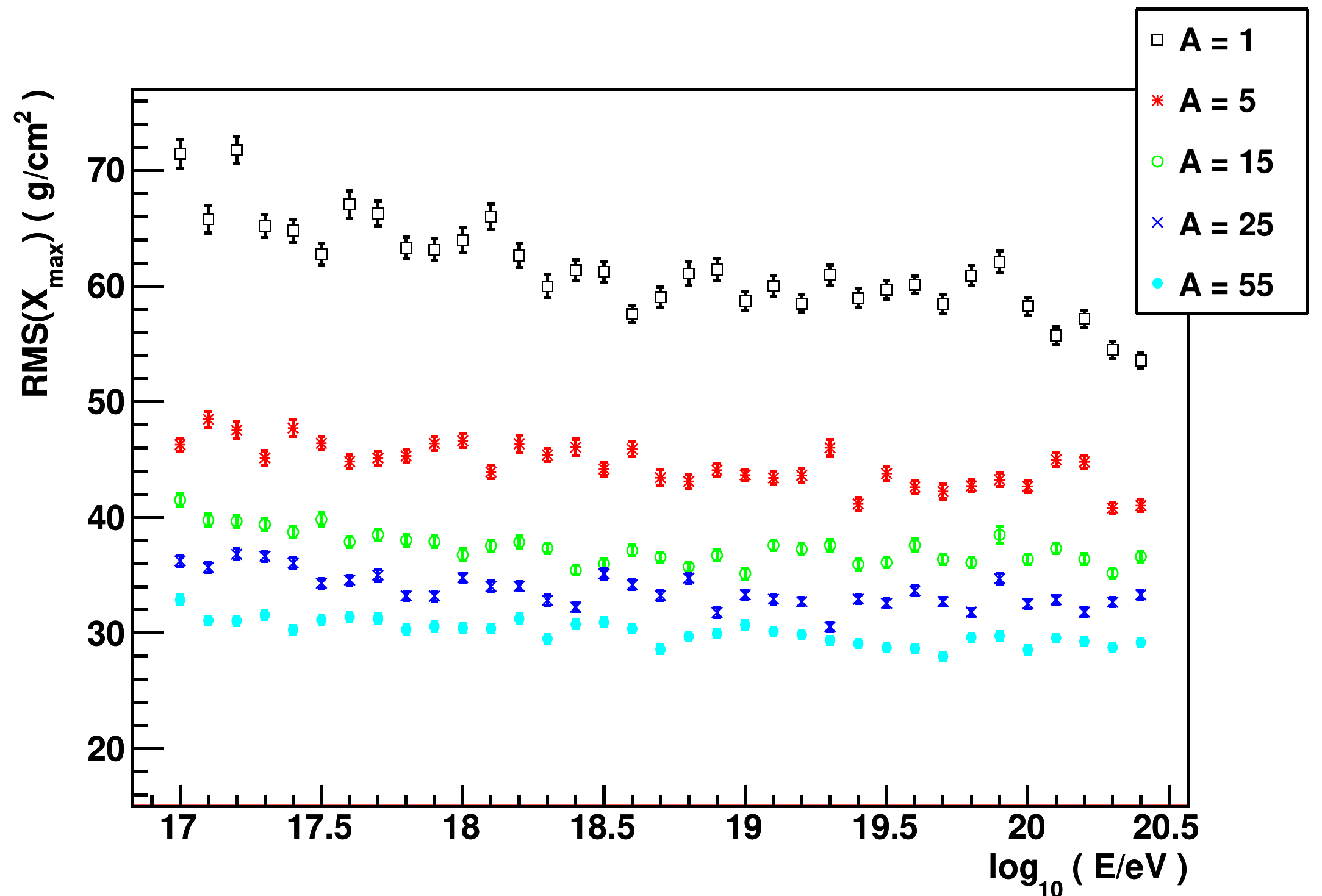}}\\

  \subfloat[\Conex - \sigmaXmax \emph{versus} mass.]{\includegraphics[width=0.4\textwidth]{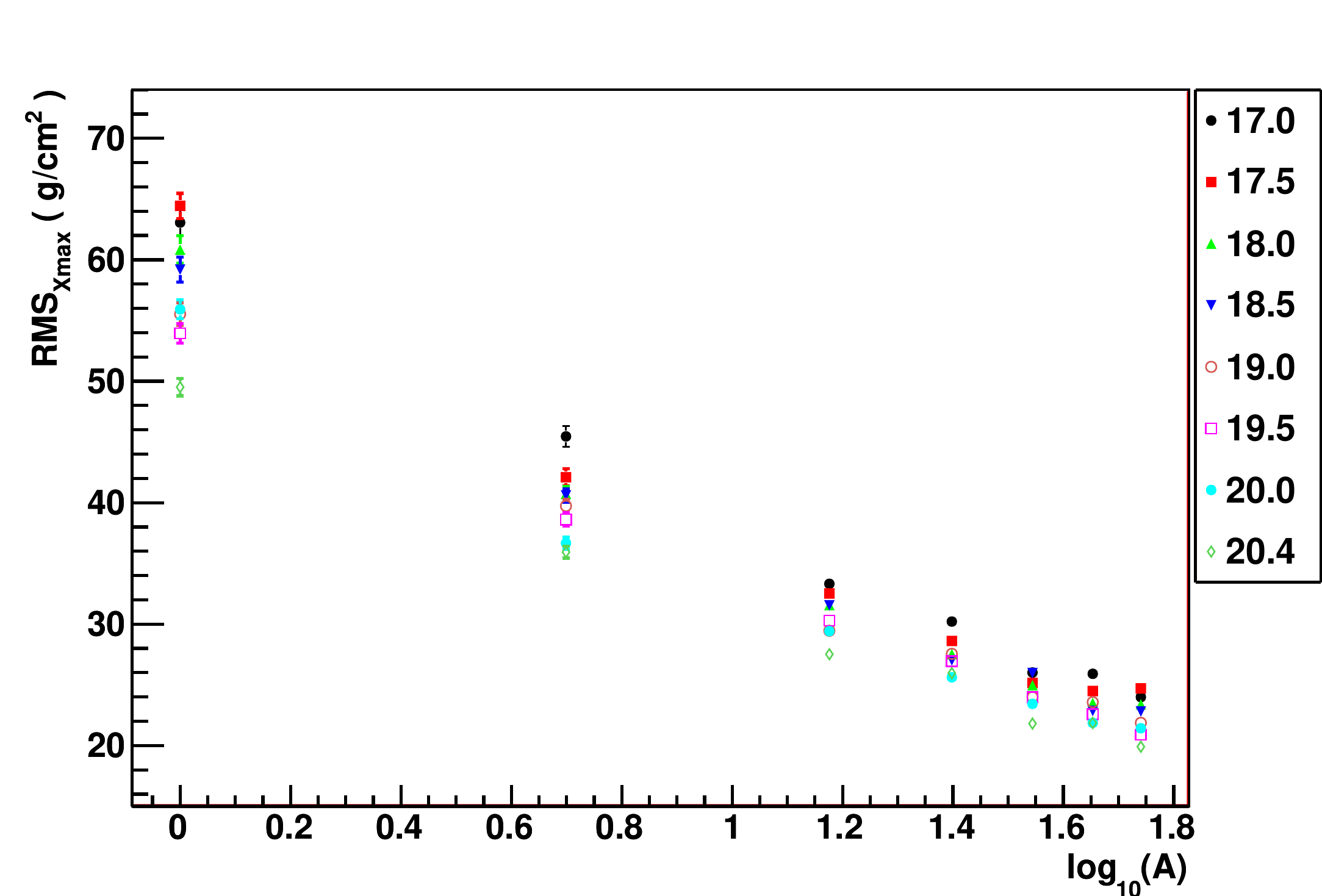}}
  \subfloat[\Corsika - \sigmaXmax \emph{versus} mass.]{\includegraphics[width=0.4\textwidth]{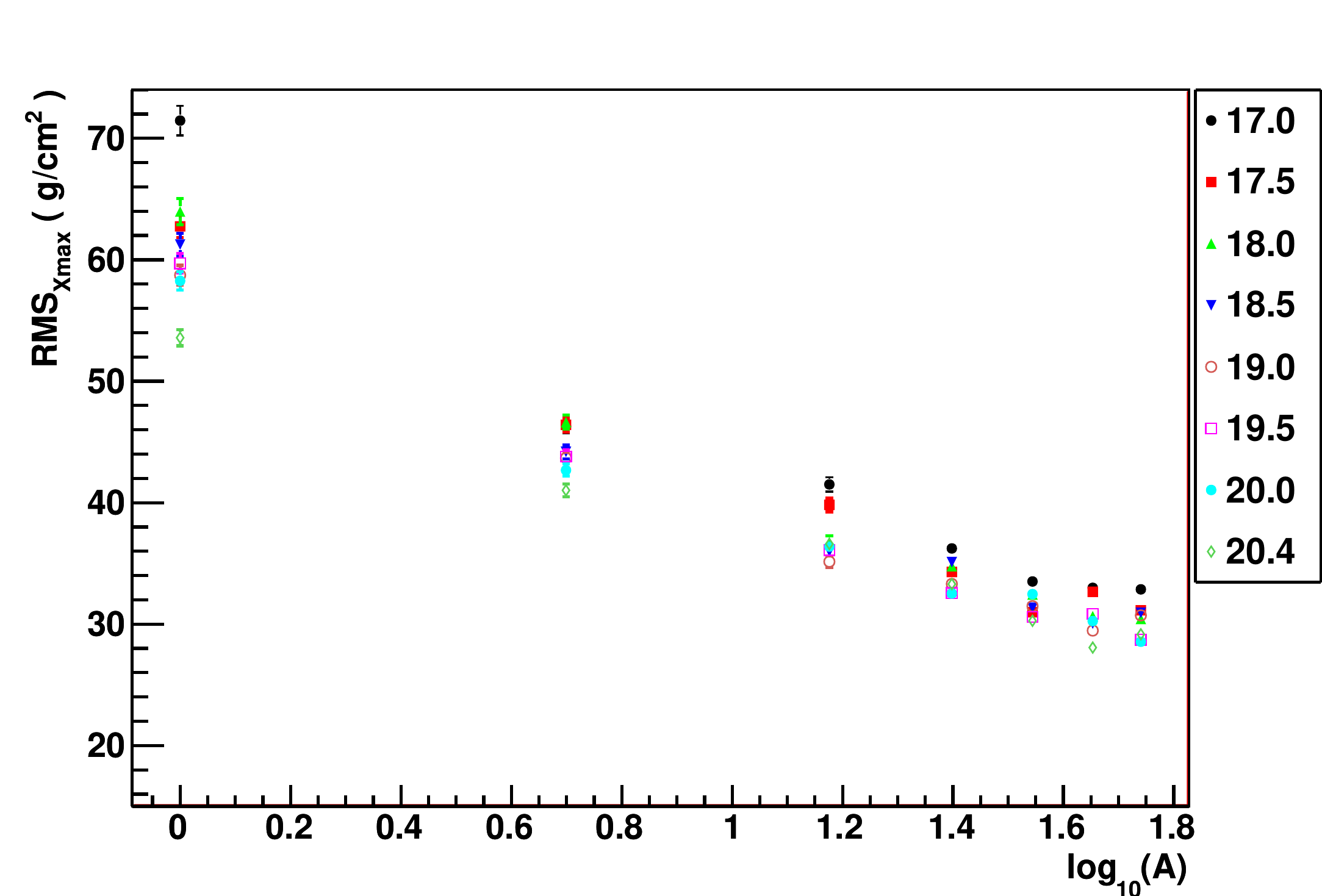}}\\

  \subfloat[Difference between \Corsika and \Conex \emph{versus}
    energy. Lines are linear fit to 1p and 55p data.]{\label{fig:rms:diff:qgsjet:energy}\includegraphics[width=0.4\textwidth]{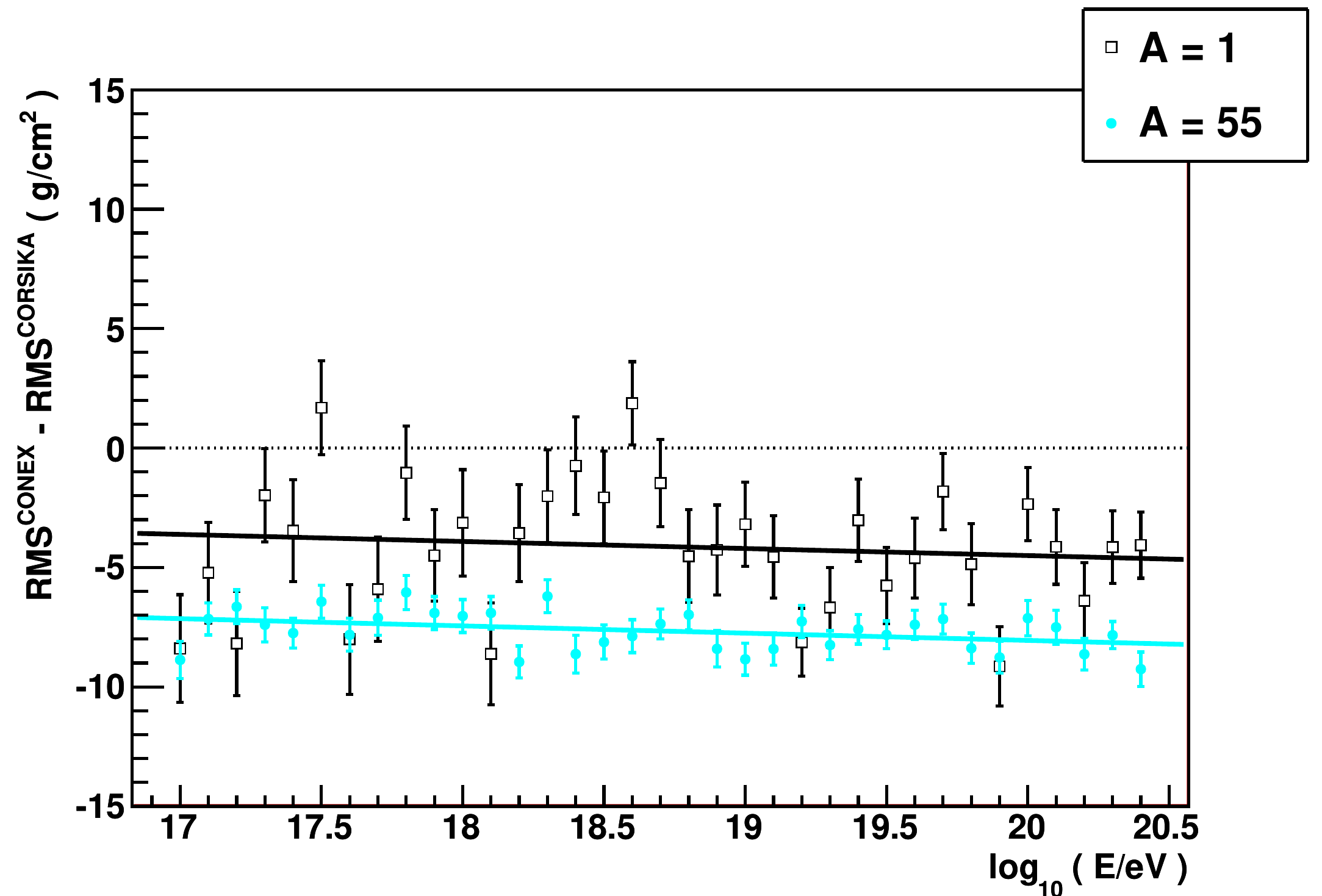}}
  \subfloat[Difference between \Corsika and \Conex \emph{versus} mass.]{\label{fig:rms:diff:qgsjet:mass}\includegraphics[width=0.4\textwidth]{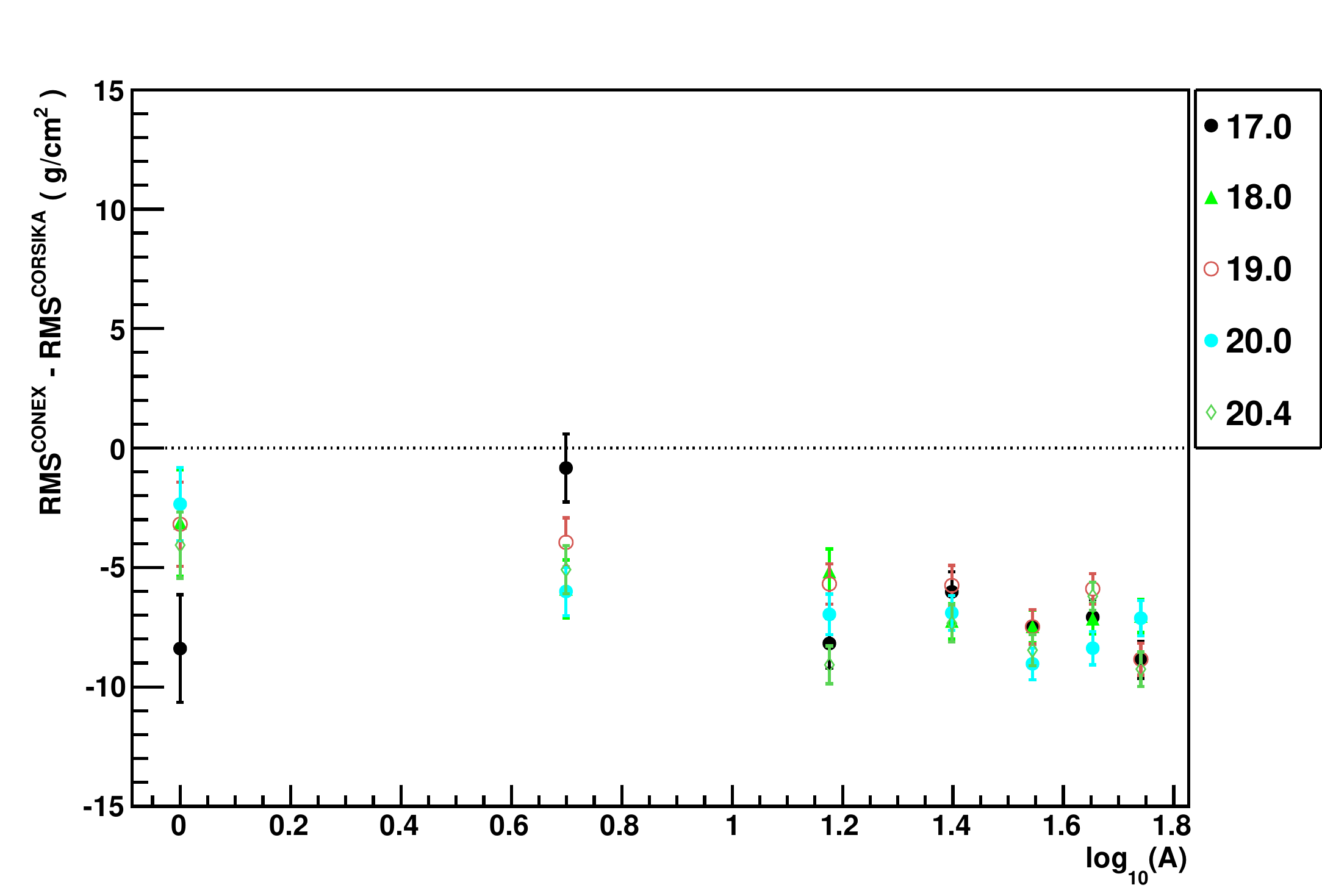}}\\

  \caption{\sigmaXmax as a function of energy and mass as calculated by
    \Corsika and \Conex using \Qgsjet. Showers have been simulated
    with primary energy ranging from $10^{17.0}$ to $10^{20.4}$ eV in
    steps of $\log_{10}{(E/eV)} = 0.1$ and primary nuclei types with
    mass: 1, 5, 15, 25, 35, 45 and 55. A set of 1000 showers has been
    simulated for each combination. Not all energies and primaries are
    shown for clarity.}
  \label{fig:rms:xmax:corsika:conex:qgsjet}

\end{figure}

Similar results are presented in figures
\ref{fig:rms:xmax:corsika:conex:sibyll} and
\ref{fig:rms:xmax:corsika:conex:qgsjet} for the \sigmaXmax. In the
parameter space studied by us the differences in the \sigmaXmax
calculated by \Corsika and \Conex are smaller than 8 \gcm. No
significant trend of the difference with energy or mass was seen.

The evolution of \sigmaXmax shown in figure
\ref{fig:rms:sibyll:conex},~\ref{fig:rms:sibyll:corsika},~\ref{fig:rms:qgsjet:conex}
and ~\ref{fig:rms:qgsjet:corsika} shows large fluctuations
apparently larger than the estimated statistical fluctuation. The
statistical fluctuation shown as error bars of \sigmaXmax is the standard
statistical variance of the variance of a distribution. No Gaussian
approximation was used. The trend of \sigmaXmax with energy is
statisticaly incompatible with a linear behavior. A linear fit of
\sigmaXmax versus energy shows a mean $\chi^2/NDOF = 123/35 \sim  3.5$.

% MEAN VERSUS RMS

\begin{figure}[!htb]
  \centerline{\includegraphics[width=0.7\textwidth]{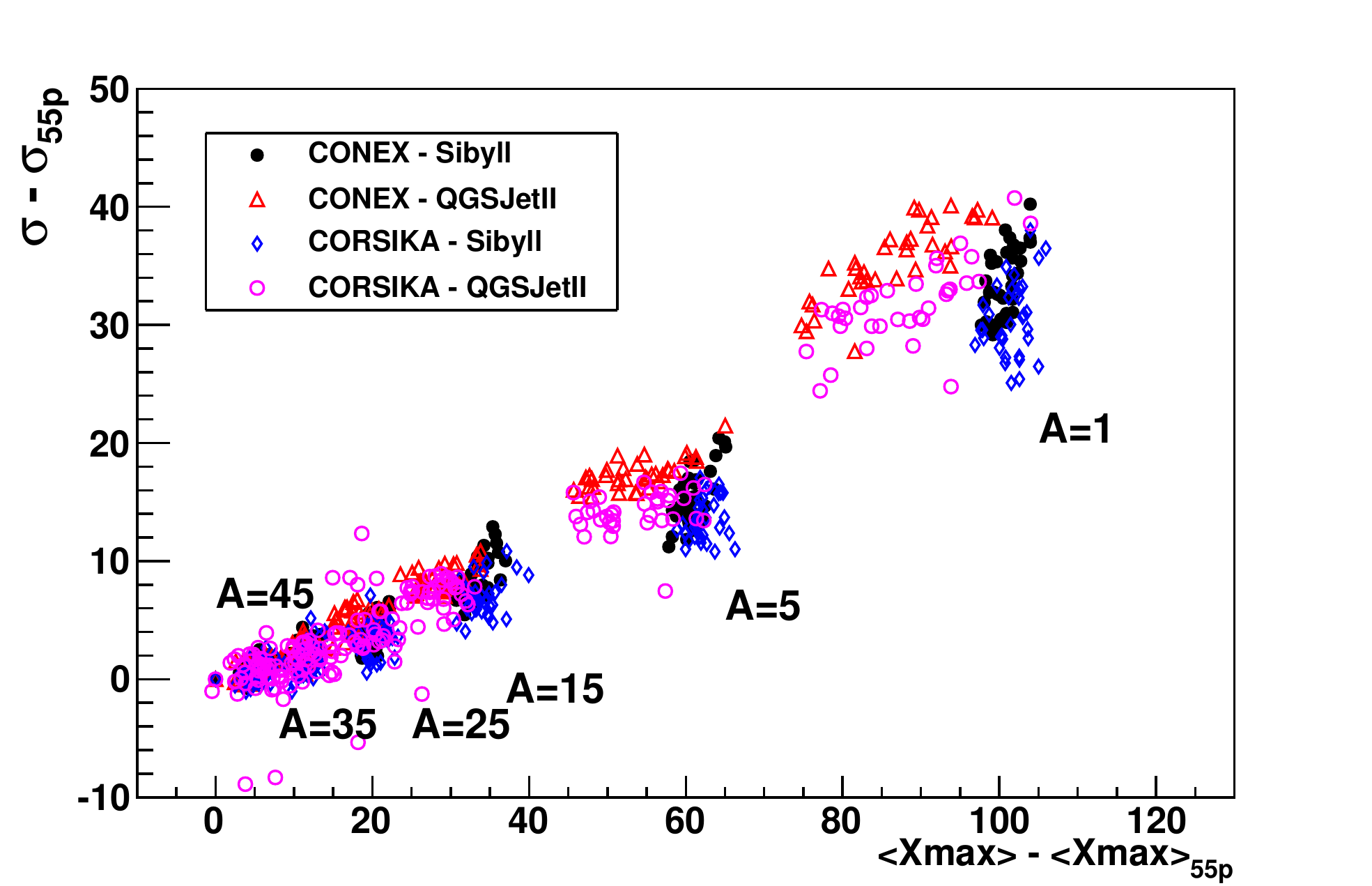}}
  \caption{\sigmaXmax \emph{versus}  \meanXmax  for all primary particles used in this work. The corresponding \sigmaXmax and
\meanXmax for a nuclei with mass 55 has been taken as reference. Each blob corresponds to the  \meanXmax and \sigmaXmax predictions for one primary particle at different energies. }
  \label{fig:mean:rms}
\end{figure}

Figure \ref{fig:mean:rms} summarizes the differences in \meanXmax and
\sigmaXmax between the simulation programs and between the hadronic
interaction models. This figure shows simultaneously the \meanXmax and
\sigmaXmax values, where the corresponding \meanXmax and \sigmaXmax
for a nuclei with mass 55 has been taken as reference (as suggested in
~\cite{Kampert2012660}). This figure illustrates the importance of
taking into account the simulation program differences into the
systematic uncertainty of the model predictions. Each blob corresponds
to the  \meanXmax and \sigmaXmax predictions for one primary particle
at different energies.

%SLOPE ANALYSIS

\begin{figure}
  \centering

  \subfloat[Slope of \meanXmax \emph{versus} mass - \Sibyll.]{\label{fig:slope:sibyll:conex}\includegraphics[width=0.4\textwidth]{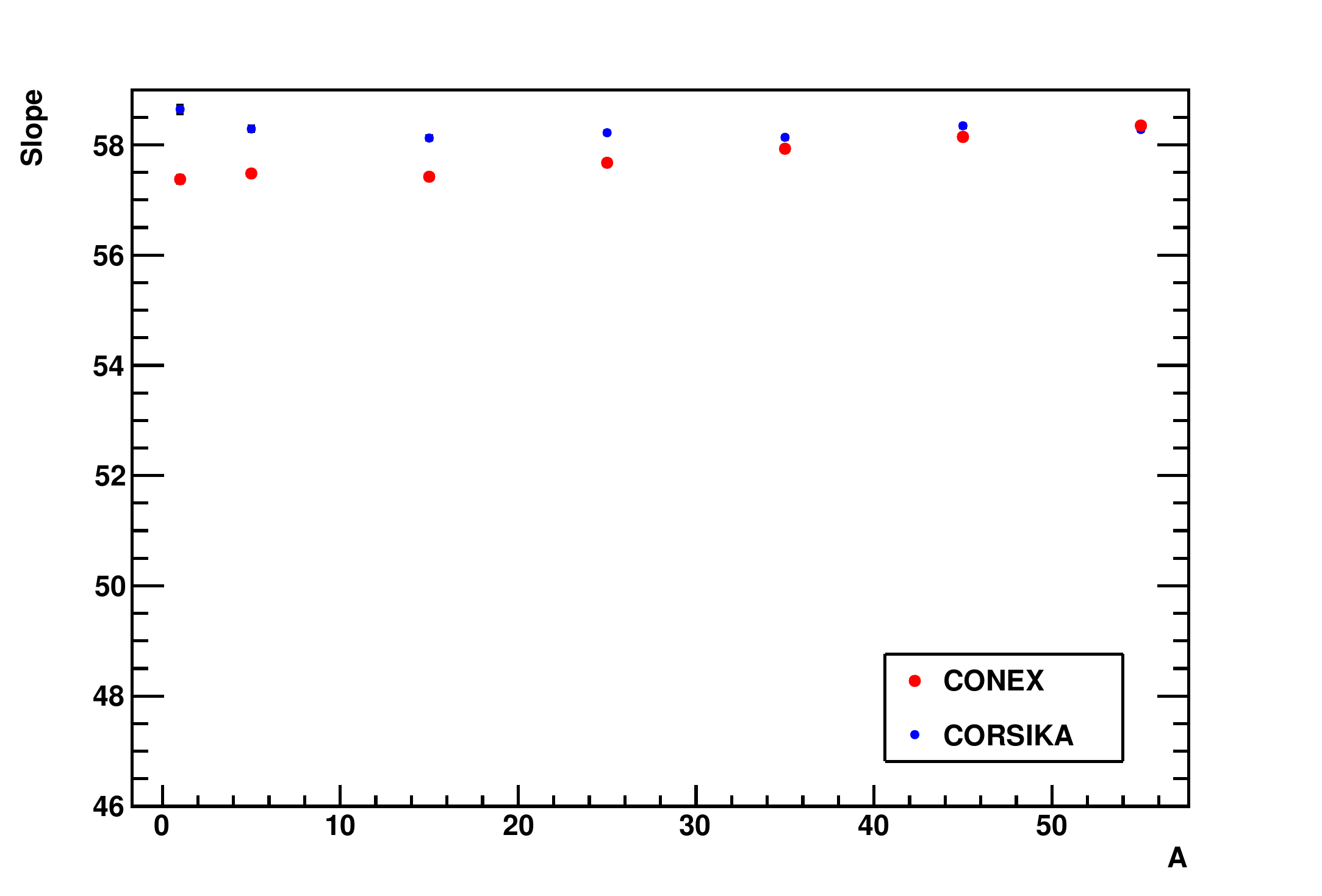}}
  \subfloat[Slope of \meanXmax \emph{versus} mass - \Qgsjet.]{\label{fig:slope:qgsjet:conex}\includegraphics[width=0.4\textwidth]{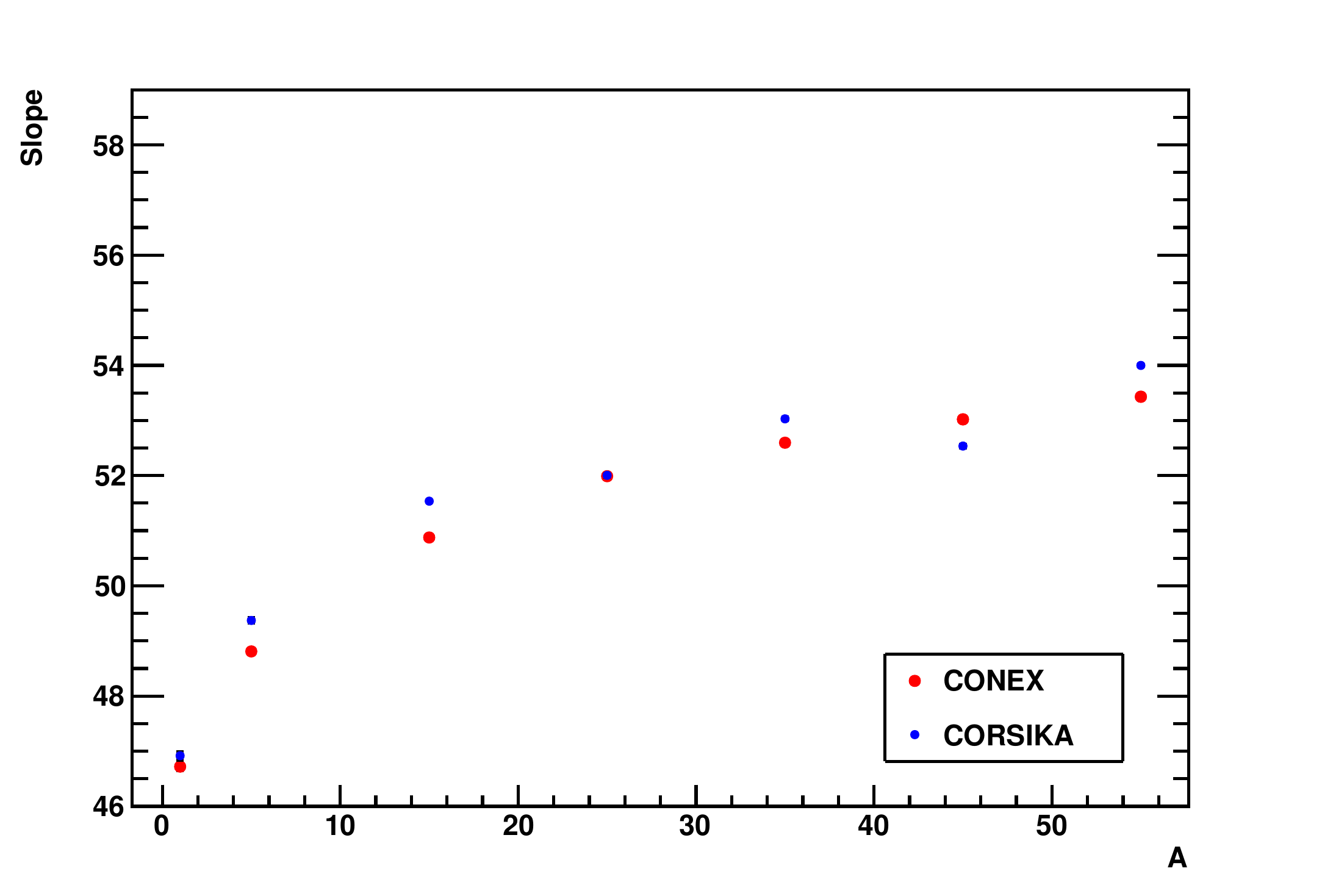}}\\

  \subfloat[Slope of \sigmaXmax \emph{versus} mass - \Sibyll.]{\includegraphics[width=0.4\textwidth]{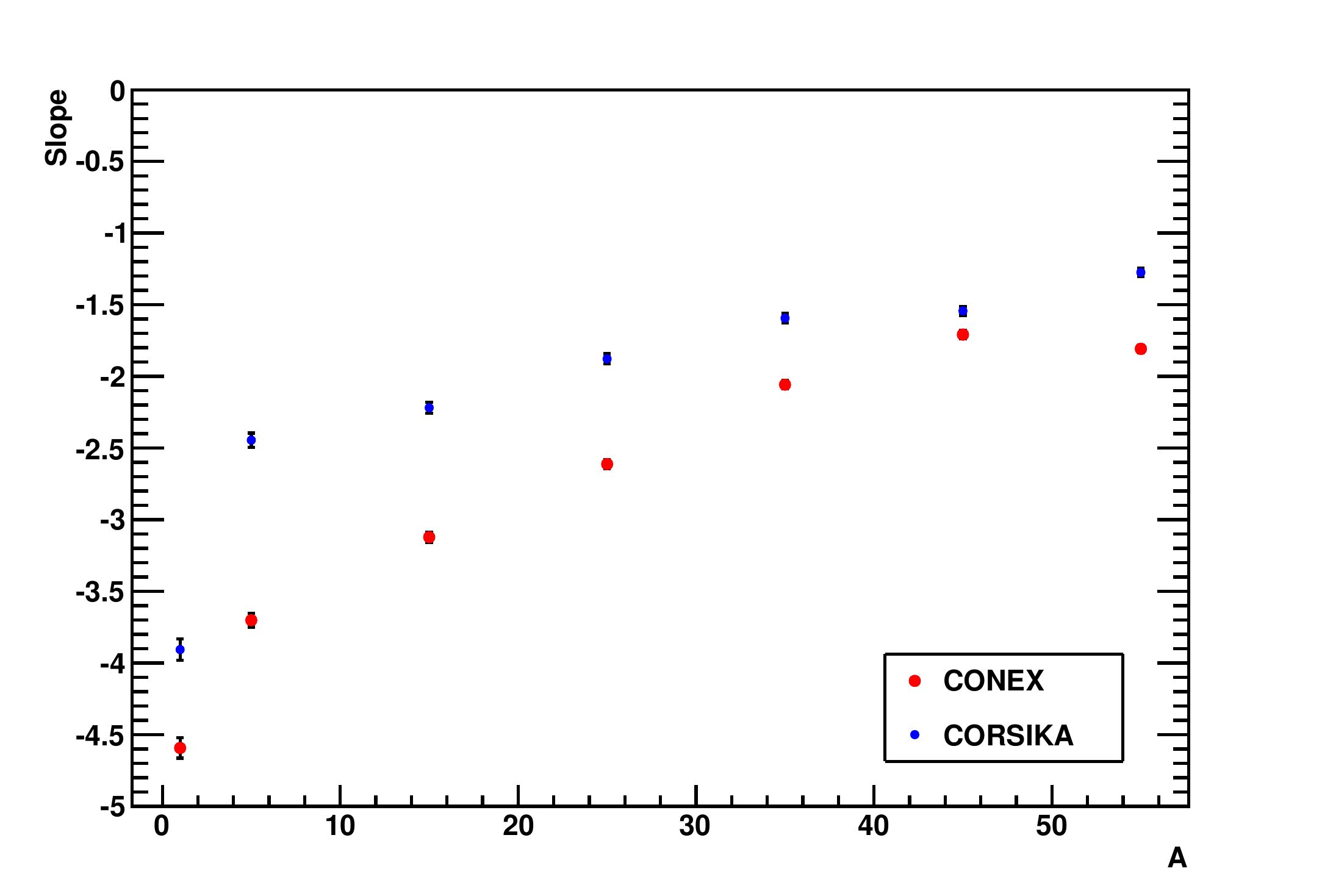}}
  \subfloat[Slope of \sigmaXmax \emph{versus} mass - \Qgsjet.]{\includegraphics[width=0.4\textwidth]{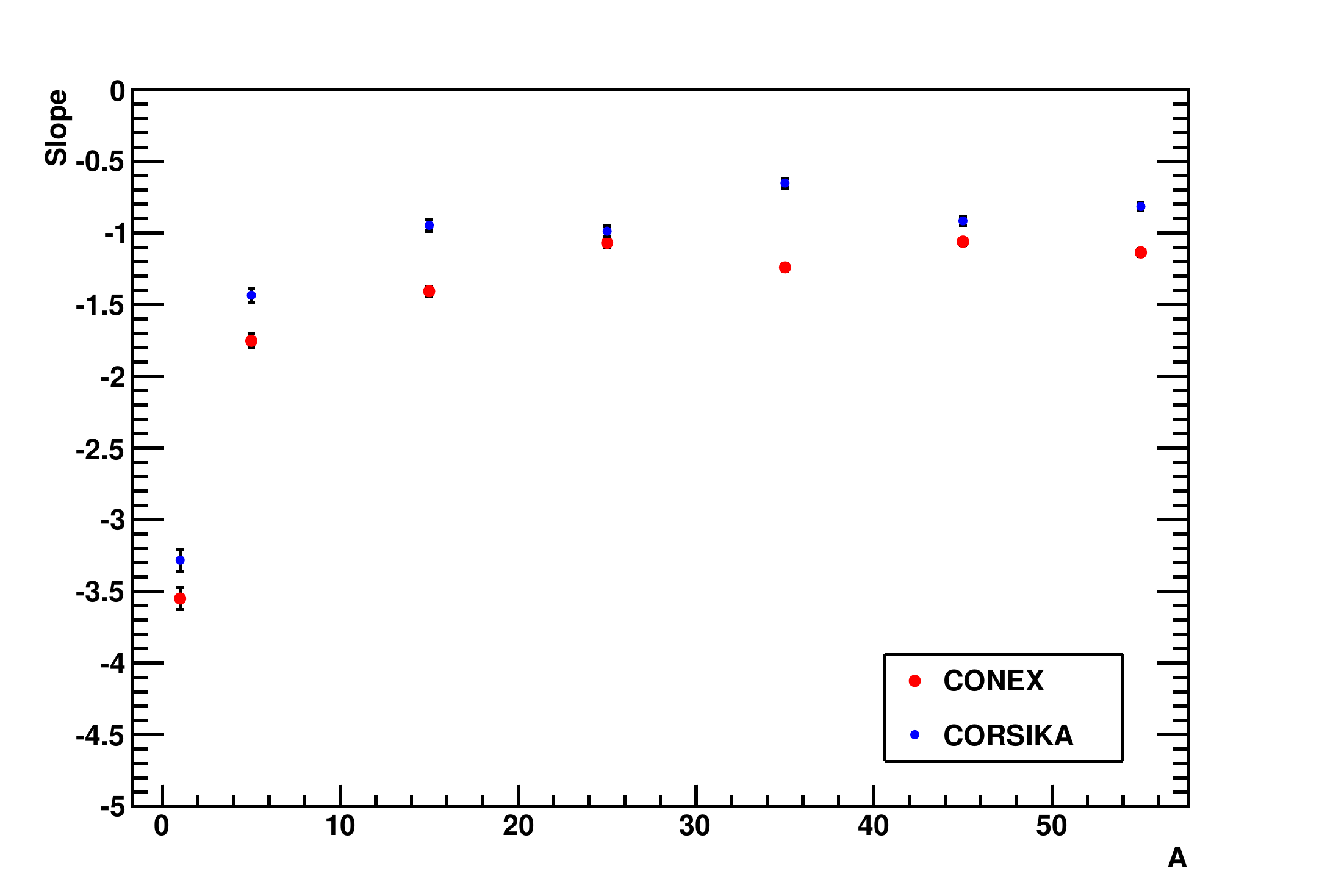}}\\

  \caption{Slope Analysis. Slope of a straight line fit to the variation of
    \meanXmax and \sigmaXmax with energy (plotted as a function of mass). Slope in units of \gcm/eV.}
  \label{fig:slope:analysis}

\end{figure}

The elongation rate
theorem~\cite{linsley__1977,t._k._gaisser__1979,linsley_validity_1981}
proposes the use of the slope of the variation of \meanXmax with
energy as a composition parameter. According to this proposal, changes
in this slope represents changes in
composition. Figure~\ref{fig:slope:analysis} show the slope of the
variation of \meanXmax and \sigmaXmax with energy as a function of
primary particle mass for \Corsika and \Conex using \Sibyll and
\Qgsjet. The slope in this figure corresponds to the fit of a straight
line in the energy range ($10^{17} \leq E \leq 10^{20.4}$ eV).
There is a good agreement between \Conex and \Corsika when
the same hadronic model is used. The slope of the \sigmaXmax when
\Sibyll is used presents the largest discrepancy between \Conex and
\Corsika (see~\ref{fig:slope:analysis}.c).

It has been shown before~\cite{bib:heck:presentation} that the
dependencies of the \meanXmax and \sigmaXmax are not strictly straight
lines however the departure of the linear dependency is very small for
energies above \energy{17} as studied here.

It is clear from figure~\ref{fig:slope:analysis} that the slope of the
\meanXmax as a function of energy is dominated by the hadronic
interaction model rather than by the shower simulation.

\section{Parametrization of the \Xmax distributions}
\label{sec:par:xmax}

 The \Xmax distributions can be described by a function which is a
 convolution of a Gaussian with an exponential~\cite{bib:gauss:expo}:

\begin{equation}
\frac{d\Xmax}{dN} =  N_f \exp{ \left( \frac{t_0-t}{\lambda} +
  \frac{\sigma^2}{2\lambda^2}\right) } Erfc \left( \frac{t_0-t+
  \sigma^2/\lambda}{\sqrt{2}\sigma }  \right)
\label{eq:gauss:expo}
\end{equation}

\begin{figure}
  \centering
  \subfloat[A = 1 - $10^{18}$ eV.]{\includegraphics[width=0.3\textwidth]{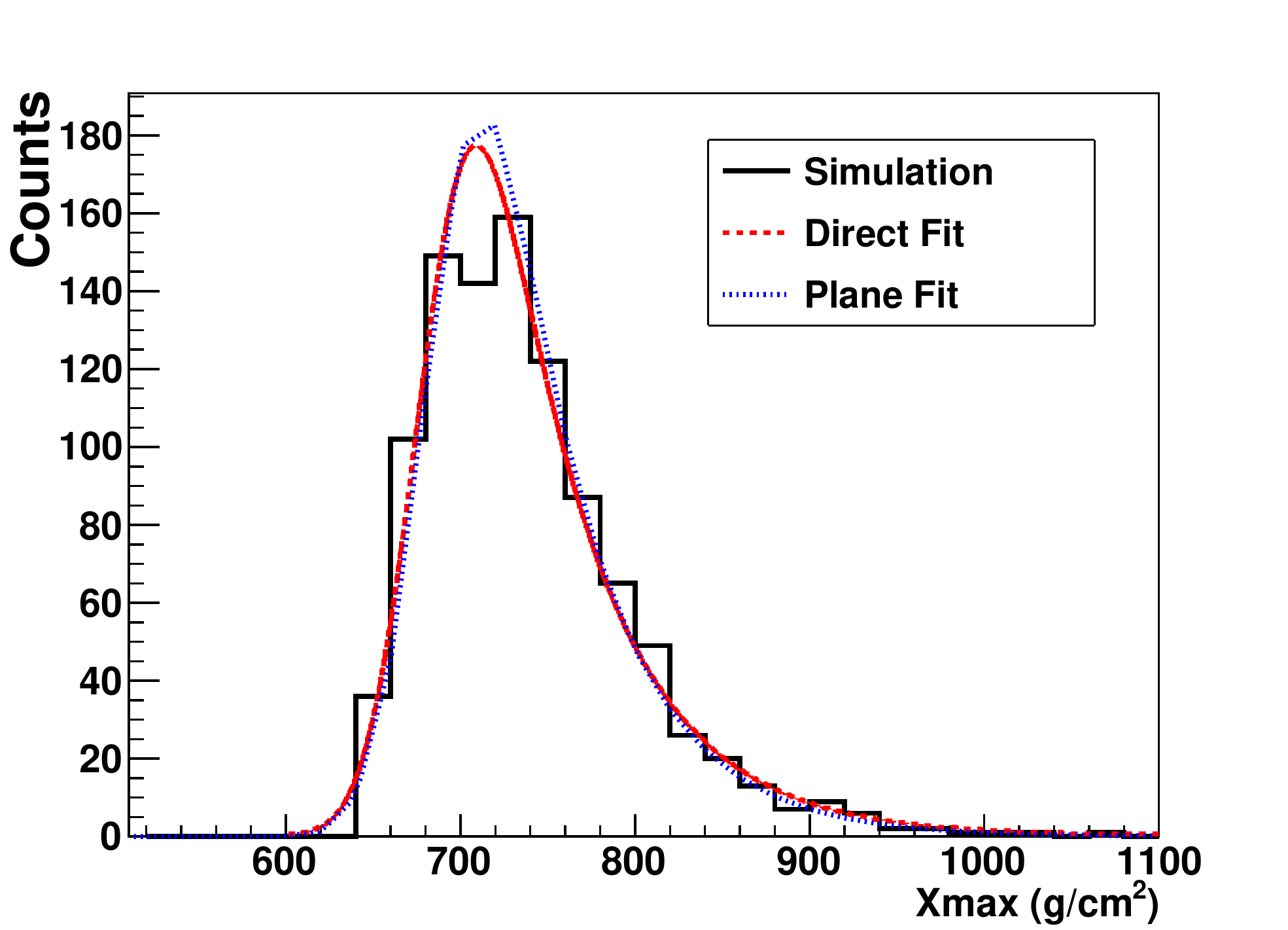}}
  \subfloat[A = 1 - $10^{19}$ eV.]{\includegraphics[width=0.3\textwidth]{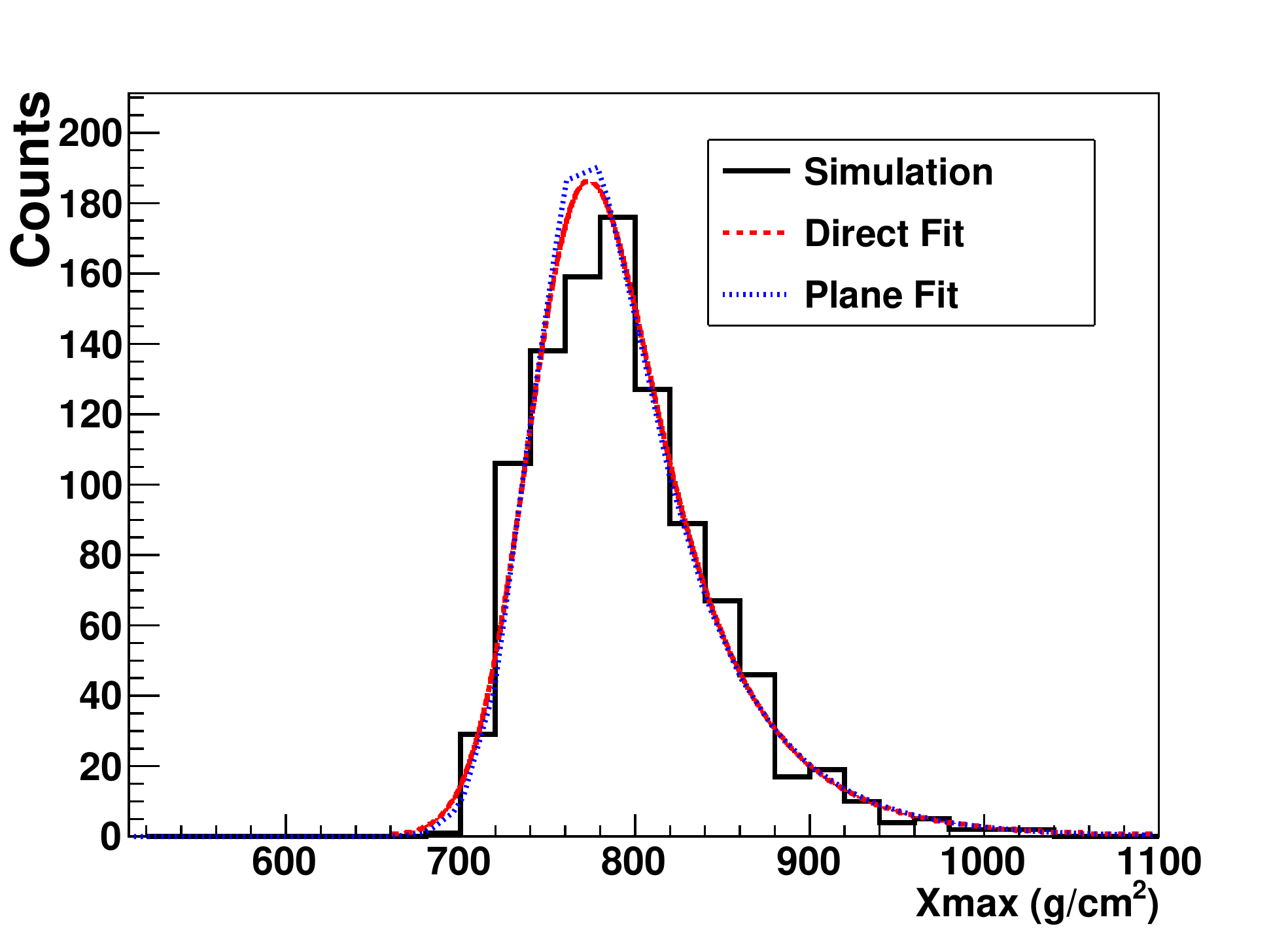}}
  \subfloat[A = 1 - $10^{20}$
    eV.]{\includegraphics[width=0.3\textwidth]{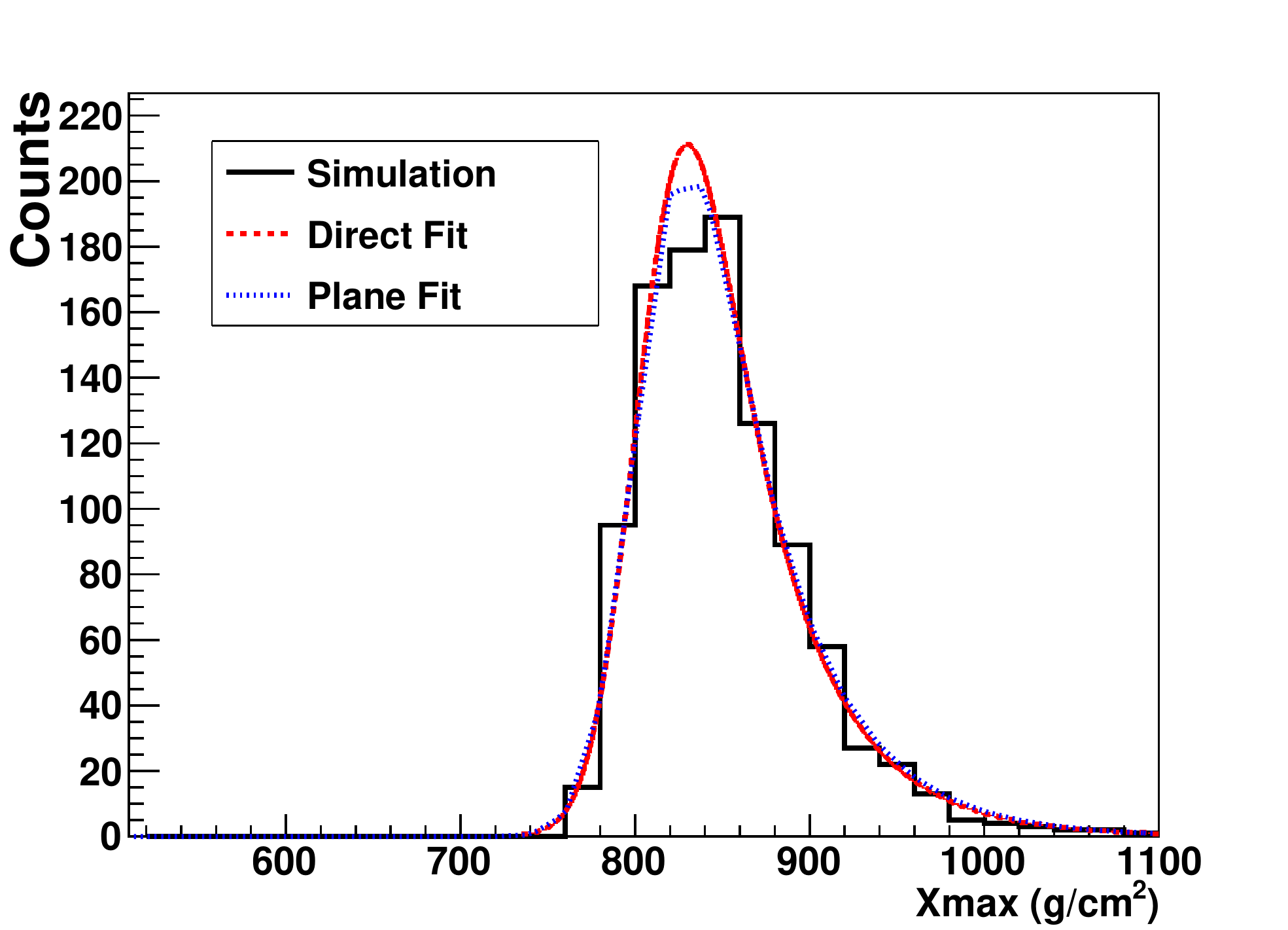}}\\
  \subfloat[A = 55 - $10^{18}$ eV.]{\includegraphics[width=0.3\textwidth]{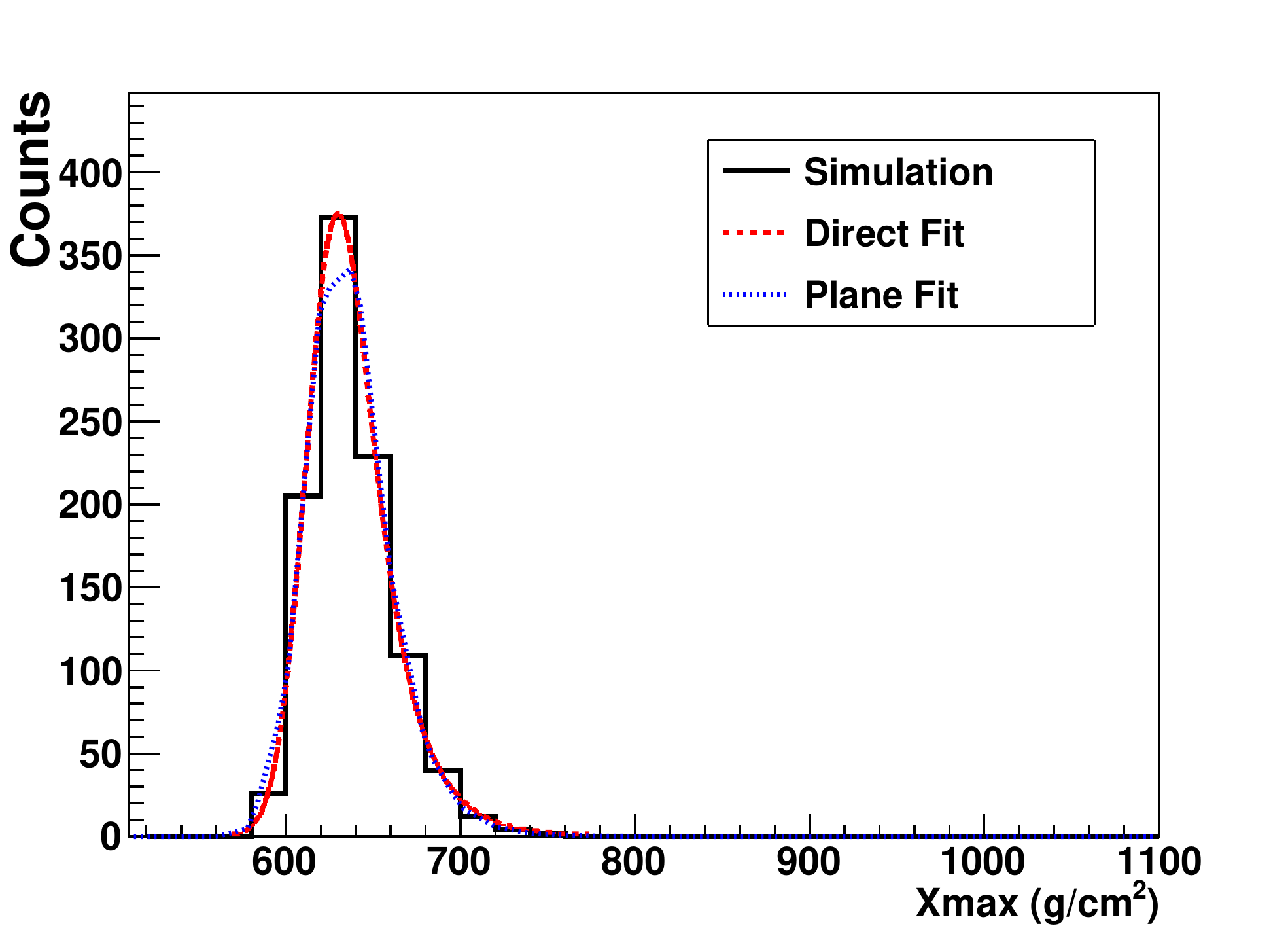}}
  \subfloat[A = 55 - $10^{19}$ eV.]{\includegraphics[width=0.3\textwidth]{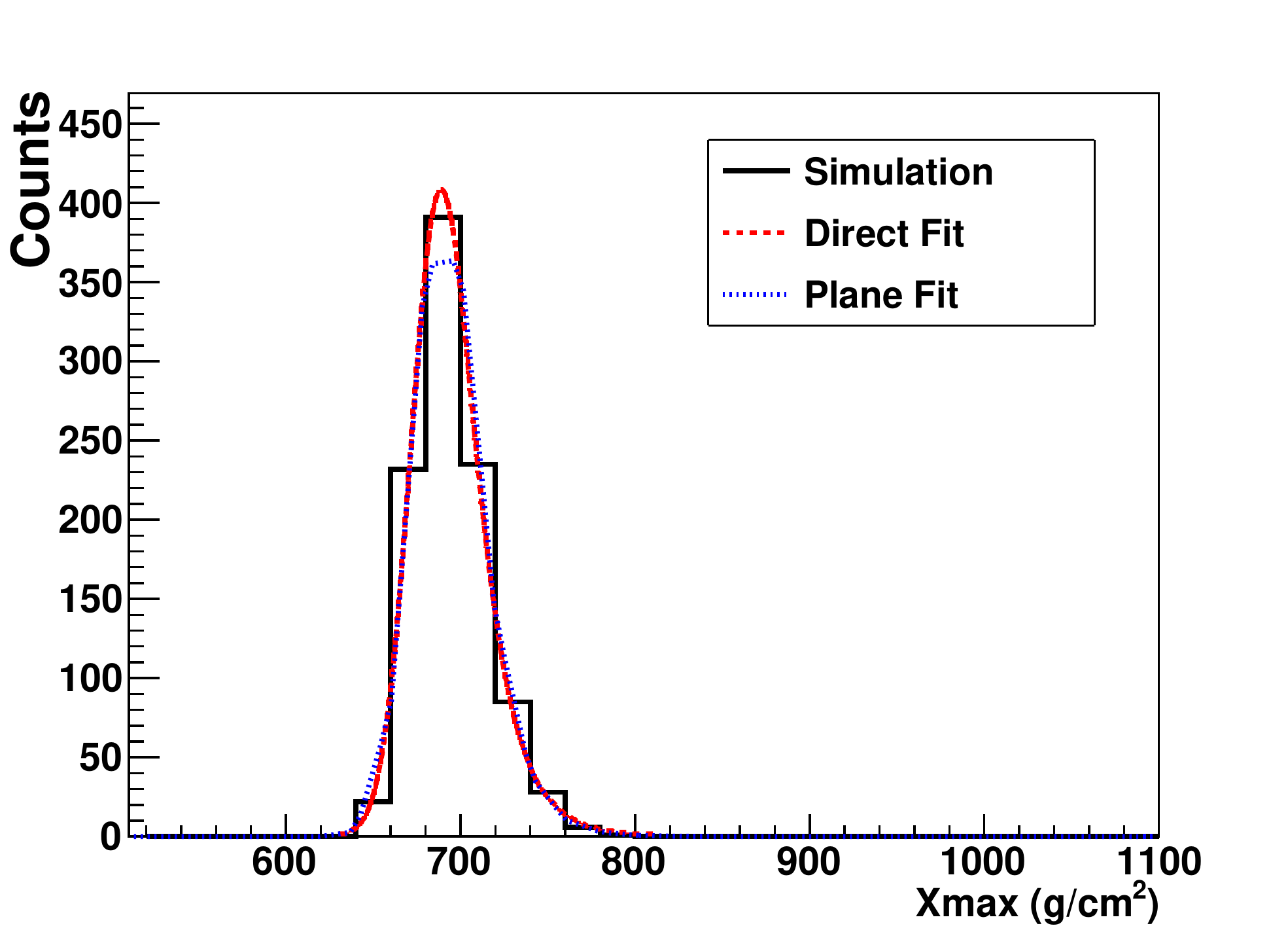}}
  \subfloat[A = 55 - $10^{20}$ eV.]{\includegraphics[width=0.3\textwidth]{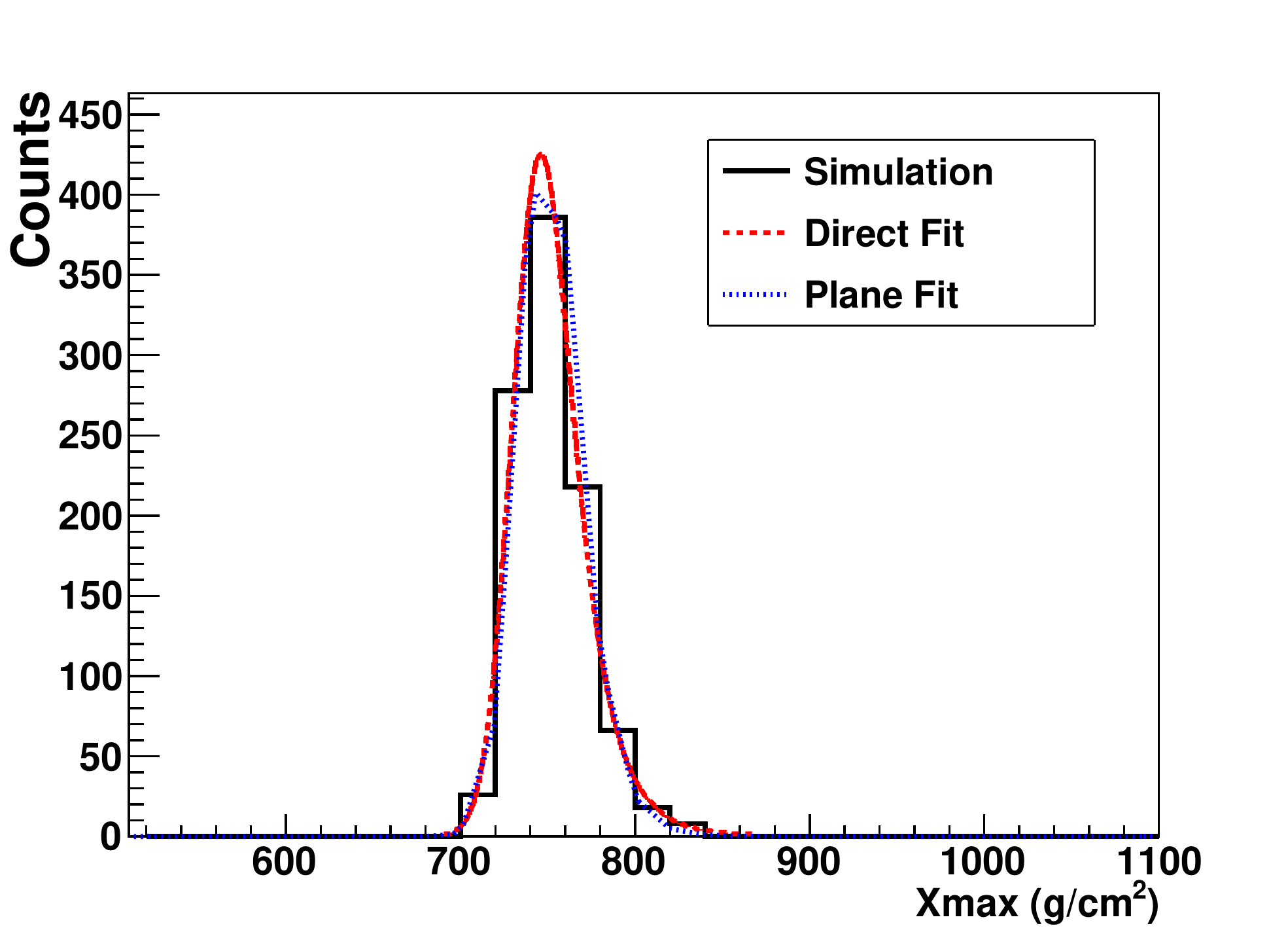}}\\
  \caption{Simulated \Xmax distribution fitted by a Gaussian
    convoluted with an exponential (equation~\ref{eq:gauss:expo}). In
    this example we show showers simulated with \Conex and
    \Sibyll. Full line shows the \Xmax distribution and dashed line
    the fit of equation~\ref{eq:gauss:expo}.}
  \label{fig:xmax:dist:fit}
  \end{figure}

This equation has four parameters. $N_f$ is a normalization factor
which gives the total number of events in the \Xmax
distribution. $\lambda$, $t_o$ and $\sigma$ are parameters which are
related to the decay factor of the exponential, the maximum of the
distribution and the width of the distribution respectively. $Erfc$ is
the error function. We used
this equation to fit the \Xmax distributions of all mass and energies
we have simulated. Figure~\ref{fig:xmax:dist:fit} shows examples \Xmax
distributions we fit with this equation. The aim of this study is to
use equation~\ref{eq:gauss:expo} to fit the \Xmax distribution and
calculate the \meanXmax and \sigmaXmax. We show below that a convolution
of a Gaussian with an exponential allows a good description of the
\meanXmax and \sigmaXmax. The proposed function is also a fairly good
description of the \Xmax distribution, see
figures~\ref{fig:xmax:dist:fit}.

\begin{figure}
  \centering
  \subfloat[$t_0$]{\includegraphics[width=0.3\textwidth]{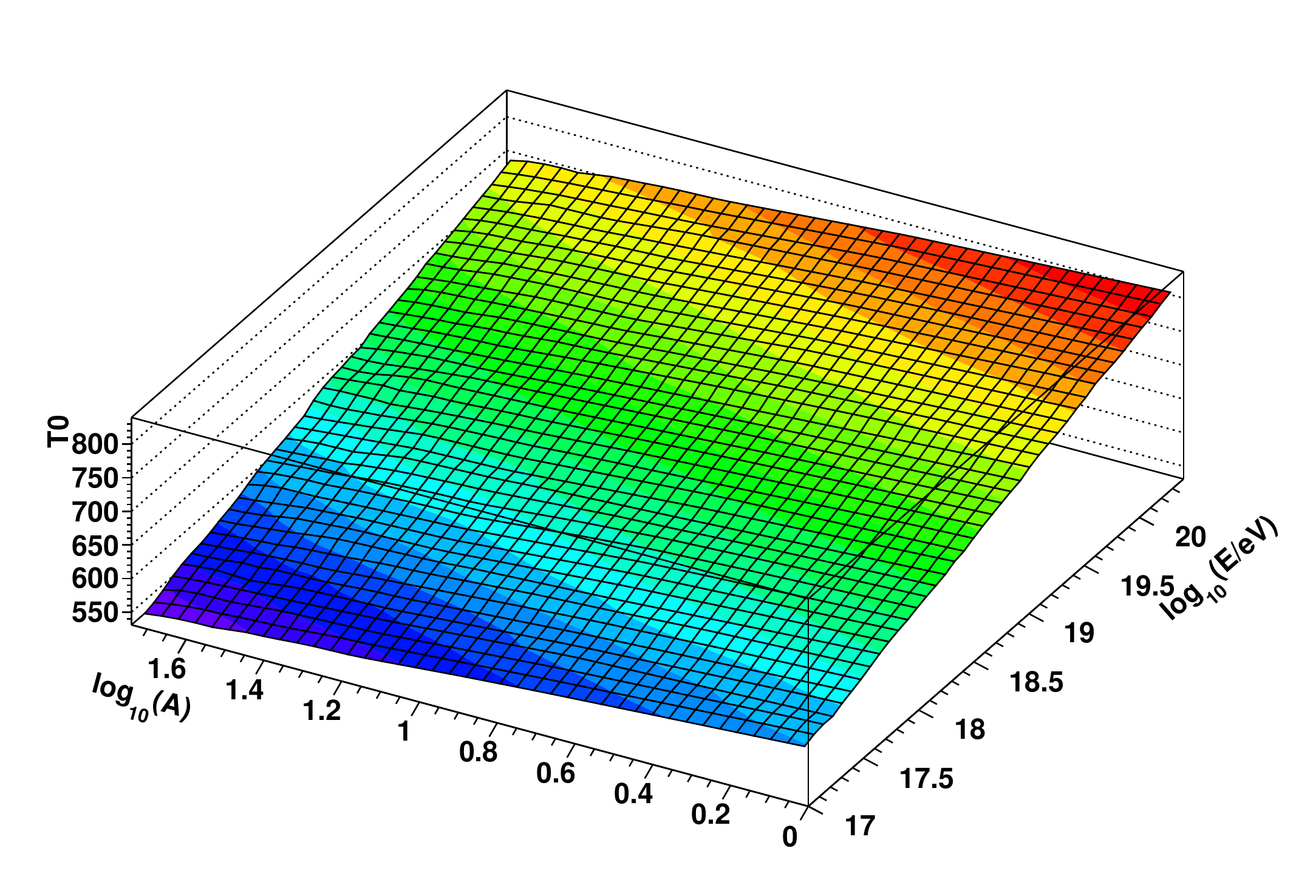}}
  \subfloat[$\sigma$]{\includegraphics[width=0.3\textwidth]{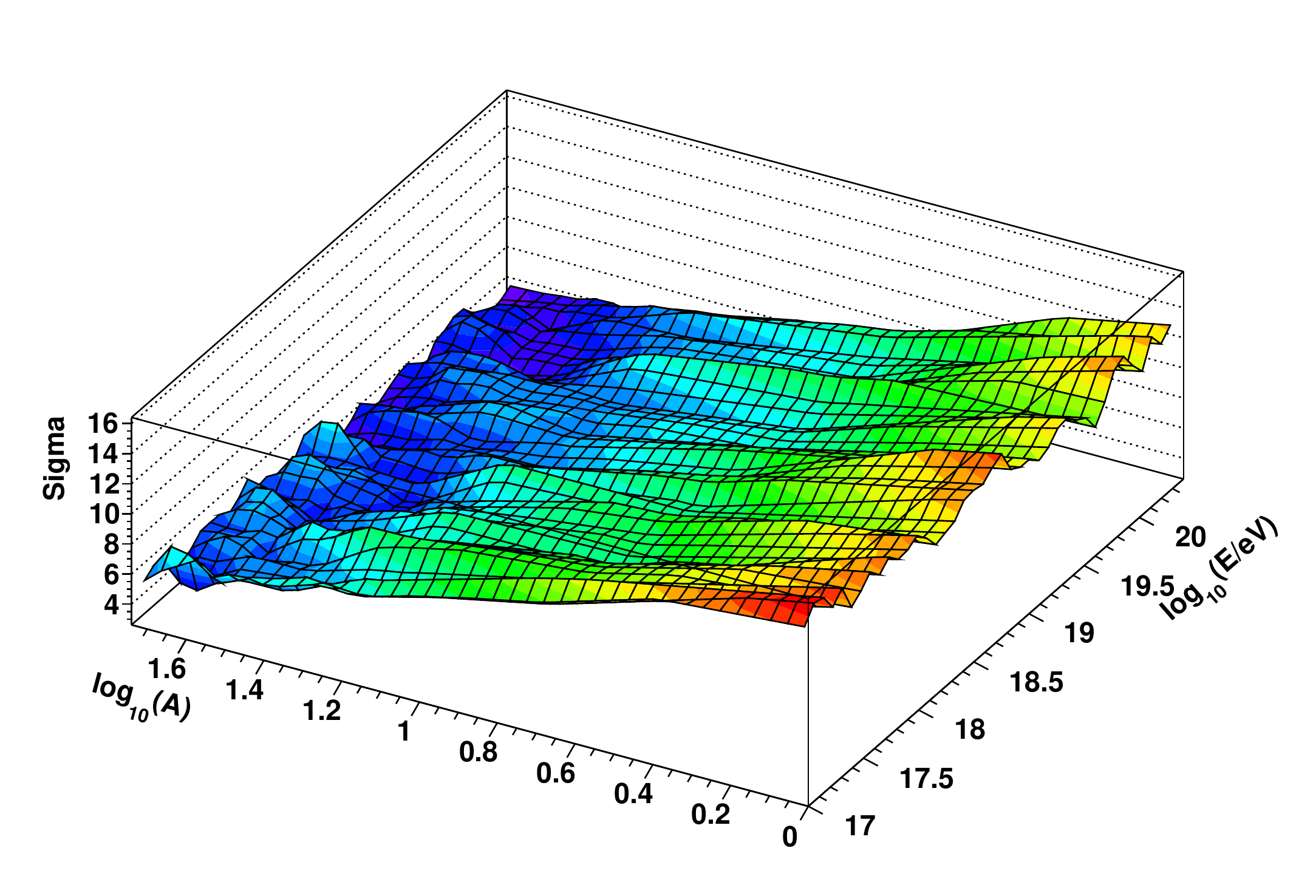}}
  \subfloat[$\lambda$]{\includegraphics[width=0.3\textwidth]{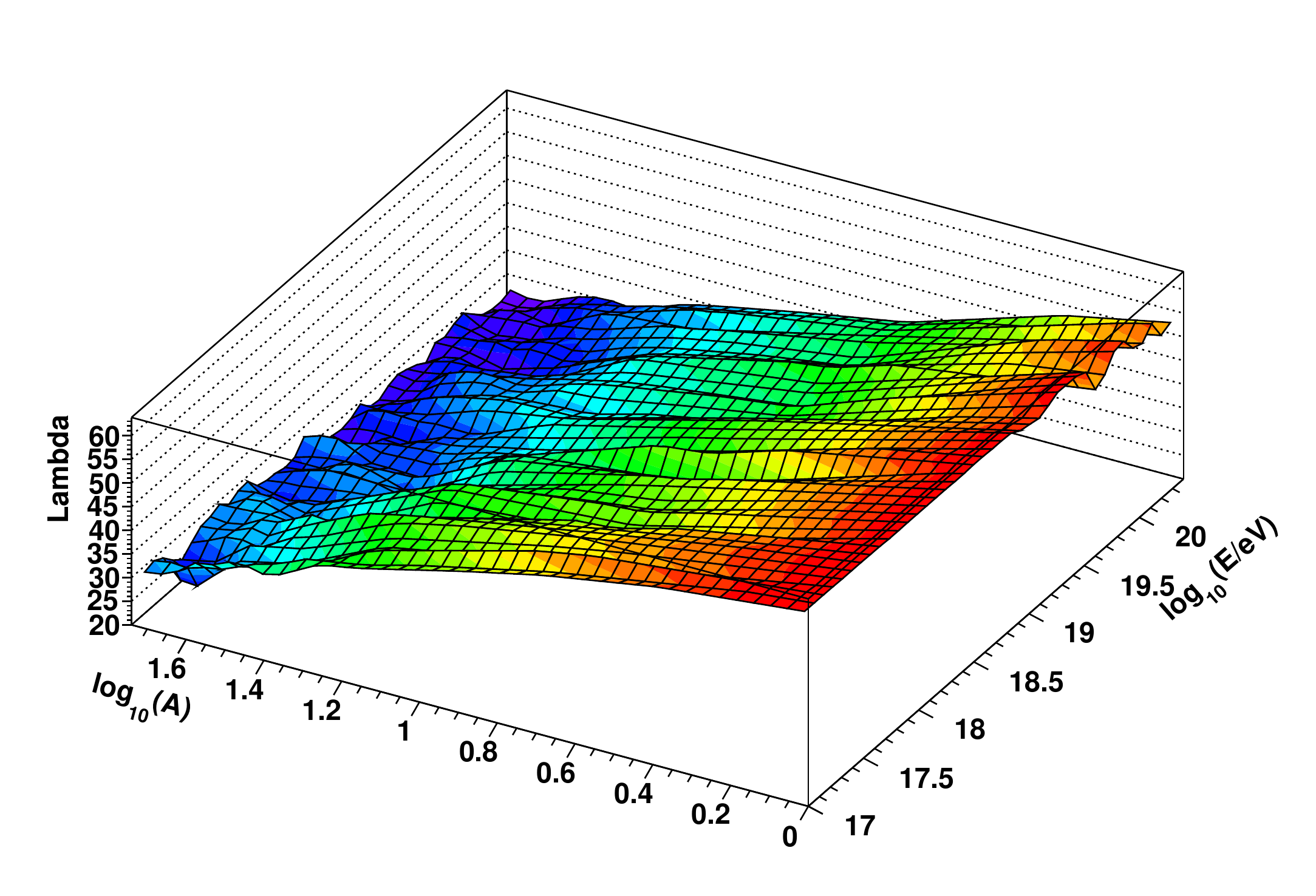}}
  \caption{Parametrization of the \Xmax distribution. \Conex -
    \Sibyll. These figures show the general behavior of the three
    parameters used to describe the \Xmax distributions as a function
    of energy and mass.}
  \label{fig:par:xmax}
\end{figure}

After that, we parametrized $\lambda$, $t_o$ and $\sigma$ as a
function of primary mass and energy using the simulated
showers. Figure~\ref{fig:par:xmax} show the evolution of the three
parameters with energy and mass.

Figure~\ref{fig:par:xmax}a shows how $t_o$ has a very smooth
dependence with mass and energy, recovering the already explored
dependence of \Xmax with mass and energy. On the other hand, $\sigma$
and $\lambda$ are not completely independent parameters. Both
parameters influence the width of the \Xmax distribution in
different ways. The parameter $\lambda$ changes the width of the \Xmax
distribution by modifying the decays of the exponential, making the
high \Xmax tail longer or shorter. $\sigma$ also changes the width of
the \Xmax distribution by modifying the width of the central
part. In fact, note that mathematically $\sigmaXmax = \sqrt{( \sigma +
\lambda )}$.

Given the degeneracy in shaping the width of the \Xmax
distribution, the parameters $\sigma$ and $\lambda$ are inversely
correlated.  The parameters $\sigma$ and $\lambda$ compensate each
other, fluctuations to higher values of $\sigma$ are correlated to
fluctuations to smaller values of $\lambda$.

We performed a fit to plots in figure~\ref{fig:par:xmax} with a linear
dependence on $log_{10}(A)$ and $log_{10}(E)$ following equation:

\begin{equation}
  \left.
   \begin{array}{l l l}
     t_0\\
     \sigma\\
     \lambda \\
   \end{array} \right\} = C_1 \times log_{10}(E/eV) + C_2 \times log_{10}(A) + C_3
\label{eq:fit:surfaces}
\end{equation}

\begin{figure}
  \centering
  \subfloat[\Conex - \Sibyll]{\includegraphics[width=0.5\textwidth]{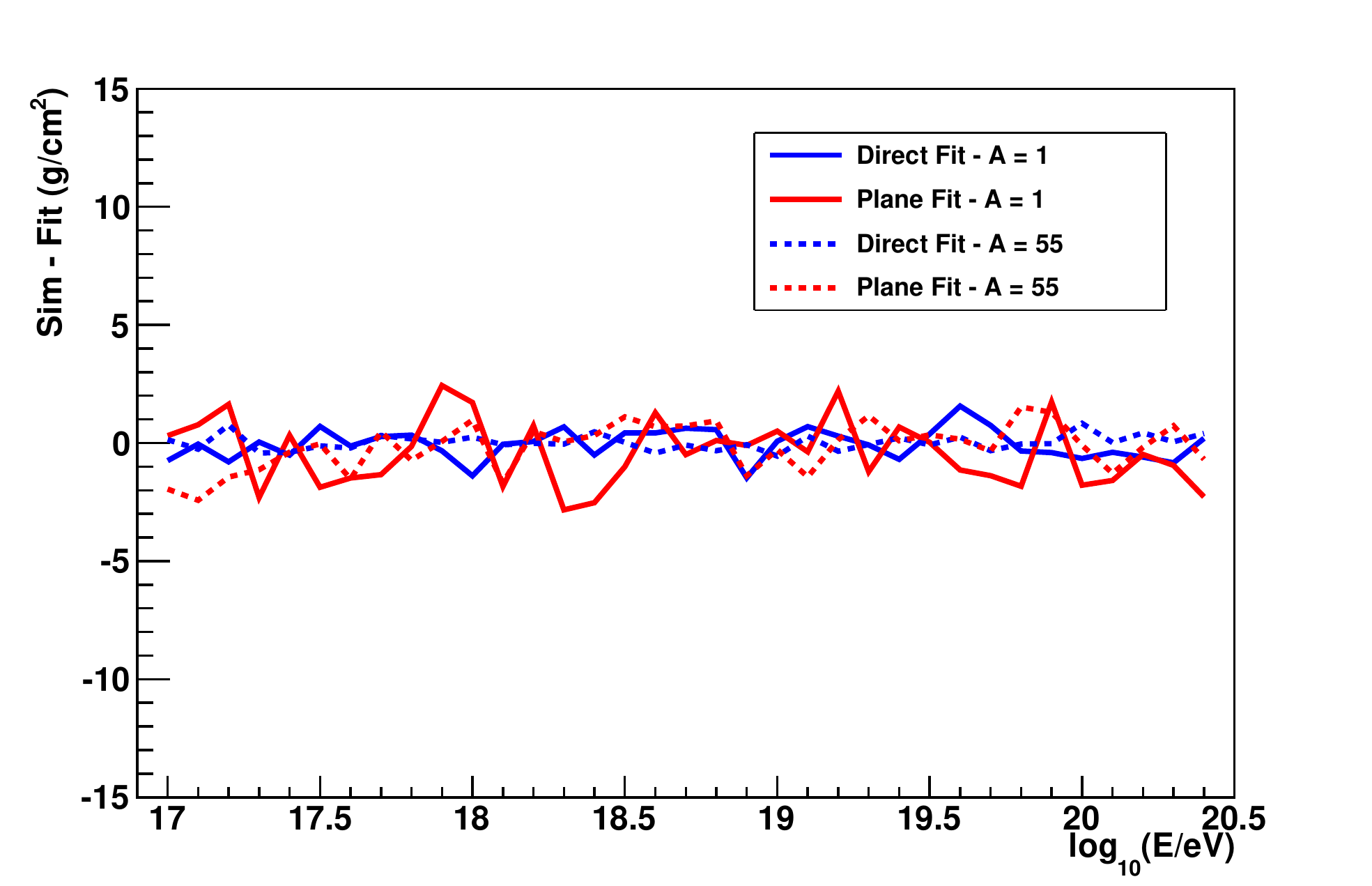}}
  \subfloat[\Conex - \Qgsjet]{\includegraphics[width=0.5\textwidth]{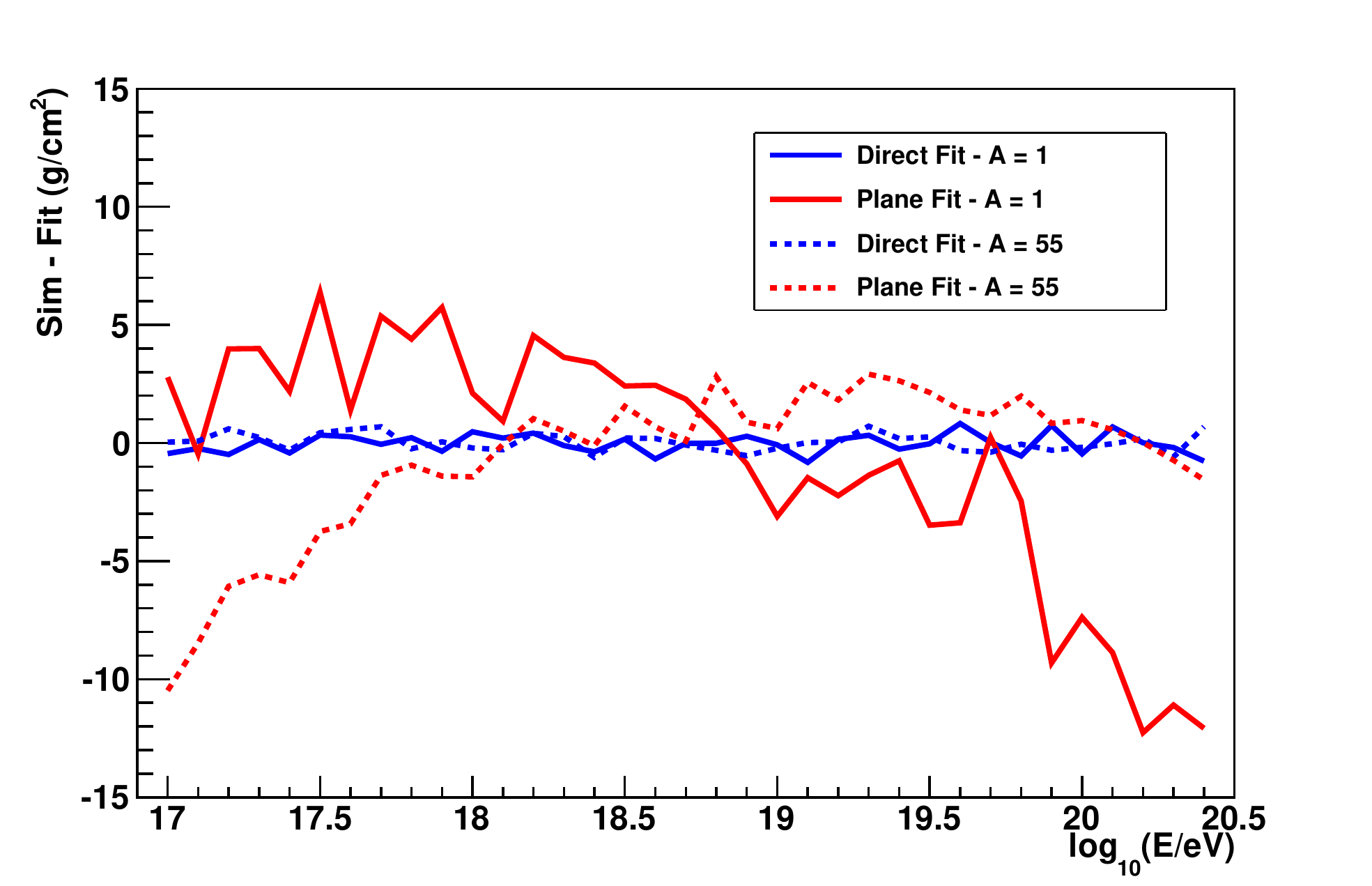}} \\
  \subfloat[\Corsika - \Sibyll]{\includegraphics[width=0.5\textwidth]{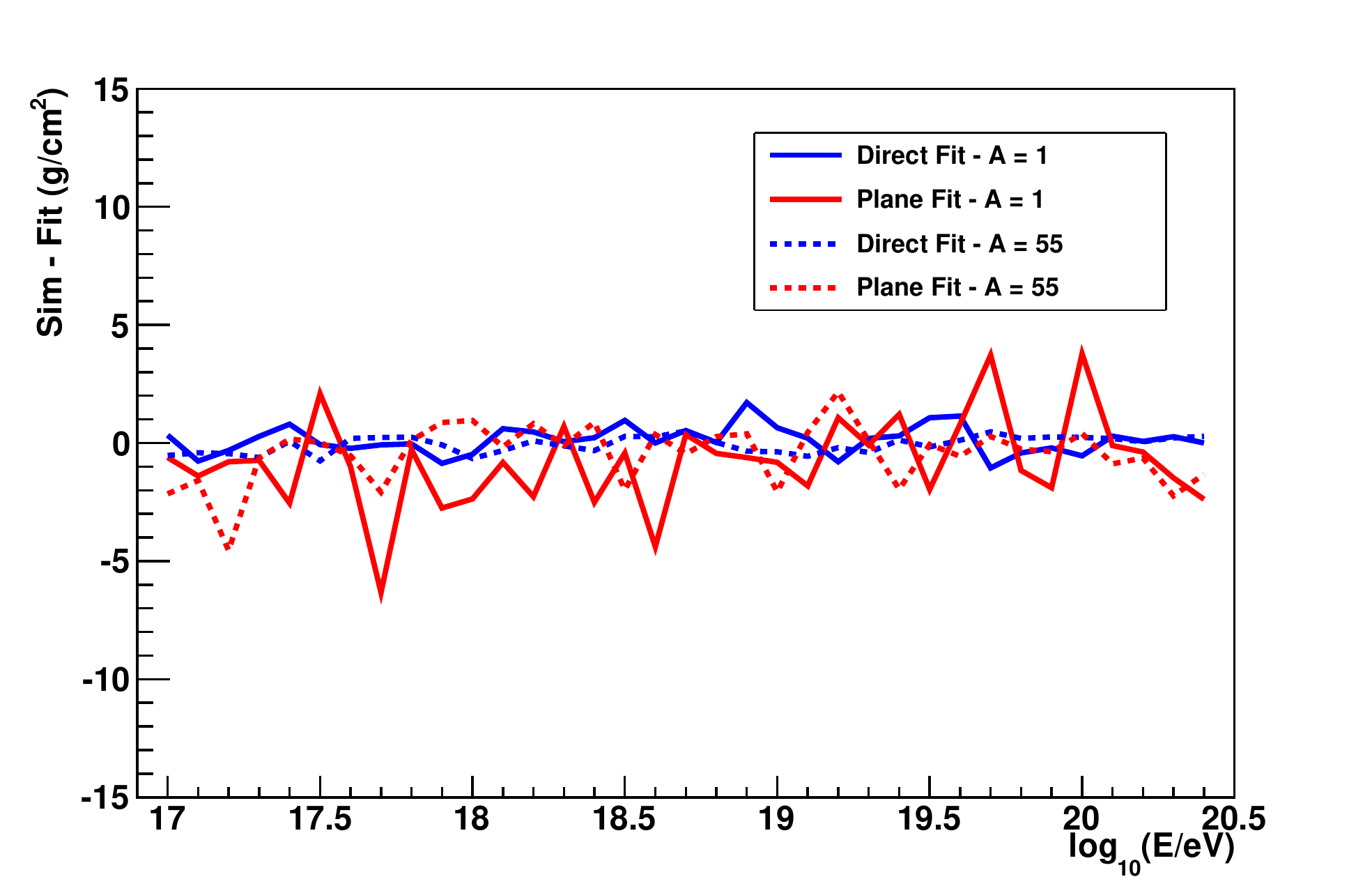}}
  \subfloat[\Corsika -  \Qgsjet]{\includegraphics[width=0.5\textwidth]{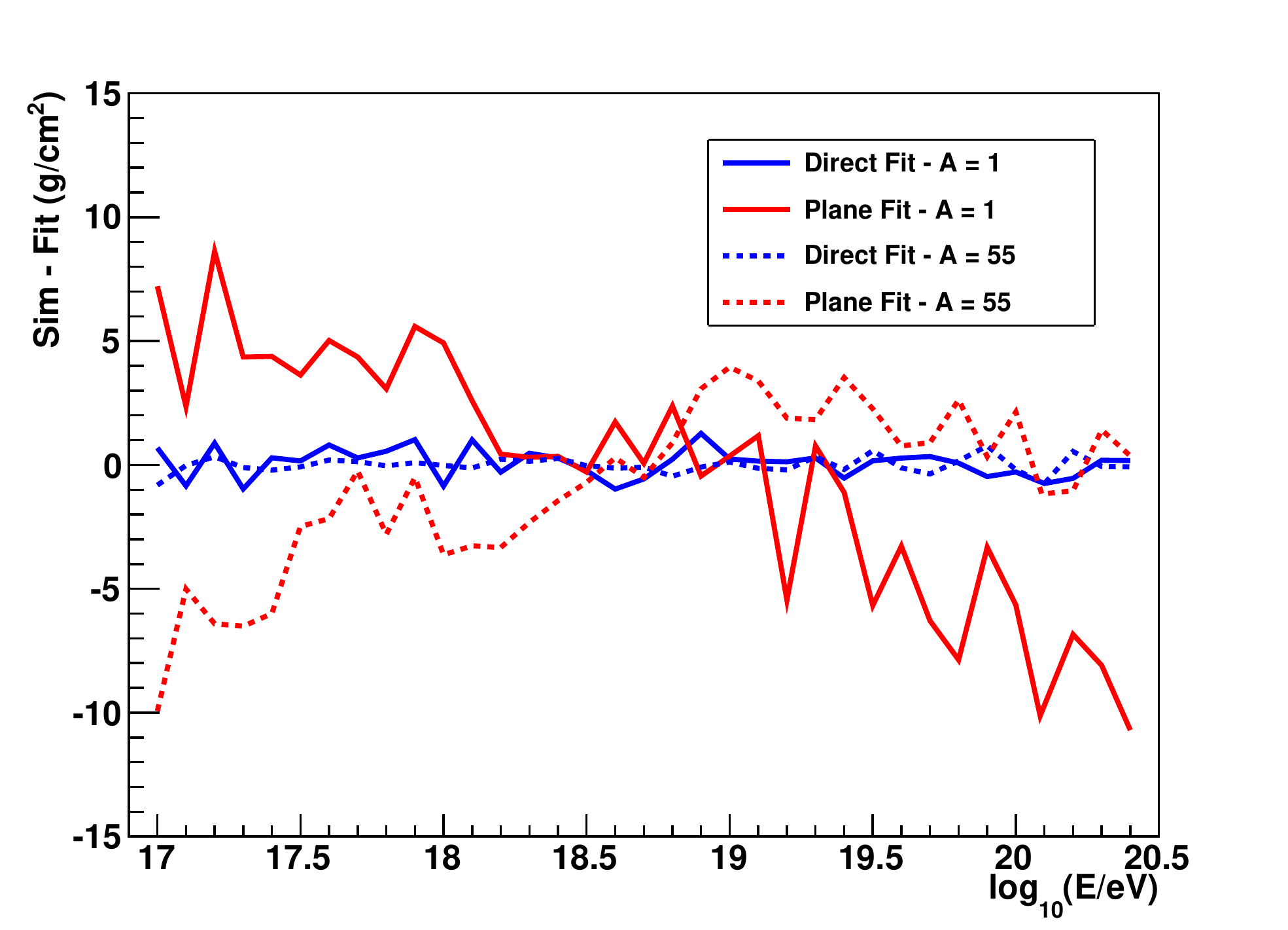}} \\
  \caption{Differences in \meanXmax \emph{versus} energy. Comparison between the simulation, the direct fit of the
    \Xmax distribution using equation ~\ref{eq:gauss:expo} and the
    calculation using equation~\ref{eq:fit:surfaces} and table~\ref{tab:par:xmax:conex}}
    \label{fig:xmax:comp:mean}
\end{figure}

\begin{figure}
  \centering
  \subfloat[\Conex - \Sibyll]{\includegraphics[width=0.5\textwidth]{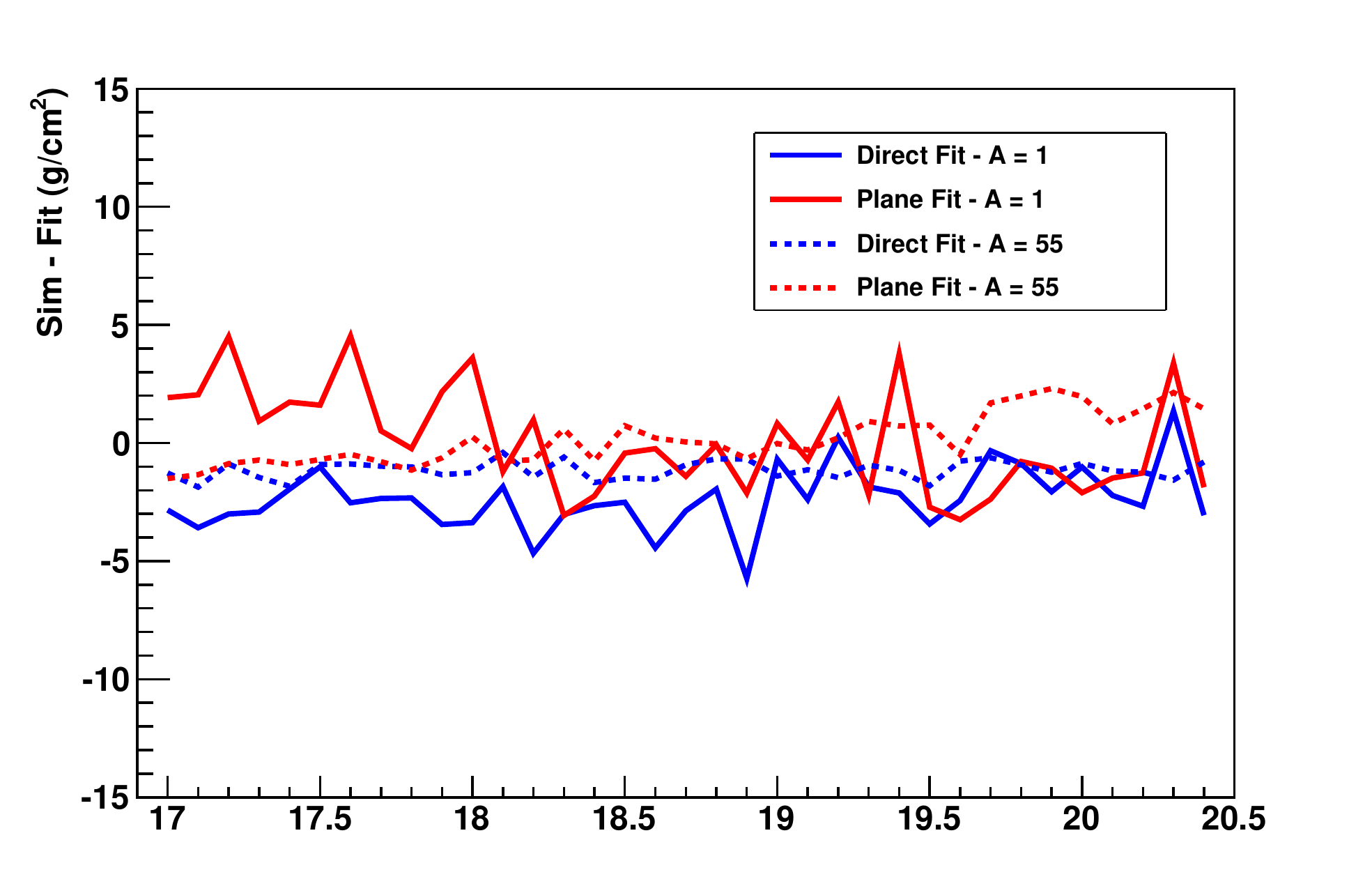}}
  \subfloat[\Conex - \Qgsjet]{\includegraphics[width=0.5\textwidth]{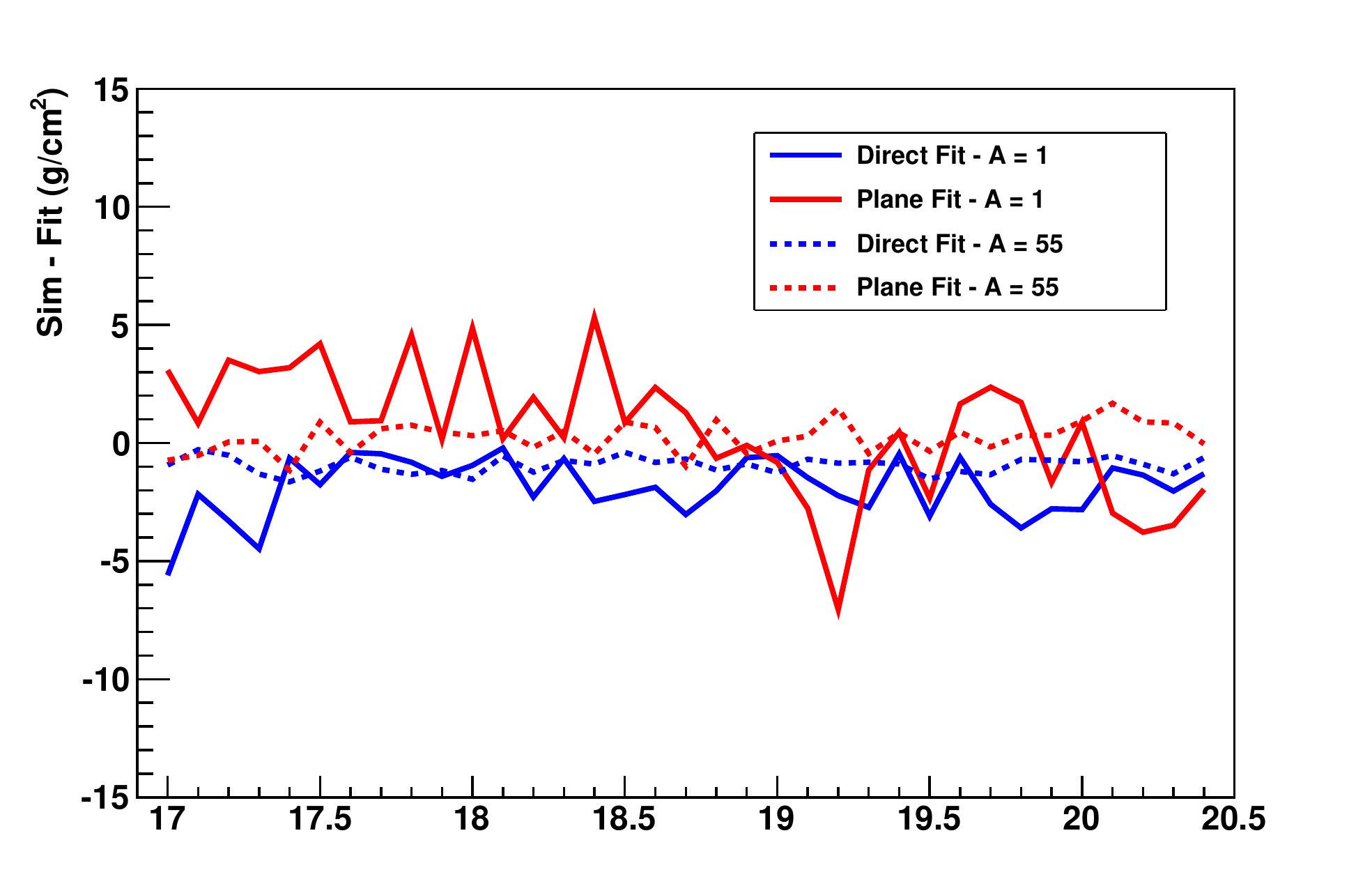}} \\
  \subfloat[\Corsika - \Sibyll]{\includegraphics[width=0.5\textwidth]{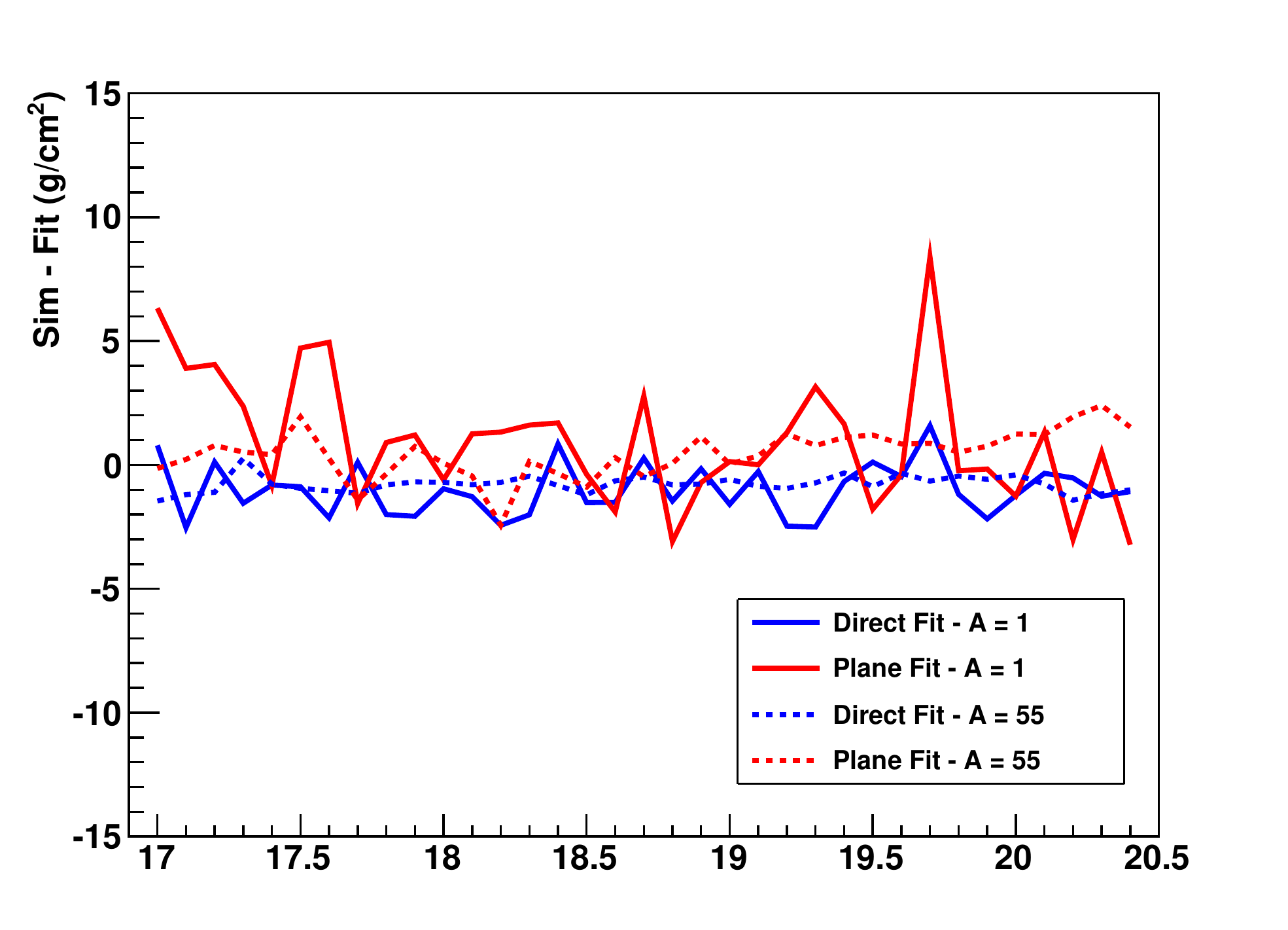}}
  \subfloat[\Corsika -  \Qgsjet]{\includegraphics[width=0.5\textwidth]{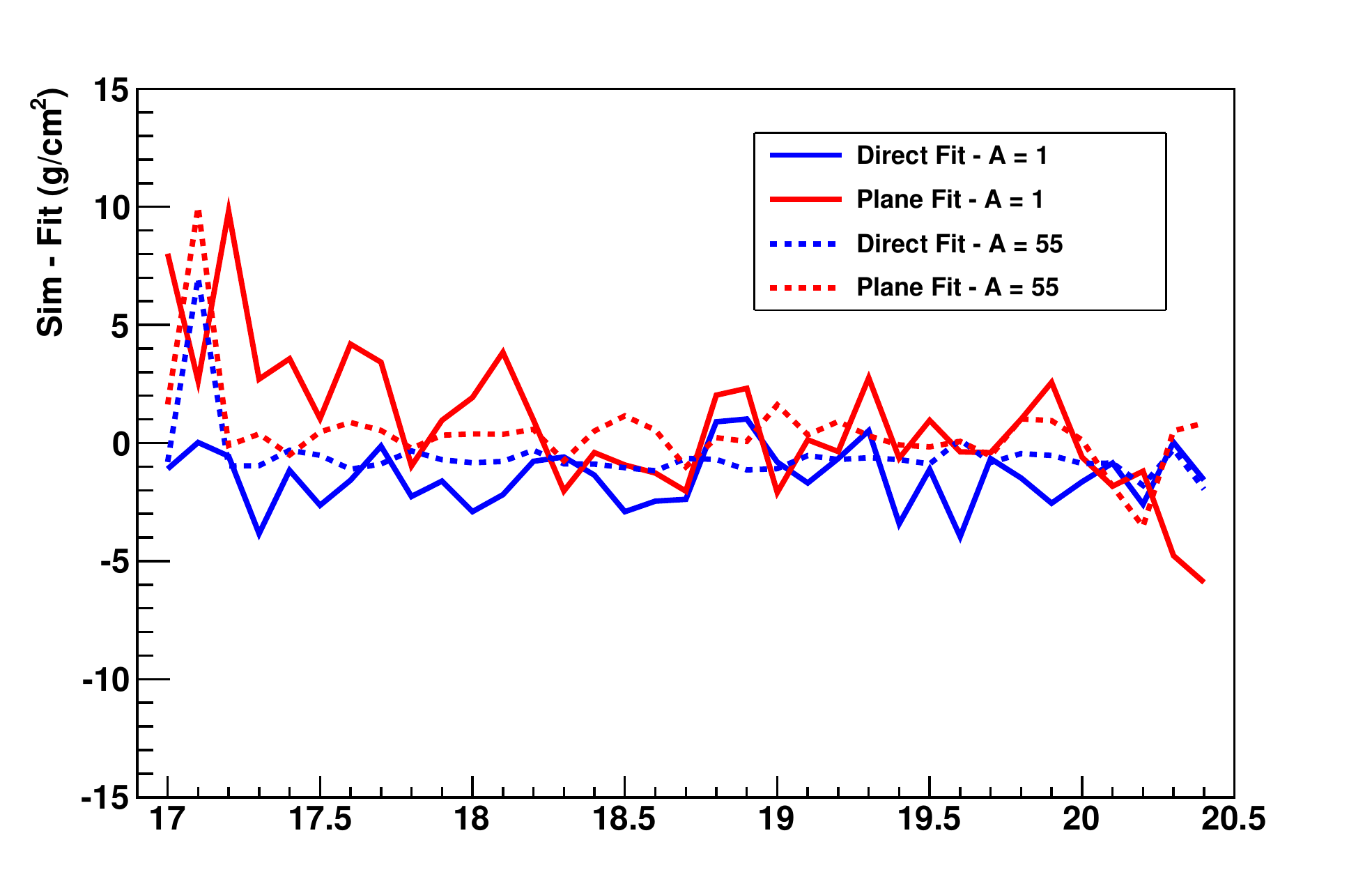}} \\
  \caption{Differences in \sigmaXmax \emph{versus} energy. Comparison between the simulation, the direct fit of the
    \Xmax distribution using equation ~\ref{eq:gauss:expo} and the
    calculation using equation~\ref{eq:fit:surfaces} and table~\ref{tab:par:xmax:conex}}
    \label{fig:xmax:comp:rms}
\end{figure}

Tables~\ref{tab:par:xmax:conex} and \ref{tab:par:xmax:corsika} show
the fitted parameters for \Conex and \Corsika respectively. Despite
the fluctuations of  $\sigma$ and $\lambda$ a linear fit in $log_{10}(A)$
and $log_{10}(E)$ is reasonably good approximation to describe the \Xmax
distribution.
This can be seen in figures~\ref{fig:xmax:comp:mean}
and~\ref{fig:xmax:comp:rms} where we show a comparison between the
simulation, the direct fit of the \Xmax distribution using
equation~\ref{eq:gauss:expo} and the calculation using
equation~\ref{eq:fit:surfaces} and table~\ref{tab:par:xmax:conex}.

It is clear that a direct fit of
the \Xmax distributions with equation~\ref{eq:gauss:expo} (blue lines) leads to a
very good description of the first and second moments of the \Xmax
distribution. For all simulations and hadronic models in the entire
energy range and for all primary particle used in our study the direct
fit resulted in a difference on the simulation smaller than  2 \gcm
for \meanXmax and smaller than 4 \gcm in the \sigmaXmax.

The fit to a plane  $log_{10}(A)$
and $log_{10}(E)$ is a reasonably good approximation to describe the \Xmax
distributions.
This can be seen in figures~\ref{fig:xmax:comp:mean}
and~\ref{fig:xmax:comp:rms} where we show a comparison between the
simulation, the direct fit of the \Xmax distribution using
equation~\ref{eq:gauss:expo} and the calculation using
equation~\ref{eq:fit:surfaces} and table~\ref{tab:par:xmax:conex}.

The parametrization as a function of energy and mass of the parameters
that describe the Xmax distributions (equation \ref{eq:fit:surfaces})
introduced some systematic errors in \meanXmax for the case of \Qgsjet
model (up to 10 \gcm). This is shown with red lines in
figure~\ref{fig:xmax:comp:mean} (right hand side plots). The reason
for this systematic errors is because the parametrization used
(equation \ref{eq:fit:surfaces}) is not the optimum one for \Qgsjet
model.

\begin{table}[!h]
  \begin{center}%\scriptsize
    \begin{tabular}{|c|l|r|r|r|}\hline
      & \multicolumn{1}{c|}{Had. Model}  & \multicolumn{1}{c|}{$C_1$ ($\pm$ err)} & \multicolumn{1}{c|}{$C_2$ ($\pm$ err)} & \multicolumn{1}{c|}{$C_3$ ($\pm$ err)}\\\cline{1-5}
     \multirow{2}{*}{$t_o$}     & \Qgsjet & 53.06 (0.05)  & -28.74 (0.12)  & -275.93 (1.18) \\\cline{2-5}
                                & \Sibyll   & 60.48 (0.07)  & -38.48 (0.13)  & -402.80 (1.22) \\\hline\hline
     \multirow{2}{*}{$\sigma$}  & \Qgsjet & -0.26 (0.06)  &  -5.63 (0.21)  &   31.68 (3.38) \\\cline{2-5}
                                & \Sibyll   & -1.09 (0.07)  &  -5.28 (0.19)  &   44.41 (1.54) \\\hline\hline
     \multirow{2}{*}{$\lambda$} & \Qgsjet & -2.68 (0.14)  & -19.50 (0.43)  &   100.32 (2.63) \\\cline{2-5}
                                & \Sibyll   & -2.61 (0.11)  & -17.89 (0.14)  &  96.28 (1.76) \\\hline\hline
    \end{tabular}\vspace{-0.5cm}
  \end{center}
  \caption{Fitted coefficients (equation~\ref{eq:fit:surfaces})-
    CONEX. All values in \gcm.}
  \label{tab:par:xmax:conex}
\end{table}

\begin{table}[!h]
  \begin{center}%\scriptsize
    \begin{tabular}{|c|l|r|r|r|}\hline
      & \multicolumn{1}{c|}{Had. Model}  &  \multicolumn{1}{c|}{$C_1$ ($\pm$ err)} & \multicolumn{1}{c|}{$C_2$ ($\pm$ err)} & \multicolumn{1}{c|}{$C_3$ ($\pm$ err)} \\\cline{1-5}
     \multirow{2}{*}{$t_o$}     & \Qgsjet & 53.32 (0.30)  & -29.47 (0.52)  & -283.93 (5.62) \\\cline{2-5}
                                & \Sibyll   & 60.77 (0.23)  & -38.88 (0.31)  & -408.88 (4.67) \\\hline\hline
     \multirow{2}{*}{$\sigma$}  & \Qgsjet & 0.06 (0.002)  &  -5.06 (0.17)  &   35.99 (3.21) \\\cline{2-5}
                                & \Sibyll   & -0.56 (0.08)  &  -4.70 (0.21)  &   44.01 (2.03) \\\hline\hline
     \multirow{2}{*}{$\lambda$} & \Qgsjet & -1.73 (0.15)  & -20.63 (0.34)  &   82.69 (3.54) \\\cline{2-5}
                                & \Sibyll   & -2.49 (0.22)  & -19.54 (0.34)  &  96.04 (3.46) \\\hline\hline
    \end{tabular}\vspace{-0.5cm}
  \end{center}
  \caption{Fitted coefficients (equation~\ref{eq:fit:surfaces}) - \Corsika. All values in \gcm.}
  \label{tab:par:xmax:corsika}
\end{table}

\section{Conclusion}
\label{sec:conclusion}

We have studied the simulation programs \Corsika and \Conex with the
hadronic interaction models \Sibyll and \Qgsjet. We have shown that
the \meanXmax and the \sigmaXmax depend slightly  on the combination
of program and hadronic interaction model chosen.  It is widely known
that \meanXmax and \sigmaXmax predicted by \Sibyll and \Qgsjet are
different mainly due to the different extrapolations of the hadronic
interaction properties to the highest energies. We have quantified
here the differences between \Corsika and \Conex by predicting the
\meanXmax and the \sigmaXmax using the same hadronic interaction
model. These differences are small, but should be considered as
systematic uncertainties of the model predictions.

\begin{figure}
  \centering

  \subfloat[\meanXmax.]{\includegraphics[width=0.5\textwidth]{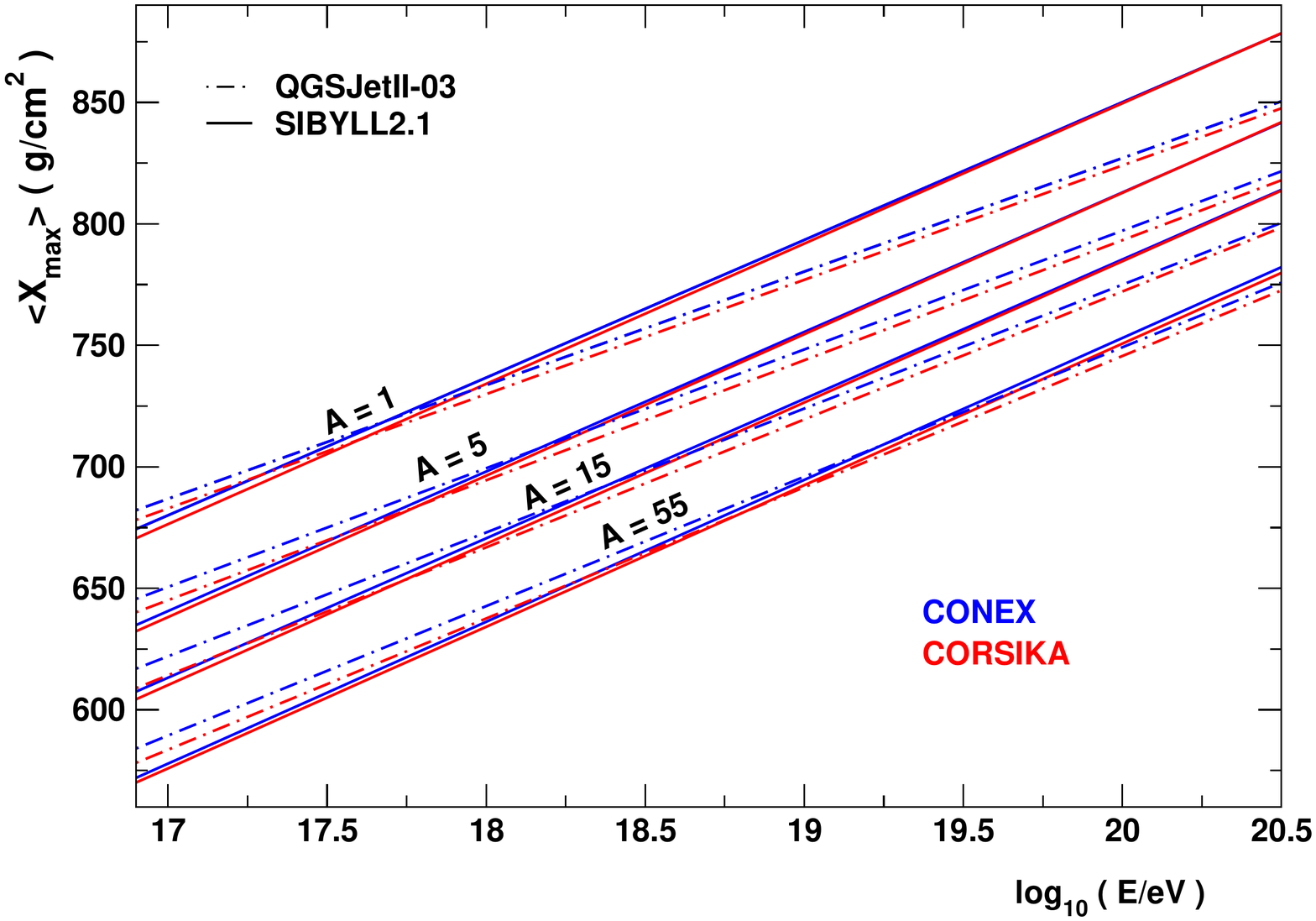}}
  \subfloat[\sigmaXmax.]{\includegraphics[width=0.5\textwidth]{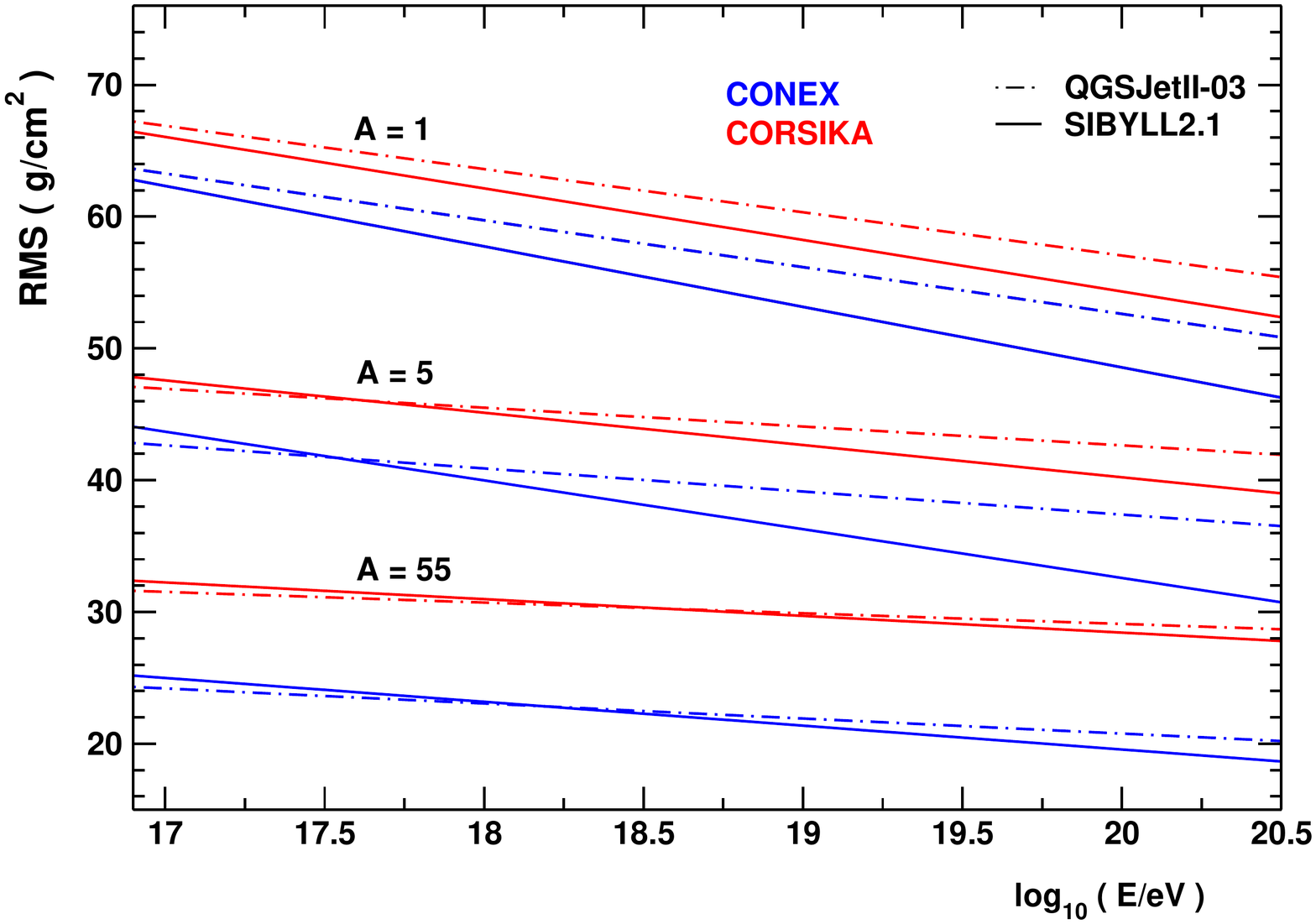}}\\
  \caption{Comparison of \meanXmax and \sigmaXmax for \Corsika and
    \Conex as a function of energy.}
  \label{fig:auger:compare}

\end{figure}

Figure ~\ref{fig:auger:compare} shows the evolution of the \meanXmax
and \sigmaXmax with energy.  No clear dependency of the difference
between \Corsika and \Conex with energy or primary particle type was
seen. When using \Qgsjet or \Sibyll, \Corsika and \Conex  predict the
\meanXmax with a difference smaller than 7 \gcm, and the \sigmaXmax
with a difference smaller than 5 \gcm. The differences in the slopes
of a linear fit to the evolution of the \meanXmax and \sigmaXmax with
energy for \Corsika and \Conex are quite small ($ < 3$ \%).

No assumption is made here for the cause of these differences. An
investigation for the possible cause could be done, but in the
meanwhile these differences between the programs should be considered
as systematics error in the analysis of the \meanXmax and \sigmaXmax
when one tries to infer the composition abundance. Table
\ref{tab:syst} shows the suggested systematic uncertainties for the
model predictions. Maximum values for the systematic
uncertainties can be extracted from figures
\ref{fig:mean:diff:sibyll:energy}, \ref{fig:mean:diff:qgsjet:energy},
\ref{fig:rms:diff:sibyll:energy} and \ref{fig:rms:diff:qgsjet:energy}.

\begin{table}[!h]
  \begin{center}\scriptsize
    \begin{tabular}{|c|c|c|c|c|}\hline
     \multirow{3}{*}{Hadronic Model} & \multirow{3}{*}{Mass (A)}  &  \multicolumn{3}{c|}{systematic uncertainty suggested for the model predictions} \\\cline{3-5}
                                &    &  \multicolumn{1}{c|}{\meanXmax} & \multicolumn{1}{c|}{\sigmaXmax}  & \multicolumn{1}{c|}{Elongation rate} \\
                                &    &     \multicolumn{1}{c|}{\gcm}   &    \multicolumn{1}{c|}{\gcm}     & \multicolumn{1}{c|}{\gcm per energy decade} \\\hline\hline
     \multirow{2}{*}{\Sibyll}   & 1 &  $2.57 - 1.05\times log_{10}(E/EeV)$  &  $-4.58  - 0.66\times log_{10}(E/EeV)$  &  $< 1.30$ \\\cline{2-5}
                                & 55   &  $ 2.57 - 0.09\times log_{10}(E/EeV)$  &  $-7.78 - 0.52\times log_{10}(E/EeV)$  &  $< 0.08$ \\\hline\hline
     \multirow{2}{*}{\Qgsjet}   & 1 &  $ 3.73 - 0.31\times log_{10}(E/EeV)$  &  $-3.82  - 0.29\times log_{10}(E/EeV)$  &  $< 0.20$ \\\cline{2-5}
                                & 55   &  $5.13 -  0.70\times log_{10}(E/EeV)$  & $-7.38  - 0.30\times log_{10} (E/EeV)$  &  $< 0.60$ \\\hline\hline
    \end{tabular}\vspace{-0.5cm}
  \end{center}
  \caption{Systematic uncertainties suggested for the model predictions.}
  \label{tab:syst}
\end{table}

\begin{figure}
  \centering

  \subfloat[Allard et al. - \Conex - \Sibyll.]{\includegraphics[width=0.4\textwidth]{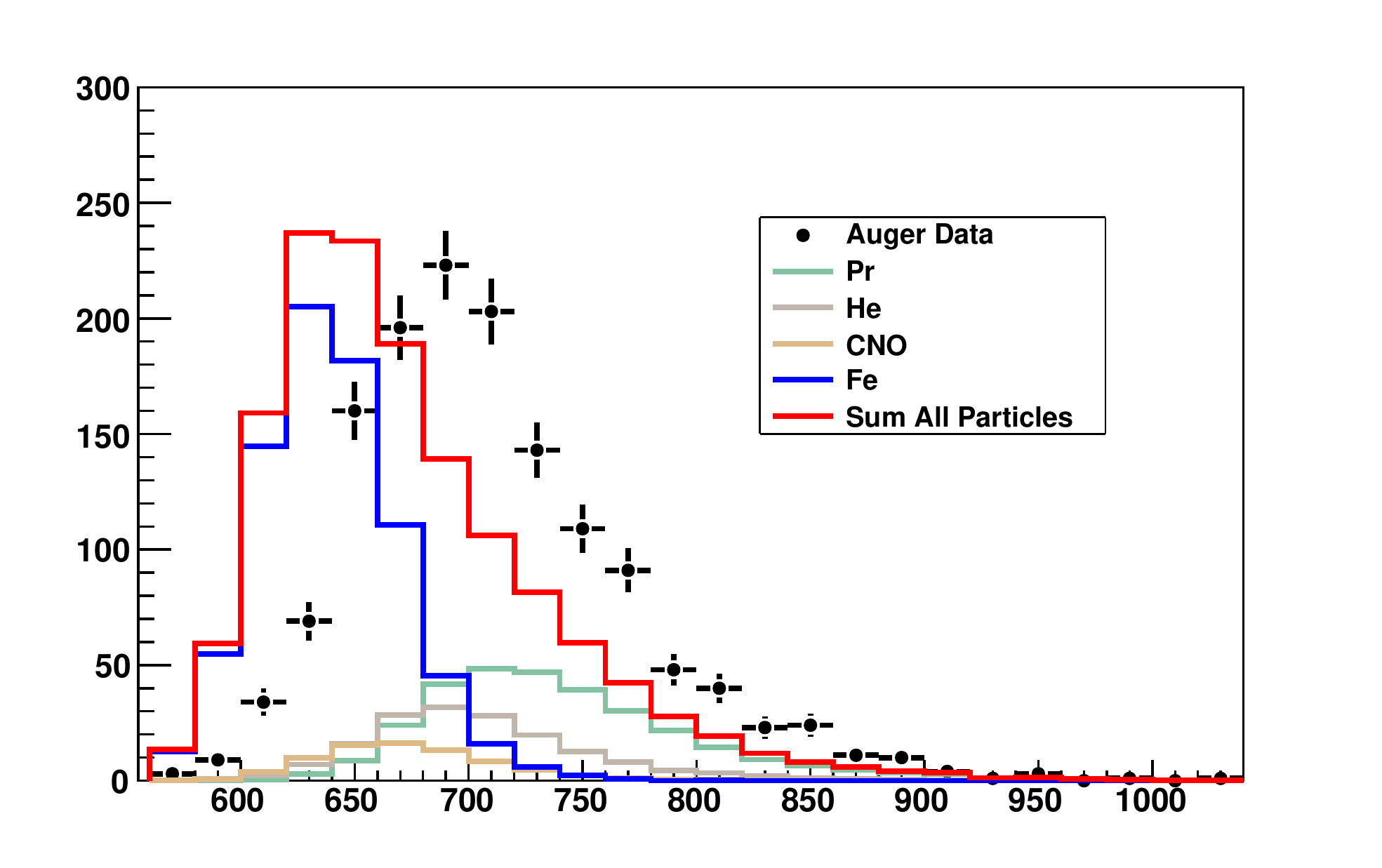}}
  \subfloat[Berezinsky et al. - \Conex - \Sibyll.]{\includegraphics[width=0.4\textwidth]{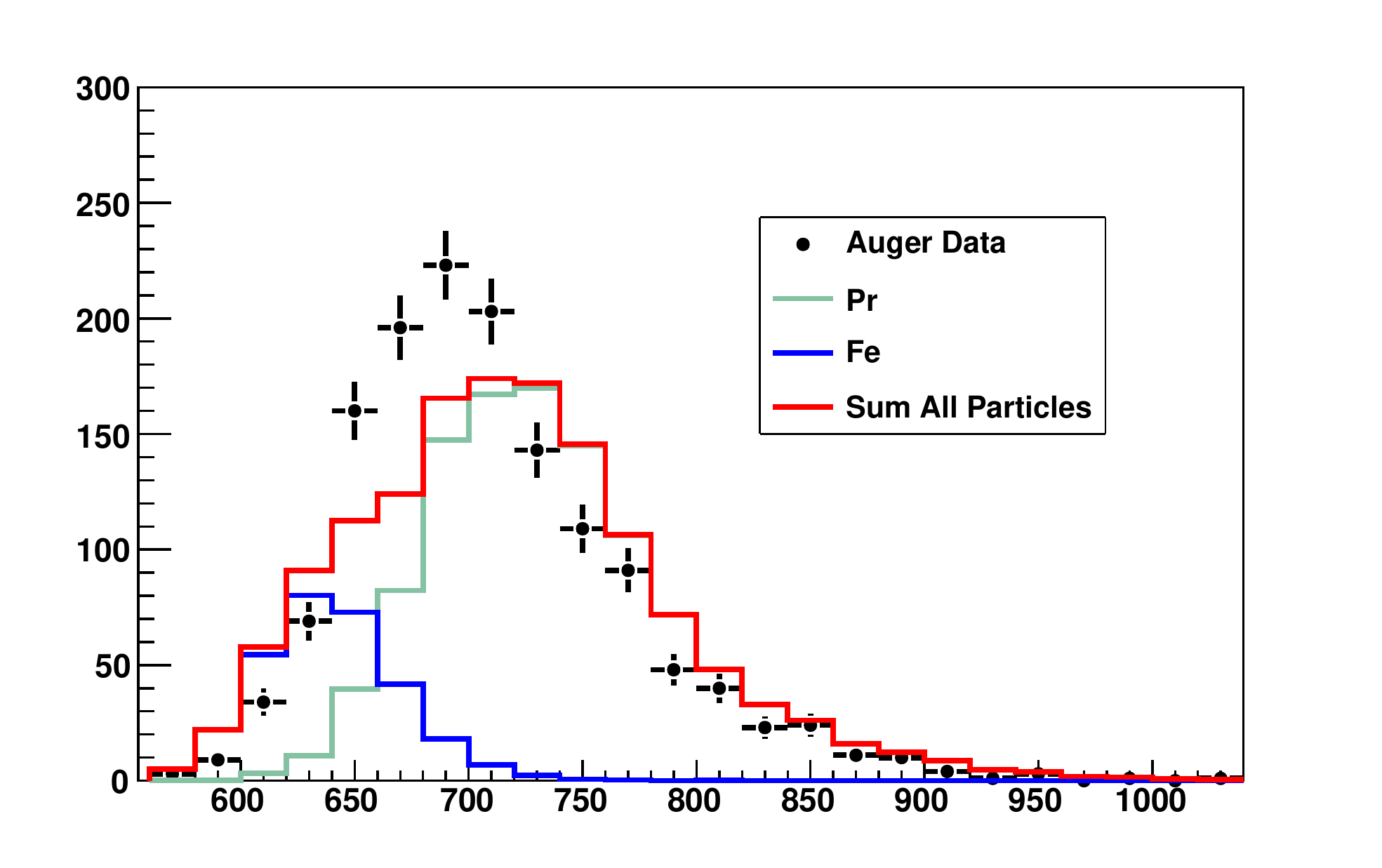}}\\
  \subfloat[Allard et al. - \Corsika - \Qgsjet.]{\includegraphics[width=0.4\textwidth]{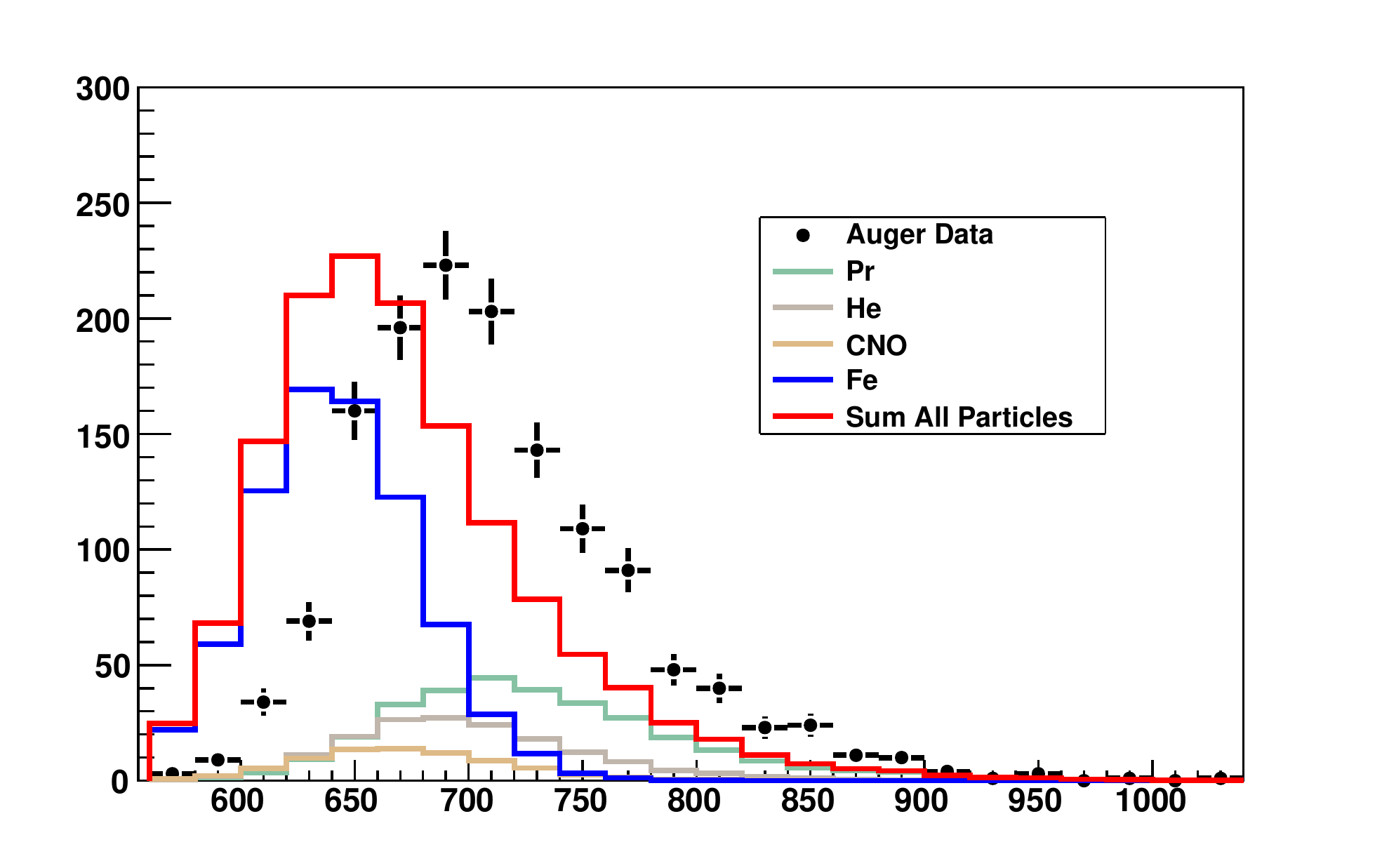}}
  \subfloat[Berezinsky et al. - \Corsika - \Qgsjet.]{\includegraphics[width=0.4\textwidth]{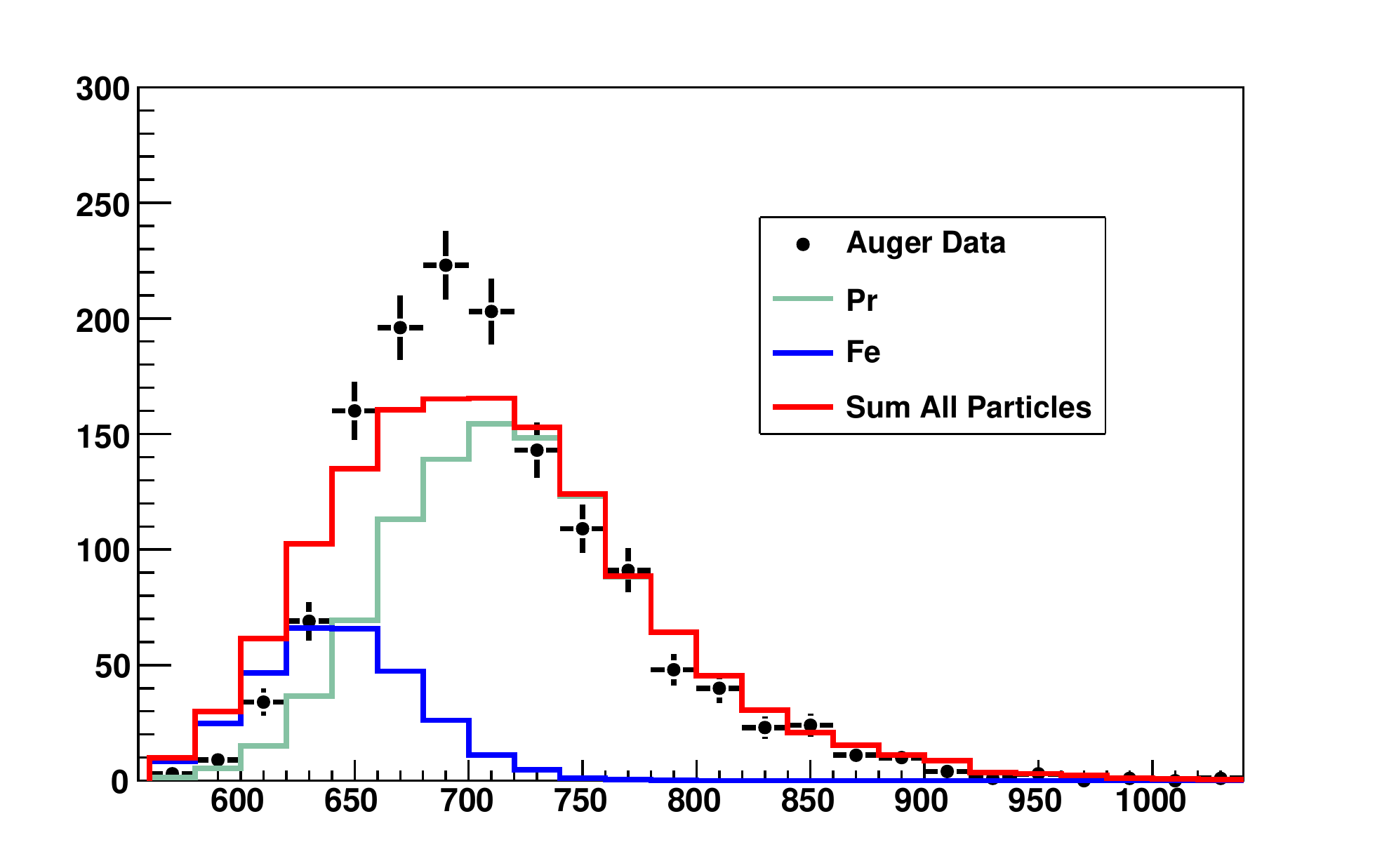}}\\
  \caption{\Xmax distributions. Data measured by the Pierre Auger
    Observatory with energy $10^{18.0} < E < 10^{18.1}$
    eV~\cite{bib:auger:xmax:icrc:2011}. Astrophysical models extracted
    from~\cite{bib:berezinsky,bib:allard}. The models have been
    calculated at $E = 10^{18.05}$ eV. The curves have been
      calculated using the parametrizations proposed above.}
  \label{fig:xmax:models}

\end{figure}

Section~\ref{sec:par:xmax} shows the parametrization of the \Xmax
distributions as a function of energy and mass. The curves shown there
can be used to estimate the first and second moments of the \Xmax distribution from abundance
calculations based on astrophysical arguments. As an example of the
usage of this parametrization we have taken the astrophysical models
developed by Berezinsky et al.~\cite{Berezinsky2004617} (Model 3) and Allard et
al.~\cite{bib:allard} (Model A) and used our paremetrization to
transform the abundance curves predicted by the models into a \Xmax
distribution. Figure~\ref{fig:xmax:models} shows a \Xmax
distributions predicted by the models in comparison to the data
measured by the Pierre Auger
Observatory~\cite{bib:auger:xmax:icrc:2011}. We have convolved the
model predictions with a Gaussian detector resolution of 20 \gcm.
The utility of the parametrization is such that the models can be
compared to the \Xmax distribution instead of only the \meanXmax and
\sigmaXmax.

\section{Acknowledgments}
We thank the financial support given by FAPESP(2008/04259-0,
2010/07359-6) and CNPq. We thank S.Ostapchenko, D. Heck, M. Unger,
A. Bueno Villar, R. Engel, P. Gouffon, R. Clay and T. Pierog for reading this manuscript
and sending us relevant suggestions.

\bibliographystyle{elsarticle-num}
\bibliography{artigo}

% THINNING ANALYSIS
\newpage

%=====================================================

%=====================================================
% ATOMIC WEIGHT ANALYSIS
\newpage

%=====================================================
%=====================================================
% MEAN - SIBYLL

\newpage

%=====================================================

%=====================================================
% MEAN - QGSJET

\newpage

%=====================================================

\newpage
% RMS - SIBYLL

%=====================================================
% RMS - QGSJETII
\newpage

%===============================================================
% MEAN - RMS

%===============================================================
%SLOPE - ANALYSIS

\newpage

%===============================================================
%PARAMETRIZATION

%================================================================================================
%EVOLUTION OF THE DIFFERENCES
\newpage

%================================================================================================
%XMAX DISTRIBUTIONS
\newpage

\newpage

\end{document}